\documentclass[a4paper,fleqn,usenatbib]{mnras}
\usepackage{newtxtext,newtxmath}
\usepackage[T1]{fontenc}
\usepackage{ae,aecompl}
\usepackage{graphicx}	
\usepackage{color}	
\usepackage{amsmath}	
\title[Symbiotic stars and the GALAH survey. I.]{The GALAH Survey and Symbiotic Stars. I.
Discovery and follow-up of 33 candidate accreting-only systems}
\author[Ulisse Munari et al.]
{U. Munari,$^{1}$\thanks{E-mail: ulisse.munari@inaf.it}
G. Traven,$^{2}$
N. Masetti,$^{3,4}$
P. Valisa,$^{5}$ 
G.-L. Righetti,$^{5}$
F.-J. Hambsch,$^{5}$ 
\newauthor
A. Frigo,$^{5}$ 
K. {\v C}otar,$^{6}$
G.~M.~De~Silva,$^{7,8}$
K.~C.~Freeman,$^{9}$
G.~F.~Lewis,$^{10}$
S.~L.~Martell,$^{11,12}$
\newauthor
S.~Sharma,$^{10,12}$
J.~D.~Simpson,$^{11,12}$
Y.-S. Ting,$^{13}$
R.~A. Wittenmyer$^{14}$
D.~B.~Zucker,$^{8,12,15}$
\\
$^{1}$INAF Astronomical Observatory of Padova, 36012 Asiago (VI), Italy\\
$^{2}$Lund Observatory, Department of Astronomy and Theoretical Physics, Box 43, SE-221 00 Lund, Sweden\\
$^{3}$INAF Osservatorio di Astrofisica e Scienza dello Spazio, via Gobetti 93/3, 40129 Bologna, Italy\\
$^{4}$Departamento de Ciencias F\'{i}sicas, Universidad Andr\'{e}s Bello, Fern\'{a}ndez Concha 700, Las Condes, Santiago, Chile\\
$^{5}$ANS Collaboration, c/o Astronomical Observatory, 36012 Asiago (VI), Italy\\
$^{6}$Faculty of Mathematics and Physics, University of Ljubljana, Jadranska 19, 1000 Ljubljana, Slovenia\\
$^{7}$Australian Astronomical Optics, Macquarie University 105 Delhi Rd, North Ryde, NSW 2113, Australia\\
$^{8}$Macquarie University Research Centre for Astronomy, Astrophysics \& Astrophotonics, Sydney, NSW 2109, Australia\\
$^{9}$Research School of Astronomy \& Astrophysics, Australian National University, ACT 2611, Australia\\
$^{10}$Sydney Institute for Astronomy, School of Physics, A28, The University of Sydney, NSW 2006, Australia\\
$^{11}$School of Physics, UNSW, Sydney, NSW 2052, Australia\\
$^{12}$Centre of Excellence for Astrophysics in Three Dimensions (ASTRO-3D), Australia\\
$^{13}$Department of Astrophysical Sciences, Princeton University, Princeton, NJ 08544, USA\\
$^{14}$University of Southern Queensland, Centre for Astrophysics, USQ Toowoomba, QLD 4350 Australia\\
$^{15}$Department of Physics and Astronomy, Macquarie University, Sydney, NSW 2109, Australia\\
}

\date{Accepted XXX. Received YYY; in original form ZZZ}

\pubyear{2020}

\begin{document}
\label{firstpage}
\pagerange{\pageref{firstpage}--\pageref{lastpage}}
\maketitle

\begin{abstract}
We have identified a first group of 33 new candidates for symbiotic
stars (SySt) of the accreting-only variety among the 600\,255 stars so far
observed by the GALAH high-resolution spectroscopic survey of the Southern
Hemisphere, more than doubling the number of those previously known.  GALAH
aims to high latitudes and this offers the possibility to sound the Galaxy
for new SySt away from the usual Plane and Bulge hunting regions.  In this
paper we focus on SySt of the M spectral type, showing an H$\alpha$ emission
with a peak in excess of 0.5 above the adjacent continuum level, and not
affected by coherent radial pulsations.  These constraints will be relaxed
in future studies.  The 33 new candidate SySt were subjected to a vast array
of follow-up confirmatory observations (X--ray/UV observations with the {\it
Swift} satellite, search for optical flickering, presence of a near-UV
upturn in ground-based photometric and spectroscopic data, radial velocity
changes suggestive of orbital motion, variability of the emission line
profiles).  According to Gaia $e$DR3 parallaxes, the new SySt are located at
the tip of the Giant Branch, sharing the same distribution in M(K$_S$) of
the well established SySt.  The accretion luminosities of the new SySt are
in the range 1$-$10~L$_\odot$, corresponding to mass-accretion rates of
0.1-1\,10$^{-9}$~M$_\odot$~yr$^{-1}$ for WDs of 1~M$_\odot$.  The M giant of 
one of the new SySt presents a large Lithium over-abundance.
\end{abstract}

\begin{keywords}
binaries: symbiotic stars -- methods: data analysis -- Galaxy: stellar content
\end{keywords}

\section{Introduction} \label{sec:introduction}

Symbiotic stars (SySt) are interacting binaries where a red giant (RG) fuels
a white dwarf (WD) or a neutron star (NS) companion via accretion (either
through Roche-lobe overflow or wind intercept).  Systems harboring NSs are a
recent addition to this class of celestial objects
\citep{2006A&A...453..295M, 2007A&A...464..277M, masetti2007b,
2011A&A...534A..89M, 2018A&A...613A..22B, 2019MNRAS.485..851Y} and
constitute a few percent of the known total \citep{2019AN....340..598M}, the
vast majority containing WDs \citep{2016MNRAS.461L...1M,
2017arXiv170205898S, 2019ApJS..240...21A}. For some time, a main
sequence star accreting from the RG at very high rates (10$^{-6}$ to
10$^{-4}$ M$_\odot$\,yr$^{-1}$) was considered a viable scenario for SySt in
general \citep{1984ApJ...279..252K}, but this has been progressively
abandoned in the light of expanding observational data, especially in the
far UV and X-rays \citep{2006ApJ...636.1002S, 2006A&A...453..279S,
2012BaltA..21...88M, 2013A&A...559A...6L, 2015A&A...573A...8S,
2020NewAR..9101547L}.  Even if we will focus on RG+WD systems, the results
of this paper are independent on the actual nature of the accreting star.

The RG+WD symbiotic stars are broadly divided into two major groups (see the
recent review by \citealt{2019MNRAS.488.5536M} for details): those {\it
accreting-only} ({\it acc}-SySt) whose optical spectra are dominated by the
RG with no or weak emission lines, and the {\it burning-type} ({\it
burn}-SySt) displaying a strong nebular continuum and a rich emission line
spectrum: they originate from the wind of the RG, which is largely ionized
by the very hot and luminous WD undergoing surface nuclear burning of
accreted material.

SySt are believed to spend most of their time in the accreting-only phase,
quietly accumulating material on the surface of the WD.  When enough has
been piled up, nuclear burning ignites.  If the accreted matter is electron
degenerate, the burning proceeds explosively.  It quickly reaches the Fermi
temperature (a matter of minutes, \citealt{Starrfieldw}) and ejects most of
the accreted envelope at high velocity (thousands of km~s$^{-1}$). 
Radio-interferometry has nicely resolved the expansion of ejecta following,
for ex., the 2006 outburst of RS Oph \citep{2006Natur.442..279O} and that of
2010 for V407 Cyg \citep{2020A&A...638A.130G}, the latter being also the
first nova event ever detected in GeV $\gamma$-rays by the Fermi satellite
\citep{2010ATel.2487....1C}.  The resulting nova outburst can repeat on
time-scales as short as years/decades if the WD mass is close to the
Chandrasekhar limit, like in the case for T CrB, RS Oph, V407 Cyg or V3890
Sgr.  If the matter accreted on the WD is instead not electron-degenerate,
the nuclear burning proceeds in thermal equilibrium, and no mass is ejected. 
The event takes a few years to reach peak brightness
\citep{1982ApJ...257..767F}, and requires many decades to a few centuries to
burn the accreted envelope and let the system return to low luminosity. 
Some of the best examples of this type are AG Peg, HM Sge, V1016 Cyg and
V4368 Sgr.

There is a clear disproportion among catalogued SySt in favor of the {\it
burn}-SySt type, and the known examples of the {\it acc}-SySt variety are
believed to be just the tip of the iceberg (\citealt{2016MNRAS.461L...1M},
hereafter Mk16).  Between 15 and 20 are currently known of the {\it
acc}-SySt type.  Symbiotic stars were originally proposed by
\citet{1992ApJ...397L..87M} as possible progenitors of type Ia supernovae,
and the viability of this single-degenerate channel obviously relies on the
total number of SySt spread throughout the Galaxy, which in turn heavily
depends on the number of {\it acc}-SySt.  The latter have been usually
discovered as counterparts of satellite UV/X-ray sources, while the {\it
burn}-SySt can be easily spotted at optical wavelengths through the whole
Galaxy and the Local Group thanks to their outstanding emission-line
spectrum (with ionization up to [FeX] and beyond).

The subtle way the optically-quiet {\it acc}-SySt have been usually
discovered is well epitomized by SU~Lyn, a $V$$\sim$8 mag, M6III giant at
650$\pm$35 pc distance \citep{2018A&A...616A...1G}.  For decades it was a
completely unnoticed field star, with just an old report about semi-regular
photometric variability which granted it a variable star name
\citep{1955AN....282...73K}.  While looking for optical counterparts of hard
X-ray sources newly discovered by the {\it Swift} satellite, it was noted
that SU~Lyn lay within the error box of one of them: 4PBC J0642.9+5528. 
Follow-up observations were organized with {\it Swift} (to refine the
position and the properties of the X-ray source) and with Asiago telescopes
to investigate if the optical spectra of SU Lyn could betray peculiarities
supporting a physical association with the {\it Swift} hard X--ray source. 
The association was proved by a large flux excess observed at bluest optical
wavelengths ($\lambda$$\leq$4000 \AA) and by the presence on high-resolution
spectra of weak and variable emission in H$\alpha$ and [NeIII] lines (cf. 
Mk16).

The limited sensitivity of current X-ray satellites (soon to change thanks
to $e$-ROSITA sky survey, \citealt{2012arXiv1209.3114M}) restricts the
serendipitous discovery of {\it acc}-SySt to those that lie within $\sim$1
kpc from the Sun, implying that many of their red giants rank among
naked-eye objects (like 4 Dra, HR 1105 or $o$~Cet).  To sample a much larger
fraction of the RG in our Galaxy, we have devised a reverse strategy, and
first exploratory results are presented in this paper: digging through large
spectroscopic surveys in search for the signatures of accretion onto a
companion to the RG, primarily the presence of Balmer emission lines having
a flux and a profile compatible with an origin in an accretion disk around a
degenerate companion.  The candidate {\it acc}-SySt identified this way can
then be subjected to follow-up confirmatory observations, including pointing
with X--ray/UV satellites.  The GALAH survey of the southern hemisphere
\citep{2015MNRAS.449.2604D} offers an ideal hunting opportunity: a large
number (aiming at $>$1 million) of randomly selected southern stars in the
$12 \leq V_{JK} \leq 14$ mag range are observed at high S/N and high
spectral resolution with the AAT 3.9m telescope + fibre-fed spectrograph
HERMES, over four distinct wavelength ranges that include H$\alpha$ and
H$\beta$ lines.

In this first paper we outline the methodologies guiding our search among
GALAH spectra for candidate {\it acc}-SySt, report on the discovery of a
first batch of 33 new candidates (about doubling the total known so far),
and describe the results of a wide range of follow-up observations aiming to
confirm their {\it acc}-SySt nature.  We encourage further observations by
the community in order to validate their classification.

\section{The GALAH survey} \label{sec:galah}

\subsection{Motivation}

The GAlactic Archaeology with HERMES (GALAH, \citealt{2015MNRAS.449.2604D}) is
an ongoing spectroscopic survey whose ambitious goal is to unveil the
formation and history of the the Milky Way.  This is the focus of the field
of Galactic Archaeology, which tries to determine how the fossil remnants of
star-forming regions and effects of ancient mergers paint the picture of our
Galaxy that we observe and measure today.  This complex endeavor can be
accomplished by studying the detailed chemical composition and other
properties of stars in distinct regions of the Milky Way.

The Galactic archaeology community argues that the complete Galaxy formed
gradually over time. This formation history can be traced back by
investigating remnants of initial building blocks or subsequent additions
through mergers, which have been disrupted in the course of evolution and
are now dispersed around the Galaxy.  The theoretical concept of chemical
tagging \citep{2002ARA&A..40..487F} demonstrates that individual galactic
components should have preserved their original chemical signature over
time.  It is therefore essential to disentangle their formation site from
migration history in order to explain the current mixture of stellar
populations.  

GALAH aims to achieve this by measuring the abundance of up to 31 chemical
elements coming from seven independent major groups with different
nucleosynthetic origin: light proton-capture elements Li, C, O;
$\alpha$-elements Mg, Si, Ca, Ti; odd-Z elements Na, Al, K; iron-peak
elements Sc,V, Cr, Mn, Fe, Co, Ni, Cu, Zn; light and heavy slow neutron
capture elements Rb, Sr, Y, Zr, Ba, La; and rapid neutron capture elements
Ru, Ce, Nd and Eu \citep{2015MNRAS.449.2604D}.

\subsection{Instrument}

The goals of the GALAH survey were the main driver for the construction of
the High Efficiency and Resolution Multi-Element Spectrograph (HERMES,
\cite{2010SPIE.7735E..09B,2015JATIS...1c5002S}), a multi-fibre spectrograph
working in tandem with the $3.9$-metre Anglo-Australian Telescope (AAT)
situated at the Siding Spring Observatory, Australia.  The spectrograph has
a resolving power of R $\sim 28,000$ (or R $\sim 45,000$ when slit mask is
used) and records spectra in four separate wavelength ranges given in
Table~\ref{tab:galah_bands}.  The total spectral coverage is therefore
approximately 1000~\AA, including the important diagnostic lines H$\alpha$
and H$\beta$, together with lines of numerous chemical elements that are
necessary to fulfil the ambitions of the GALAH survey.

The AAT uses the Two Degree Field (2dF) robotic positioning system with two
identical plates that are used to precisely position fibres at designated
locations.  This configuration allows HERMES to simultaneously record
spectra from up to 392 fibres distributed over a $2^\circ$ diameter field of
the night sky, with an additional 8 fibres used for the telescope guiding. 
During the exposure with the first plate, the robotic positioner places
fibres on the second plate, where each complete process of fibre allocation
takes about half an hour per plate.  HERMES can typically achieve a signal
to noise ratio (SNR) $\sim100$ per resolution element at magnitude V=14 in
the red arm during a 1-hour long exposure.  To achieve as high SNR as
possible, minimize atmospheric diffraction, and in order not to loose light
due to the fibres' field of view of only 2\arcsec, all observations are
ideally carried out close to the meridian.

\subsection{Selection function}

The main selection function of the GALAH survey is relatively simple: it
avoids the crowded regions of the Galactic plane ($|b|$~>~$10^\circ$),
accounts for the nominal telescope operations ($-80^\circ \leq \delta \leq
+10^\circ$) and restricts observations by apparent magnitude ($12 < V_{JK} <
14$), while being colour independent.  There is an additional requirement
for field selection: the density of stars has to be at least 400 per $\pi$
square degrees to match the number of fibres and field of view of the fibre
positioner.  The input catalogue for observations is based on the Two Micron
All-Sky Survey \citep[2MASS ][]{2006AJ....131.1163S}, with $V_{JK}$
approximating standard $V$ magnitude and being computed as \citep[for more
details see][]{2017MNRAS.465.3203M}:

\begin{equation}
V_{JK} = K + 2(J-K + 0.14) + 0.382 \mathrm{e}^{(J-K - 0.2)/0.5}
\end{equation}

The majority of observations are performed with the above described
selection function.  Additionally, GALAH employs a bright mode ($9 < V_{JK}
< 12$) during twilight or poor observing conditions, and a faint mode ($12 <
V_{JK} < 14.5$) when fields with normal or bright configuration are not
available to be observed.

The spectroscopic data used in this work is further complemented by
observations of the K2-HERMES survey \cite{2018AJ....155...84W} and the
TESS-HERMES survey \cite{2018MNRAS.473.2004S}, which are carried out with
the same observational and data reduction setup as the GALAH survey, albeit
with their peculiar selection functions focusing on K2
\citep{2014PASP..126..398H} and TESS \citep{2015JATIS...1a4003R} targets,
respectively.

Together, the above described observations yield a dataset of 625\,757
successfully reduced stellar spectra, of which a small fraction belongs to
repeated observations of 600\,255 unique stars observed by GALAH at the time
of writing of this paper.  They represent dwarf as well as giant stars,
the former consisting of mostly nearby stars (closer than 2 kpc) while the
latter probing distances as far as the Galactic center.

   \begin{table}
        \centering
        \caption{Short and long wavelength limits (\AA) of the four spectral
        intervals recorded by the GALAH survey.}
        \label{tab:galah_bands}
	\begin{tabular}{lcc}
		\hline
	interval & start & end \\
		\hline
   blue    & 4718 & 4903 \\
   green   & 5649 & 5873 \\
   red     & 6481 & 6739 \\
   far red & 7590 & 7890 \\
		\hline
        \end{tabular}
   \end{table}

    \begin{figure*}
	\includegraphics[angle=270,width=16.8cm]{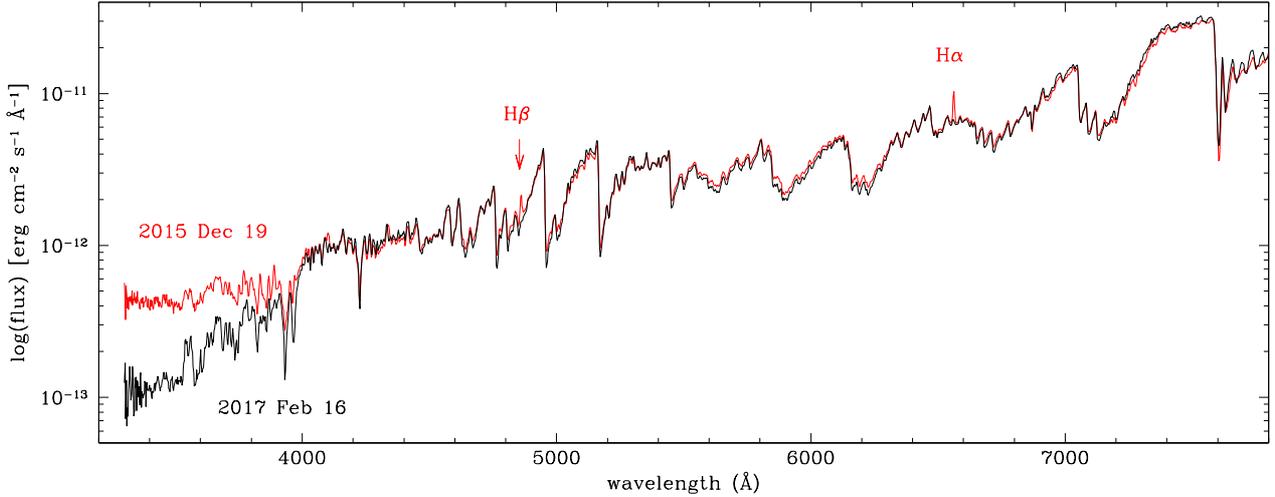}
        \caption{Two spectra of SU Lyn, a prototype of the accreting-only
         symbiotic stars, to illustrate the differences between low (black)
         and high (red) accretion rates (Asiago 1.22m telescope).  A higher
         rate causes a brightening of the accretion disk around the white
         dwarf, which manifests as an excess near-UV brightness
         ($\lambda$$\leq$4000~\AA) and detectable emission lines (primarily
         hydrogen Balmer lines).}
    \label{fig:sulyn_spec}
    \end{figure*}

\subsection{Data reduction and analysis}

Every stellar spectrum recorded by the HERMES spectrograph and used in this
work is homogeneously reduced by an automatic reduction pipeline, thoroughly
described by \cite{2017MNRAS.464.1259K}.  We hereby provide only a brief
explanation of this reduction procedure.

The path from a 2D image recorded in each of the HERMES arms to a
one-dimensional, continuum-normalized and zero-RV-shifted spectrum starts with
the following steps: raw image cosmetic corrections, spectral tracing,
optical aberrations correction, scattered light and apertures cross-talk
removal, wavelength calibration, sky subtraction, and telluric absorption
removal.  Afterwards, spectra are continuum-normalized and shifted into
their rest-frame by cross-correlating them with a set of 15 AMBRE model
spectra \citep{2012A&A...544A.126D}.  This procedure also yields initial
radial velocities, while initial stellar parameters ($T_{\rm eff}$, $\log
g$, $\mathrm{[Fe/H]}$) are determined by a (larger) grid of $16\,783$ AMBRE
spectra. These, internally also called GUESS parameters, were reported in
the first GALAH data release \citep{2017MNRAS.465.3203M}. The
spectra are then directed to the analysis pipeline with the aim of
delivering precise and accurate stellar atmospheric parameters and
individual elemental abundances, which constitute the core of results 
reported in subsequent GALAH data releases.

\subsection{The new SySt probed by GALAH}

Most of the known SySt have been discovered during photographic
objective-prism surveys of the Galactic plane and Bulge (e.g. 
\citealt{1950ApJ...112...72M,1973ApJ...185..899S,1976ApJS...30..491H}), many
of which were initially classified and studied as planetary nebulae.  Their
true nature was generally unveiled by infrared observations catching the
presence of a cool giant in the system
\citep{1982ASSL...95...27A,1984PASAu...5..369A,1987A&AS...67..541S}.  The
same objective-prism technique was applied also to the Magellanic Clouds and
several SySt were discovered there too
\citep{1983MNRAS.203...25W,1992MNRAS.258..639M}.  Searches for SySt in
external galaxies has steadily progressed over the years
\citep{2014MNRAS.438...35A, 2008MNRAS.391L..84G, 2012MNRAS.419..854G,
2015MNRAS.447..993G, 2014MNRAS.444..586M, 2017MNRAS.465.1699M,
2018A&A...618A...3R, 2009MNRAS.395.1121K, 2019IAUS..339..291D}, and the
current total of known extra-galactic SySt amounts to $\sim$70 systems
\citep{2019ApJS..240...21A,2019Ap&SS.364..132C,2019RNAAS...3...28M}.  New
surveys aiming to SySt within the Local Group \citep{2019AJ....157..156A}
promise to increase such numbers, especially for the {\it burn}-SySt type.

Systematic searches for new SySt have also been performed in recent years
through our Galaxy.  The IPHAS survey \citep{2005MNRAS.362..753D} targeted
the plane of the Milky Way for emission-line objects, among which many new
candidate SySt were identified by \citet{2008A&A...480..409C,
2010A&A...509A..41C, 2011A&A...529A..56C} and \citet{2014A&A...567A..49R}. 
SySt were also hunted toward the Bulge by \citet{2013MNRAS.432.3186M} and
\citet{2014MNRAS.440.1410M}.  The current total of confirmed Galactic SySt
is about 250 \citep{2019ApJS..240...21A, 2019RNAAS...3...28M}.
 
No recent survey has extensively probed the Galaxy for SySt toward
directions other than low on the Plane or toward the Bulge.  This is a gap
we intend to fill with the help of the GALAH survey, that is exploring the
Galaxy primarily at high latitudes.  In this way a more complete census of
SySt will be possible toward all stellar populations, allowing a more
informed comparison with the predictions about the total number of SySt
spread throughout the Galaxy.  Such estimates vary over a wide range: from
1,200$-$15,000 of \citet{2006MNRAS.372.1389L}, or 3,000$-$30,000 of
\citet{1995ApJ...447..656Y} and \citet{2010NewA...15..144Z}, to 3,000 of
\citet{1984PASAu...5..369A}, 30,000 of \citet{1993ApJ...407L..81K}, and
300,000 of \citet{1992ApJ...397L..87M}.  The large spread is in part
accounted for by differences in the adopted binary evolution codes and
stellar population synthesis, or by the type of SySt considered (burning or
accreting only), and also by the ratio between SySt currently in an active
state (and therefore detectable to observations) and those dormant (and
indistinguishable from single, field cool giants).  The populations of SySt
containing a NS is generally estimated at 100$-$1,000 (eg. 
\citealp{2012MNRAS.424.2265L}), or about 1\% of those harboring a WD.

    \begin{figure*}
	\includegraphics[width=16.8cm]{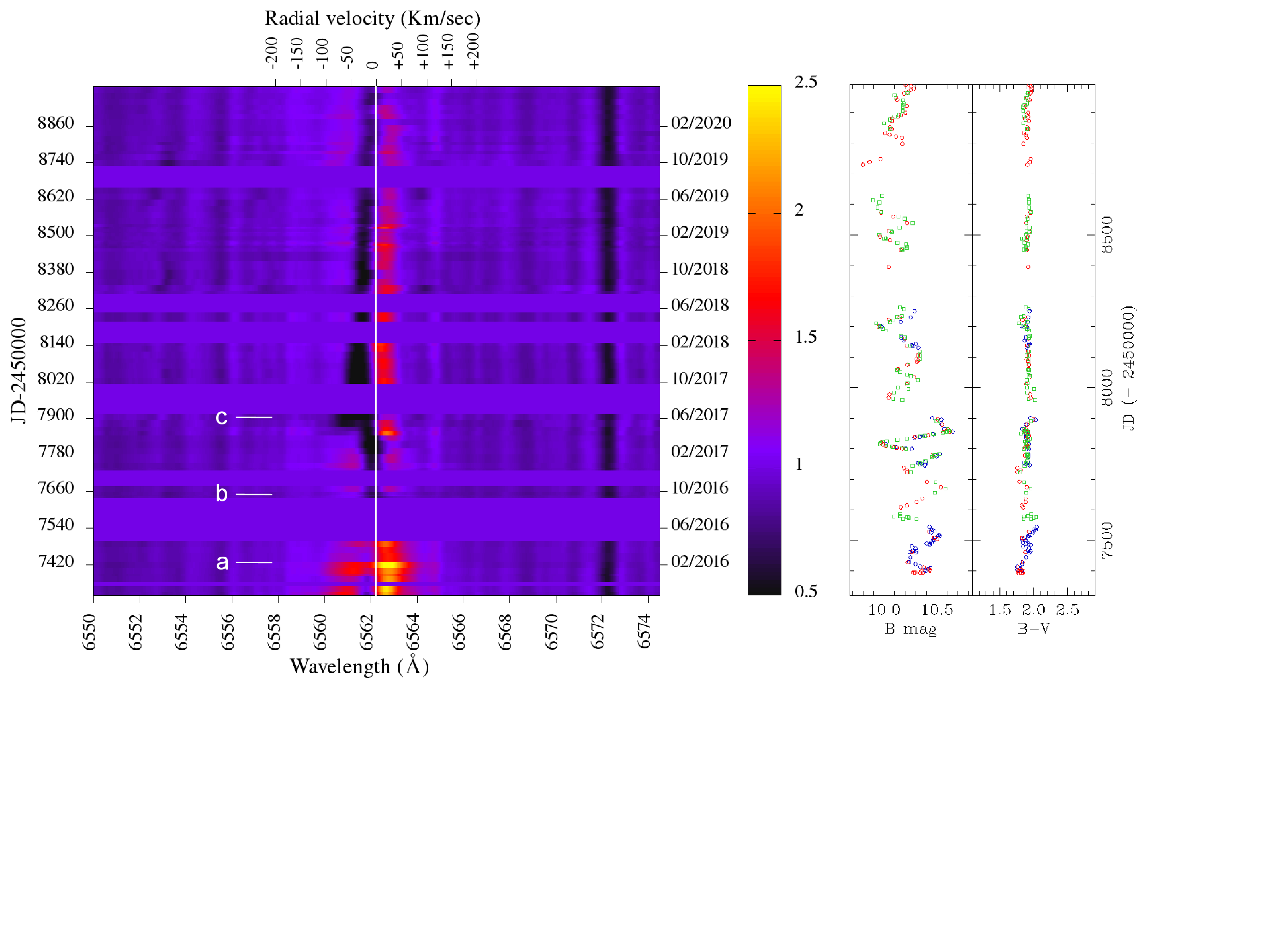}
        \caption{Temporal sequence from Asiago 1.82m and Varese 0.84m
        Echelle spectra (aligned in heliocentric wavelength) illustrating
        the great changes observed in the H$\alpha$ profile of prototype
        SU~Lyn since its recognition in late 2015 as an accreting-only
        symbiotic star.  The vertical line marks the H$\alpha$ rest
        wavelength at the stellar systemic velocity ($-$24~km~s$^{-1}$). 
        The epochs $a$, $b$ and $c$ are discussed in the text
        (Section~\ref{sec:sulyn}).  On the right the corresponding $B$,$V$
        lightcurve of SU Lyn as built from ANS Collaboration observations.}
    \label{fig:sulyn_temp}
    \end{figure*}

\section{SU Lyncis, a path-finder} \label{sec:sulyn}

As remarked earlier, several {\it acc}-SySt have been known for quite a
while.  A few of them present emission lines bright enough to be
recognizable even on objective prism plates or low resolution optical
spectra.  They are well epitomized by EG~And, an M2III giant orbited every
481 days by a WD, whose accretion luminosity is a few 10$^2$~L$_\odot$
\citep{2016ApJ...824...23N}.

The discovery of SU Lyn by Mk16 revealed a ''{\em hidden and potentially
large population}" of {\it acc}-SySt accreting at much lower rates, with
accretion luminosities a couple of orders of magnitude lower than in EG~And,
and therefore not detectable by conventional observations at optical
wavelengths.  This paper aims to specifically discover more of these SySt
characterized by low accretion luminosities (L$_{\rm
acc}$$\sim$1-10~L$_\odot$).  As the properties of SU~Lyn will guide us as a
path-finder throughout this paper, and since no paper has so far reviewed
this object at optical wavelengths, we will start with a quick overview of
spectroscopic and photometric properties of SU Lyn.  These are based on the
very intensive monitoring we are keeping of the object since the preparatory
phase leading to the Mk16 discovery paper, and whose results will be
presented and discussed in detail elsewhere.  The properties of SU~Lyn that
will be reviewed here are those that we will look for in the dataset of
GALAH stars with the aim to discover the new SySt.

The two most obvious distinctive characteristics of optical spectra of SU
Lyn (and {\it acc}-SySt of that type) are shown in
Fig.~\ref{fig:sulyn_spec}: ($a$) a near-UV ultraviolet excess ($\lambda \leq
4000$ \AA), and ($b$) weak emission lines, primarily from the hydrogen
Balmer series.  They both originate in the accretion disk forming around the
degenerate companion, fed by the material accreted from the mass-losing RG
companion.

Both can vary greatly in time, as the two epochs compared in
Fig.~\ref{fig:sulyn_spec} illustrate well: in just over one year SU Lyn
passed from an {\it active} state (Dec 19, 2015), when its SySt nature was
quite obvious, to a {\it quiet} phase (Feb 16, 2017) during which an optical
spectrum would not have betrayed its binary nature.  This needs to be kept
in mind when comparing GALAH spectra with ancillary data taken at very
different epochs.

The most important role of SU Lyn in guiding our search among GALAH stars is
however the observed profile for emission lines, H$\alpha$ in particular. 
Fig.~\ref{fig:sulyn_temp} presents a temporal sequence from Asiago 1.82m and
Varese 0.84m telescopes + Echelle spectrographs illustrating the great
variety of profiles observed for H$\alpha$ in this star from 2015 to 2020
(for a more traditional presentation of some of these same profiles for key
epochs see fig.  3 of \citealt{2019arXiv190901389M}).  The orbital plane of
SU Lyn is probably close to face-on conditions and this accounts for the
negligible changes in radial velocity observed for the M6III absorption
lines ($\leq$2 km~s$^{-1}$).  The observed spectroscopic and photometric
changes are therefore not related to changes in the aspect angle of the
binary.

The epoch marked {\textit b} in Fig.~\ref{fig:sulyn_temp} corresponds to the
lowest accretion rate and it is noteworthy for the absence of emission in
H$\alpha$ and a null near-UV excess: at optical wavelengths SU Lyn appears
like a normal and single M6III, not as an interacting binary.  Only the
reduced equivalent width of the H$\alpha$ absorption (only about half the
photospheric value of typical M6III giants) and its narrow FWHM (again half
the value for normal M6III giants) could cause doubt.  The white vertical
line in Fig.~\ref{fig:sulyn_temp} represents the photospheric velocity of
the cool giant: on epoch {\textit b}, the velocity of the H$\alpha$
absorption almost equals that of the M6III giant.  This is not the case for
most of the time, however: as clearly shown by Fig.~\ref{fig:sulyn_temp},
the H$\alpha$ absorption is normally blue-shifted compared to photospheric
absorptions as if originating in a gentle wind blowing off the inner regions
of the accretion disk (and thus appearing superimposed onto its emission),
or forming in the expanding wind of the RG that engulfs the whole binary
system.

Epoch {\textit c} in Fig.~\ref{fig:sulyn_temp} marks the sudden appearance
of a second, distinct and faster moving absorption component, that gradually
reduced its velocity during the following months and finally merged with the
slower, pre-existing absorption component.  At the time of maximum
equivalent width of this second absorption component, the emission in
H$\alpha$ briefly vanished.

Finally, epoch {\textit a} represents the condition of highest accretion
rate, with a strong near-UV excess and a prominent and structured emission
in H$\alpha$, with still superimposed the blue-shifted absorption component
mentioned above.  It is a lucky circumstance that when we first observed
SU~Lyn in preparation of Mk16 paper, the star was exhibiting a clear
emission in H$\alpha$ and the strongest near-UV up-turn: if at the time, it
had appeared as in epoch {\textit b} of Fig.~\ref{fig:sulyn_temp}, we could
not have recognized it for what it really was.

    \begin{figure*}
	\includegraphics[angle=270,width=16.8cm]{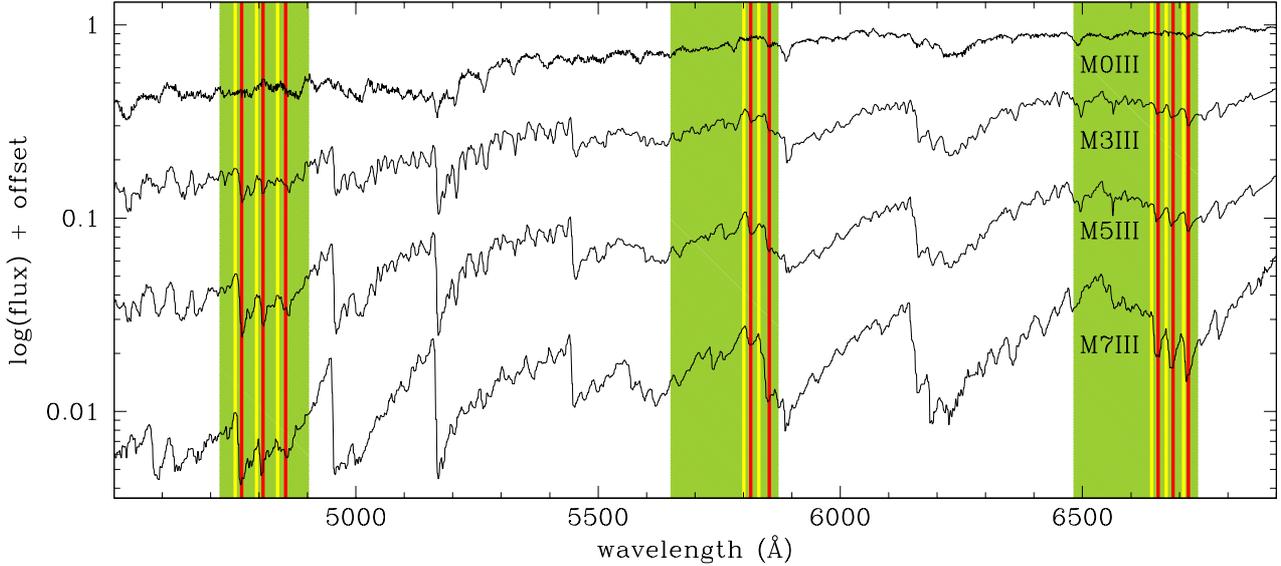}
        \caption{Definition of TiO classification indexes.  The shadowed
        regions show the wavelength ranges covered by the GALAH blue, green
        and red channels.  The yellow and red bands mark the A-B and C-D
        wavelength intervals, respectively, used to compute the $b$, $g$ and
        $r$ ratios according to Eq.(2) and Table~\ref{tab:tio_intervals}. 
        In background some Asiago 1.22m spectra illustrating the progress in
        strength of TiO bands moving along the sequence of M giants (pay
        attention to the log-scale of the ordinates).}
    \label{fig:tio_bands}
    \end{figure*}

Another important point illustrated by Fig.~\ref{fig:sulyn_temp} concerns
the photometric activity of SU Lyn.  The star is too bright to be recorded
unsaturated by surveys which monitor the heaven on a nightly basis in search
for transients (eg.  ASAS-SN, MASTER, ZTF, etc.), therefore we monitored it
ourselves with small instruments, and fully transformed the observations to
the \citet{1992AJ....104..340L} standard UBVRI system.  As illustrated by
the $B$,$V$ lightcurve on the right panel of Fig.~\ref{fig:sulyn_temp}
(synchronized to the same temporal ordinates as the spectra to the left),
SU~Lyn varies by up to 0.7 mag amplitude in a random fashion, with no
persistent periodicity.  There is no associated change in colour and no
correspondence with the behavior of the accretion disk as traced by
H$\alpha$ intensity and profile.  Thus, the variability is not caused by
brightening and fading of the accretion disk (hotter than the M6III), but
must originate with the cool giant.  All cool giants exhibit variability to
some extent \citep{1985vest.book.....H,2005lcvs.book.....S}, given the
unstable nature of their convective outer layers
\citep{2015MNRAS.454.2344P}, and their tenuous atmosphere that extends
across the temperatures for molecule formation and through those for dust
condensation.  The absence of radial velocity variability on the left panel
of Fig.~\ref{fig:sulyn_temp} precludes an origin of the variability seen in
SU Lyn with the coherent, large scale, persistent radial pulsations of the
type observed in Mira long-period variables.  The latter show a
sinusoid-like radial velocity curve of $\sim$10-15 km~s$^{-1}$ in amplitude,
in phase with the equally sinusoid-like lightcurve that can span up to 10
magnitude in amplitude, and characterized by bluer colour at maximum and
redder at minimum (see fig.  27 in \citealp{1985vest.book.....H}).

The distinction with radially-pulsating giants is important: their outer
layers experience large scale motions that lead to shocks, which in turn
power emission lines.  In our search for {\it acc}-SySt we aim to avoid the
contamination from radially-pulsating giants and their deceiving emission
lines.  The selection criteria for removing false positives from the final
{\it acc}-SySt sample are outlined in the rest of the paper.

\section{Selection steps} \label{sec:selection}

Our search for {\it acc}-SySt among GALAH targets started with isolating the
cool giants from the rest.  In this paper we focus on giants of the M
spectral types, those most abundant ($\sim$70\%) among the known SySt (see
the catalogs by \citealp{1984PASAu...5..369A}, \citealp{2000A&AS..146..407B},
\citealp{2019ApJS..240...21A}, \citealp{2019RNAAS...3...28M}).  SySt
containing other types of cool giants will be explored in follow-up papers.

\subsection{Colour}

The intrinsic colour of M0 giants in the Solar neighborhood is
$(J-K)_o$=0.97 \citep{1970ApJ...162..217L,1983A&A...128...84K}, so we first
apply a colour selection of $(J-K_s)$$\geq$0.90 to GALAH targets to both
account for the natural spread with the breadth of the M0 spectral type and
the slight difference in wavelength-baseline between the Johnson's $(J-K)$
and 2MASS $(J-K_s)$ indices.

\subsection{Parallax}

On the HR diagram, stars with an M spectral type are either very low-mass
main sequence objects (down to the limit for stable H-burning in the core),
or giants resulting from the evolution of more massive progenitors.  The
stars in our Milky Way have not lived long enough to significantly populate
the range in between.  M-dwarfs are intrinsically very faint and over the
$12 < V_{JK} < 14$ mag range covered by GALAH targets, they represent the
stars closest to the Sun, while the opposite is certainly true for the
giants.  In Table~\ref{tab:parallax} we compare (under negligible reddening)
the expected parallax of dwarfs and giants with an M spectral type at both
ends of the GALAH magnitude range.  No giant has a parallax larger than 0.4
mas, or a dwarf smaller than 10mas.  Therefore a second criterion was set
imposing a Gaia DR2 \citep{2018A&A...616A...1G} parallax $\pi$$\leq$3 mas. 
This selection criterium is fully confirmed by the improved Gaia $e$DR3
parallaxes \citep{2020yCat.1350....0G}, that became available only after
this paper was originally submitted.
 
\begin{table}
    \centering
    \caption{Parallax for dwarf and giant M stars (under negligible reddening)
    at the bright ($V$=12) and faint ($V$=14) end of the GALAH magnitude
    range.}
    \label{tab:parallax}
    \begin{tabular}{cccc}
    \hline
         & M$_V$ & \multicolumn{2}{c|}{parallax (mas)}\\
        &       &V=12 & V=14\\ \hline
    M0V & 8.90  &  24 & 10  \\
    M5V & 12.30 &115  &46   \\
        &       &     &     \\
    M0III & $-$0.10 & 0.38 & 0.15 \\
    M5III & $-$0.90 & 0.26 & 0.10 \\
    \hline
    \end{tabular}
    \end{table}

\subsection{TiO absorption bands}

The above colour and parallax criteria combined together returned an initial
sample of 29\,514 GALAH stars.  This sample could contain hotter giants of G
and K spectral types affected by a large reddening or cool giants of a
spectral type other than M (like the Carbon giants).  To prune them out and
retain only those with an M-type spectrum, we looked for signatures of their
distinctive TiO absorption bands in the GALAH spectra.  This may have
picked-up also some S-type giants whose spectra, in addition to TiO bands,
show also molecules involving $s$-type elements (ZrO in particular;
\citealp{1985aads.book.....T}).  Some percentage of known SySt do indeed
possess S-type giants \citep{2003ASPC..303...25J}.

Fig.~\ref{fig:tio_bands} presents a progression of M giant spectra with
over-plotted in green the wavelength range of the blue, green and red GALAH
channels.  Within their span lie strong TiO bands, and the later the
spectral type, the deeper such bands become.  To confirm the presence of the
TiO bands and objectively derive their depth, we defined eight narrow TiO
wavelength bins (marked in red in Fig.~\ref{fig:tio_bands}) and an equal
number of control wavelength bins (marked in yellow in
Fig.~\ref{fig:tio_bands}).  The control bins are all placed to the blue of
their respective TiO bins given the fact that absorption bands of the TiO
molecule all degrade to the red and present a steep band-head to the blue
\citep{1976ims..book.....P}.  We then defined eight {\it ratios} (b1, b2, b3
in the blue channel, g1, g2 in the green, and r1 , r2 and r3 in the red) by
integrating the flux within the control bin and dividing by it the TiO bin
as:

\begin{table}
        \centering
        \caption{Short and long wavelength limits (\AA) of the intervals 
        used to compute the ratios in Eq. (2) and illustrated in
        Fig.~\ref{fig:tio_bands}.}
        \label{tab:tio_intervals}
	\begin{tabular}{ccccc}
		\hline
	ratio & $\lambda_A$ & $\lambda_B$ & $\lambda_C$ & $\lambda_D$ \\
		\hline
b1 & 4751 & 4756 & 4765 & 4770 \\  
b2 & 4798 & 4803 & 4807 & 4812 \\  
b3 & 4835 & 4840 & 4850 & 4855 \\ 
g1 & 5801 & 5806 & 5812 & 5817 \\   
g2 & 5829 & 5834 & 5850 & 5855 \\   
r1 & 6641 & 6646 & 6653 & 6658 \\ 
r2 & 6670 & 6675 & 6684 & 6689 \\   
r3 & 6706 & 6711 & 6716 & 6721 \\
		\hline
        \end{tabular}
\end{table}

\begin{equation}
{\rm ratio} = \frac{\int_{\lambda_A}^{\lambda_B} f(\lambda)d\lambda}{\int_{\lambda_C}^{\lambda_D} f(\lambda)d\lambda}
\end{equation}
where the respective wavelength intervals A$-$B and C$-$D are listed in
Table~\ref{tab:tio_intervals}.  These ratios were computed on spectra before
continuum normalization but shifted to rest frame by the radial velocity
listed in GALAH DR3 as rv\_guess \citep{2020arXiv201102505B}.  Absence of
absorption by TiO returns a value around 1.0 for any of the ratios, and a
progressively larger value with increasing M spectral subtype (M0
$\rightarrow$ M10).  Finally, we have combined the eight ratios into a
single one computed as

\begin{equation}
f = \frac{[0.3(<b> - 1) + 1] + <g> + <r>}{3}
\end{equation}
where $<$b$>$ stands for the arithmetic average of b1, b2 and b3, and similarly
for the rest.  The $<$b$>$ average ratio stretches over 3.3$\times$ the range
of the other two, so we have consequently reduced its impact to make it
equal to the others. This is justified by the lower S/N of the spectra
recorded in the blue channel compared to green and red ones.

The depth of the TiO bands has been already used in the past to classify the
M giants of symbiotic stars, e.g.  by \citet{1987AJ.....93..938K}
that used to this aim the bands at 6180 and 7100 \AA\ (both outside the GALAH
wavelength range).  In these studies, equivalent widths of TiO bands were
measured over their entire wavelength span (hundreds of \AA), too wide
compared to the breadth of GALAH channels, and therefore inapplicable in the
present paper (and for that matter also to more conventional,
high-resolution Echelle spectra covering too small a wavelength interval
over a single order).

\begin{table}
    \centering
    \caption{Intervals in combined $f$-ratio (Eq.  3) defining the template
    bins for the selected 15\,824 M giants.  $(J-K_s)_\circ$ is the median
    value within the bin for the reddening corrected $(J-K_s)$ from 2MASS,
    and M($K_s$) is the median value of the absolute magnitude computed 
    from Gaia $e$DR3 parallaxes satisfy the condition $\sigma(\pi)$/$\pi$$<$0.1. 
    The last column lists the number of stars in each bin.  The template
    spectra for each bin are plotted in Fig.~\ref{fig:figA1}.}
    \label{tab:templates}
    \begin{tabular}{cccccr}
    \hline
&&\\
bin &   left limit & right limit   & $(J-K_s)_\circ$ & M($K_s$) & N. stars\\
&&\\
      00 &       1.0200 &        1.0303 &  0.904  & $-$3.78    & 3419  \\ 
      01 &       1.0304 &        1.0419 &  0.935  & $-$4.01    & 2309  \\ 
      02 &       1.0420 &        1.0561 &  0.968  & $-$4.27    & 1732  \\ 
      03 &       1.0562 &        1.0736 &  0.994  & $-$4.58    & 1424  \\ 
      04 &       1.0737 &        1.0950 &  1.017  & $-$4.81    & 1282  \\ 
      05 &       1.0951 &        1.1214 &  1.043  & $-$4.99    & 1200  \\ 
      06 &       1.1215 &        1.1539 &  1.065  & $-$5.18    & 1048  \\ 
      07 &       1.1540 &        1.1938 &  1.088  & $-$5.42    & 919   \\ 
      08 &       1.1939 &        1.2428 &  1.110  & $-$5.69    & 917   \\ 
      09 &       1.2429 &        1.3030 &  1.141  & $-$5.94    & 849   \\ 
      10 &       1.3031 &        1.3770 &  1.166  & $-$6.12    & 638   \\ 
      11 &       1.3771 &        1.4679 &  1.182  & $-$6.29:   & 77    \\ 
      12 &       1.4680 &        3.6559 &  1.139  &            & 10    \\ 
&&\\
    \hline
    \end{tabular}
    \end{table}

    \begin{figure}
	\includegraphics[width=\columnwidth]{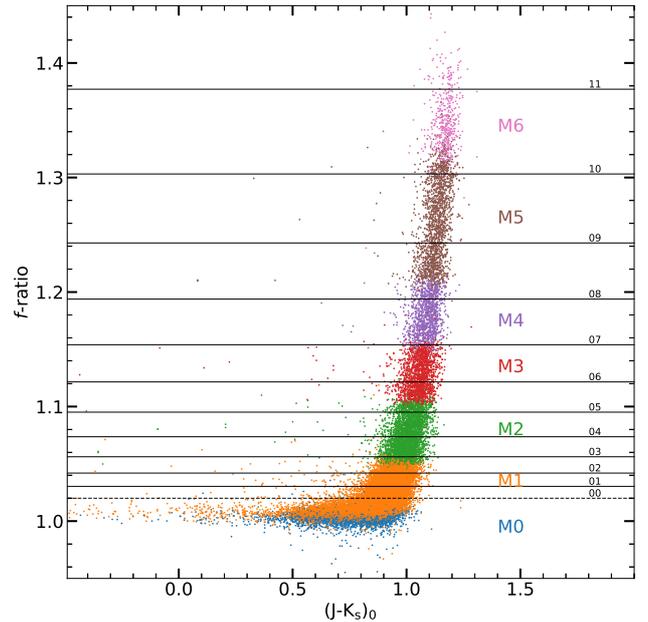}
        \caption{Relation between the reddening-corrected 2MASS
        $(J-K_s)_\circ$ colour index and the combined $f$-ratio from Eq.(2)
        for the 29\,514 GALAH stars in our initial sample.  The coloured
        bands mark the approximate location of the MKK spectral subtypes
        (M0III, M1III, ...).  The finer, log-scale subdivision according to
        Table~\ref{tab:templates} is given by the horizontal lines and it is
        marked by the numbers to the right (00, 01, ...).}
    \label{fig:fratio}
    \end{figure}

The way our ratios are defined, makes them insensitive to spectral fluxing,
continuum normalization, or reddening.  The relation between $f$-ratio and
reddening corrected $(J-K_s)_\circ$ index for the initial sample of 29\,514
GALAH stars is illustrated in Fig.~\ref{fig:fratio}.  The latter shows a
number of highly reddened stars contaminating the sample.  To filter them
out we took all stars laying above the knee at $f$$>$1.02, and considered
the remaining 15\,824 objects (the 15k-sample for short) as representing the
true M-giants present among the 600\,255 stars observed by GALAH at the time
of writing of this paper.  The $(J-K_s)_\circ$ colour index has been
calculated from 2MASS photometry and the full extinction as given by
\citet{2011ApJ...737..103S}.  The choice of taking the full value of the
tabulated extinction is justified by considering the great distance to our
targets (typically several kpc, beyond which Gaia DR2 or $e$DR3 parallaxes
become inaccurate) and their large galactic latitude, which makes their
sight-lines exit the Galactic dust slab well before reaching them.
  
\subsection{H$\alpha$ in emission}

The next step has been searching the 15k-sample of true M giants for those
showing H$\alpha$ (and H$\beta$) in emission.  All the symbiotic stars
considered in this paper show H$\alpha$ emission directly on the recorded
spectrum, well before any reference spectrum is subtracted (cf. 
Fig.~\ref{fig:example_HaHb}).  To properly reconstruct the profile of the
emission line, it is however necessary to subtract the underlying spectrum
of the M giant.  To define the latter, we grouped the M giants by similar
strength of TiO bands, and computed the median of the continuum normalized
spectra within each bin.  The spectral progression of the templates around
H$\alpha$ and H$\beta$ is compared in Fig.~\ref{fig:figA1} of the Appendix.

We initially selected to define a unique template for each of the eleven
M0\,III$\leftrightarrow$M10\,III spectral subtypes of the original MKK
classification scheme \citep{1943assw.book.....M,1973ARA&A..11...29M}, but
quickly realized that at the high S/N and high resolution of GALAH spectra,
any such sub-type spans too wide a range in stellar properties for a single
template to adequately represent them all.  Similarly, equally spaced bins
in $f$-ratio are not satisfactory, as they oversample at low $f$ values. 
Therefore, we decided for a logarithmic progression, with
Table~\ref{tab:templates} listing the 13 final bins, their interval in terms
of $f$-ratio, and the number of true M giants contained in that bin.  Their
subdivision of the giant branch is shown in Fig.~\ref{fig:fratio}
(horizontal lines).  The median of all the spectra in a given bin defines
the template spectrum for that bin.  Taking the median value eliminates any
disturbance from cosmic ray hits or rare presence of emission components. 
In addition, the median computed over thousands of individual spectra has an
extremely high S/N and it is completely insensitive to chemical
peculiarities of individual stars.

The template is finally subtracted from continuum normalized spectra of the
true M giants within the given bin, and the residuals inspected for emission
in H$\alpha$.  In this first paper we limit ourselves to consider only the M
giants showing an H$\alpha$ emission profile rising above a threshold of 0.5
on the subtracted spectrum.  This returned a total of 223 stars.

\subsection{Visual inspection}

The spectra of these 223 stars were visually inspected for
inconsistencies that could cause the H$\alpha$ to appear in emission on the
subtracted spectrum.

Some spectra were found to suffer from a mismatch in radial velocity with
the template.  In these few cases, the radial velocity rv\_guess from GALAH
DR3, which is used to shift spectra in all four GALAH channels to rest
frame, provides correct results for the blue, green, and infrared, but not
for the red channel, resulting in an excess flux around H$\alpha$ in the
difference spectrum, which is obviously spurious.

Another, more frequent cause for false positives was the presence of an
emission component probably originating from the sky background that has not
been fully cancelled-out by the sky-subtraction procedure.  This may happen
e.g.  when the target star is located in a region of the sky affected by
diffuse background emission, like emission nebulae and HII regions.  Such
cases are easy to recognize because the H$\alpha$ and H$\beta$ emission
lines are ($a$) typically quite sharp, single-component and symmetric, ($b$)
frequently accompanied by $[$NII$]$ 6548, 6584 nebular emission lines (which
do not form in the high density environment of an accretion disk), and ($c$)
usually displaced in velocity with respect to the M-giant spectrum by an
amount larger than expected from orbital motion in a symbiotic binary.  In
multi-fibre spectroscopic surveys like GALAH, only a limited number of
fibres are assigned to record the sky background.  They are distributed
over the field of view away from the position of known stars.  This is
perfectly fine for a great fraction of the sky away from the Galactic plane. 
There are however, here and there, regions of the sky affected by background
emission due to diffuse nebulosity.  The intensity of such emission may vary
appreciably over limited angular distances, and the same may be the case for
the internal nebular gas dynamics and therefore the resulting wavelength for
H$\alpha$ and H$\beta$ emitted lines.  Any sparse mapping realized by a
limited number of sky fibres of such a complex 2D pattern can naturally
result in non-null residuals of the sky-background subtraction, in addition
to other caveats of sky subtraction as explained in
\cite{2017MNRAS.464.1259K}.

There was also a group of stars which presented an equal excess 
emission in H$\alpha$ and H$\beta$ on the subtracted spectrum, always with
the same characteristic profile (steep red and blue ends, flat and rounded
top), as if originating from a hydrogen deficiency compared with the
template spectrum.  These stars were also disregarded in the current
selection process.

After a further few rejections for miscellaneous reasons, we were left with a
final selection of 83 M giants showing genuine emission in H$\alpha$ with an
intensity in excess of 0.5 in the subtracted spectrum (the 83-sample
hereafter).  Stars with an emission in H$\alpha$ weaker than 0.5, which are
inherently more subtle to treat, will be investigated in a follow-up paper.

\subsection{Filtering out the radial pulsators} \label{sec:filtering_radialp}

Not all the objects in the 83-sample are necessarily valid {\it acc}-SySt.  The
main false-positives are expected to be the (large-amplitude) radial pulsators
(e.g.  Miras or SRa variables), that may give origin to emission lines
(primarily hydrogen Balmer) deep in their atmosphere where shocks form.

It is of course not precluded that a Mira pairs with a compact companion to
form a symbiotic star.  For example V407 Cyg, prior to its 2010 nova
outburst, was primarily studied as a Mira suffering from possible
dust-obscuration episodes \citep{1990MNRAS.242..653M,1998AstL...24..451K}
with little (or none) spectroscopic evidences for binarity
\citep{2013ApJ...770...28H}.  {\it Mira} itself (= $o$~Cet) is a well known
{\it acc}-SySt: {\sc Hubble} in the ultraviolet and {\sc Chandra} at X-rays
have spatially resolved its WD companion, the accretion disk and the stream
fueling it \citep{2006ESASP.604..183K}.  Such a plethora of multi-wavelength
information is however not available for a typical GALAH anonymous star, and
to be on the safe side we decided - at the initial stage of our project
represented by this paper - to filter out the radial pulsators.  Such a
pruning is based on converging photometric (lightcurve) and spectroscopic
(H$\beta$/H$\alpha$ emission ratio) criteria, described below.

    \begin{figure*}
	\includegraphics[width=16.8cm]{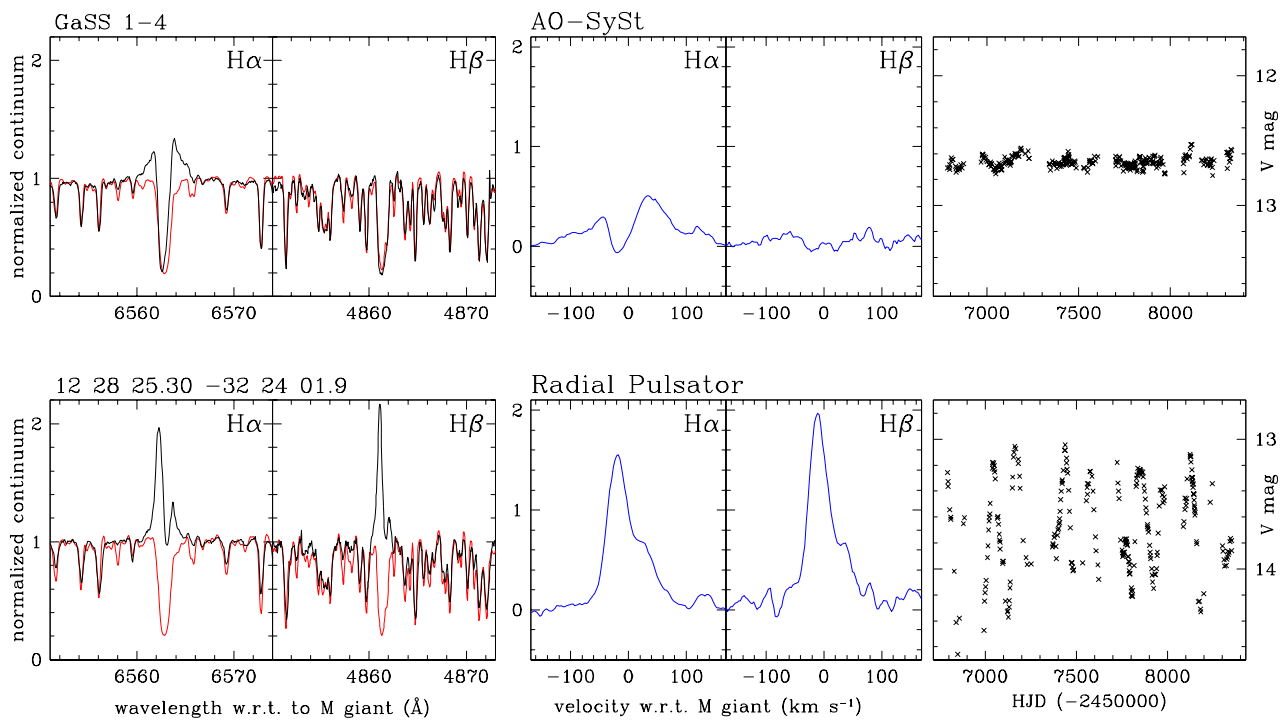}
        \caption{{\it Left-most panels}: example H$\alpha$ and H$\beta$
        profiles (in black) from RV-zeroed GALAH spectra of a symbiotic star
        (top row) and of a radial pulsator (bottom row), compared with those
        of the template for the same $f$-ratio bin (in red, cf. 
        Table~\ref{tab:templates}).  {\it Center panels}: the result of
        subtracting the (red) template spectrum from the (black) object
        spectrum from the left panels.  {\it Right-most panel}: the
        corresponding $V$-filter lightcurve from ASAS-SN sky patrol data.}
    \label{fig:example_HaHb}
    \end{figure*}

We decided not to rely on existing compilations of known variables (as for
example collected in the
VSX\footnote{http://vizier.u-strasbg.fr/viz-bin/VizieR-3?-source=B/vsx/}
catalog), because the adopted classification criteria (as well as amplitude
and periods) are sometimes in disagreement among themselves and also
in conflict with the naming conventions listed in the IAU General Catalog of
Variable Stars\footnote{http://www.sai.msu.su/gcvs/}.  We preferred instead
to analyze in a homogeneous way their photometric behavior ourselves.  To
this aim we downloaded the available photometry for the 83-sample from the
ASAS-SN sky patrol survey \citep{2014ApJ...788...48S,2017PASP..129j4502K} and
examine ourselves the resulting lightcurves, which are usually composed of
hundreds of individual observations in either $g$ or $V$ bands distributed
over a time-interval of a few years, and therefore adequate to reveal the
presence of long and stable periodicities related to radial pulsation. 
ASAS-SN sky patrol covers the whole accessible sky on both hemispheres every
night, and the $12 < V_{JK} < 14$ mag range of our targets is ideally placed
at the center of ASAS-SN dynamic range, away from the bright saturation
limit or the noisy detection threshold.

The lightcurve of a Mira shows regular, long period (from several months to
longer than a year) and large amplitude (from a few to $>$10 magnitudes)
sinusoid-like variations (e.g. 
\citealp{1985vest.book.....H,2005lcvs.book.....S}).  The large scale radial
pulsation is betrayed by the in-phase variation of the radial velocity of
the absorption spectrum, with amplitudes of the order of $\sim$10-15 km/s
\citep{1926ApJ....63..281J,1985vest.book.....H}.  Our path-finder SU~Lyn
shows a significant variability (0.6 mag amplitude in $V$), but it lacks the
larger amplitude and the regular beat of a Mira.  As clearly illustrated by
Fig.~\ref{fig:sulyn_temp}, the M giant in SU~Lyn does not radially pulsate
in any significant way (the absorption lines are stable in radial velocity
to better than 2 km~s$^{-1}$), and therefore no shocks form in the
atmosphere which could lead to emission in the Balmer lines.  Armed with
these considerations, and to stay on the cautionary side, we selected to
flag (in the 83-sample) all the stars showing ASAS-SN lightcurves with an
amplitude in excess of 0.7 mag and a stable and clean sinusoid-like shape
(cf.  Fig.~\ref{fig:example_HaHb}).  Such a lightcurve could however arise
also from orbital motion, for example from ellipsoidal distortion when the
cool giant fills its Roche lobe and/or its side facing the companion is
irradiated and heated up (e.g.  as in the accreting-only and recurrent nova
T CrB; \citealp{2016NewA...47....7M}).  A spectroscopic confirmation of the
radially pulsating nature is therefore required, and this needs to be
accommodated within the single-epoch spectra (i.e.  no revisit) of the GALAH
survey.

In radially-pulsating cool giants, outward moving material collides against
the gas lifted during the previous cycle and now falling back inward.  The
resulting shock is hot enough to excite emission in the Balmer lines of
hydrogen.  Such shocks develop deep within the atmosphere, interior to the
outer layers where absorption by TiO molecules occurs.  The absorption by
TiO begins around 4200~\AA\ and, by superposition of successive bands (they
all decline toward the red), their combined absorption grows rapidly
stronger with wavelength (Fluks et al.  1994).  As a consequence, emission
in H$\delta$ (4101 \AA) and higher Balmer terms exits the atmosphere
unscathed by TiO, and their flux declines with increasing upper $n$-quantum
number in the usual way.  What happens to the red of H$\delta$?  Going from
H$\gamma$ to H$\beta$ and then H$\alpha$, we move into deeper and deeper
overlapping absorption by TiO bands and the {\it observed} emission in the
line becomes more and more absorbed.  Saying it a different way
\citep{1926ApJ....63..281J,1977aars.book.....Y}, in pulsating M giants the
strongest line is H$\delta$ and the intensity of successive Balmer emission
lines declines {\it either} going to the blue or to the red (see exemplary
spectrum for LQ Sgr presented by \citealp{1995A&A...297..759B}).  In most
astrophysical environments, the observed emission in H$\alpha$ is invariably
stronger than in H$\beta$, while in Miras the opposite is the case
\citep{1940slpv.book.....M}, because of the TiO absorption in the outer
atmospheric layers.  This offers a clear distinction between the emission
originating from the accretion disk in an {\it acc}-SySt (in terms of
equivalent widths: EW(H$\alpha$) $>$ EW(H$\beta$)) from that expected to
come from the internal shock regions of an M-type radial pulsator
(EW(H$\alpha$) $\leq$ EW(H$\beta$)).

GALAH high-resolution spectra show a further distinction between the two
types.  In {\it acc}-SySt systems, a sharp absorption component appears
superimposed to the broad H$\alpha$ emission profile, generally {\it
blue-shifted} by about 5$-$25 km~s$^{-1}$ and half as wide as the
photospheric H$\alpha$ absorption line (FWHM$\sim$15$-$25 km~s$^{-1}$ as
opposed to 45$-$50 km~s$^{-1}$), that - by analogy with the prototype SU~Lyn
or other {\it acc}-SySt discussed by \citet{2012BaltA..21...39J} and
\citet{2012BaltA..21..165G} - is believed to originate in the gentle wind
blowing off the accretion disk or in the out-flowing wind of the RG that
engulfs the whole binary system.  In the GALAH spectra of radially pulsating
stars, the narrow absorption superimposed to the emission is instead
generally {\it red-shifted} as if coming from cooler material falling back
toward the star after being lifted higher up during the previous pulsation
cycle.

These photometric and spectroscopic criteria to separate {\it acc}-SySt and
radially pulsating objects in the 83-sample are illustrated in
Fig.~\ref{fig:example_HaHb}, which presents line profiles and lightcurves
well typical of the rest of the sample: high-amplitude and regular beating
for the lightcurve of a pulsator, with strong H$\beta$ and red-shifted
narrow-absorption; lower amplitude and less regular lightcurve for the {\it
acc}-SySt, an H$\beta$ much weaker than H$\alpha$ and blue-shifted
narrow-absorption.  By applying these criteria, 30 radial pulsators are
pruned from the 83-sample, reducing the candidates {\it acc}-SySt to 53. 
Hereafter, we refer to them as the 53-sample.

\subsection{Avoiding contamination by T Tau stars}

T~Tau pre-main sequence variables have been considered as possible
contaminants in previous large-scale searches for new SySt, especially those
based on IPHAS $r$,$i$,H$\alpha$ photometric survey
\citep{2008A&A...480..409C,2010A&A...509A..41C,2011A&A...529A..56C,2014A&A...567A..49R}. 
T~Tau stars are distributed over a wide range of spectral types, from mid-F
down to the coolest M-types, may be heavily reddened, and present emission
lines originating from the circumstellar disk, the spotted surface and the
magnetic-confined accretion columns on the magnetic poles
\citep{2016ARA&A..54..135H}.  T~Tau stars can therefore mimic the red
$r$$-$$i$ colour of SySt and similarly stand out in the $r$$-$H$\alpha$
photometric index.  T~Tau stars are less prone to be confused with SySt if
spectra are available, especially if they cover a broad wavelength range (as
those we have collected in Section~\ref{sec:uv_upturn} below for our program
stars): the spectra of T~Tau stars show the absorption features typical of a
main sequence star, not of a giant, and the emission lines are rather
different both in assortment (CaII H \& K doublet usually quite strong in
T~Tau stars, while generally absent in SySt) and in their profile (in T~Tau
stars sporting narrow jet components superimposed on a broad pedestal coming
from the rotating disk; \citealp{2019A&A...631A..44G}).  In addition,
LiI~6707~\AA\ is in strong absorption in most T~Tau stars, whereas it is
generally absent in SySt.

In the context of our search for new SySt, T~Tau stars are however of no
concern, because they cannot be confused with cool giants when the
segregation is carried out on Gaia parallaxes, as we did.  Of relevance to
the present paper are only T~Tau stars with an M spectral type (M-TTau for
short).  The intrinsically brightest among M-TTau are those with the
earliest M-types, M0V and M1V.  The absolute magnitude of M0V and M1V stars
are M(V)=8.8 and 10.8, respectively, while that of M0III and M1III giants is
M(V)=$-$0.5 \citep{2000asqu.book..381D,2007AJ....134.1089S}.  The inflated
photospheric radius (not yet settled on the main sequence) and the
contribution by the circumstellar disk may raise the intrinsic brightness of
M-TTau above that of single and isolated dwarfs of the same spectral type,
but they will never fill the huge $\geq$10mag gap between main-sequence and
giant stars of the M spectral type.

In confirmation of this, we have retrieved the Gaia $e$DR3 parallaxes for
the brightest known M-TTau (selected from the compilation by
\citealp{2012ApJ...744..121Y}) and computed their M(V) magnitudes by
adopting the photometric calibration of the Gaia Consortium
\citep{2018A&A...616A...4E}.  The absolute M(V) magnitudes of M-TTau turned
out to be all fainter than +6.5, confirming that in no way an M-TTau may
have slipped into our sample of M giants segregated on the base of Gaia
parallaxes (see also Section~\ref{sec:evolutionary_stat} below).  As a
further check, we took the compilation of T~Tau stars listed in SIMBAD and
VSX, and checked if any matched the M-giants of the initial 15k-sample, and
verified that none did.

\subsection{Comparison with VVCep binaries} \label{sec:vvcep}

There are binaries which host cool supergiants and massive hot
companions, of which a comprehensive catalog of 108 Galactic objects has
been recently published by \citet{2020RNAAS...4...12P}.  The 23 entries from
this catalog with an M spectral type for the cool giant are summarized in
Table~\ref{tab:vvcep}, where we add distances, absolute magnitudes and
height above the galactic plane as derived from Gaia~$e$DR3 parallaxes and
the 2MASS JHK$_S$ survey.  Such objects are usually named VVCep stars, from
the best known member of the group, and while not related to SySt are
sometimes discussed in parallel.  We have not included $\delta$~Sge in
Table~\ref{tab:vvcep} as it seems more related to other types of objects
like TX~CVn or 17~Lep than to VVCep stars
\citep{2002ApJS..143..513G,2015MNRAS.454.2344P}.

    \begin{figure}
	\includegraphics[width=\columnwidth]{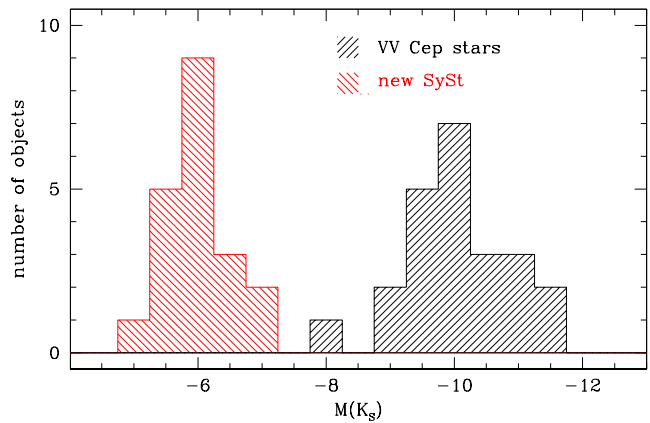}
        \caption{Distribution in M(K$_S$) for
        the program new SySt compared to that of VVCep stars.}
    \label{fig:syst_vs_vvcep_mk}
    \end{figure}

    \begin{figure}
	\includegraphics[width=\columnwidth]{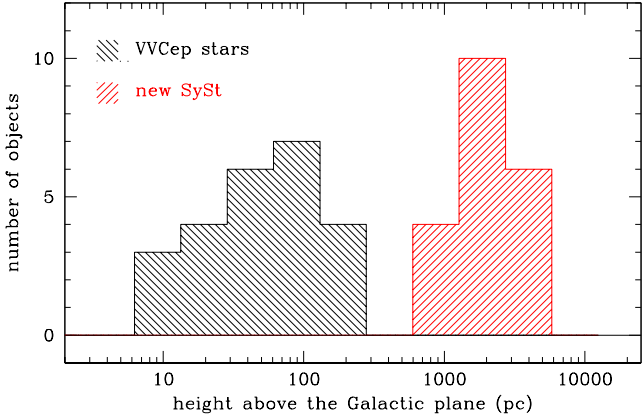}
        \caption{Distribution in height over the plane of the Galaxy for
        the program new SySt and for VVCep stars.}
    \label{fig:syst_vs_vvcep_z}
    \end{figure}

The VVCep stars all have very massive components on very wide orbits.  The
orbital solution for VV Cep itself gives an orbital period of 20.3~yrs, an
orbital separation of 24.8 AU, and individual masses of 18.2~M$_\odot$ and
18.6~M$_\odot$ for the M supergiant and the B-type companion, respectively
\citep{2004ASPC..318..222B}.  Of interest to us is the fact that in VVCep
stars, the Balmer lines, and H$\alpha$ in particular, may be seen in
emission, with a line profile reminiscent of that seen in Be stars (in which
the emission comes from an equatorial ring-like shell formed by material
leaving the hot star which is rotating close to break-up velocity).  While
the hot component in VVCep dominates the spectrum in the blue, that of the M
supergiant takes over at red wavelength.  With an H$\alpha$ in emission
superimposed, it could be questioned whether any of our new SySt may indeed
belong to VVCep stars.  This, however, can be safely excluded on two
independent grounds.

    \begin{figure*}
	\includegraphics[width=16.8cm]{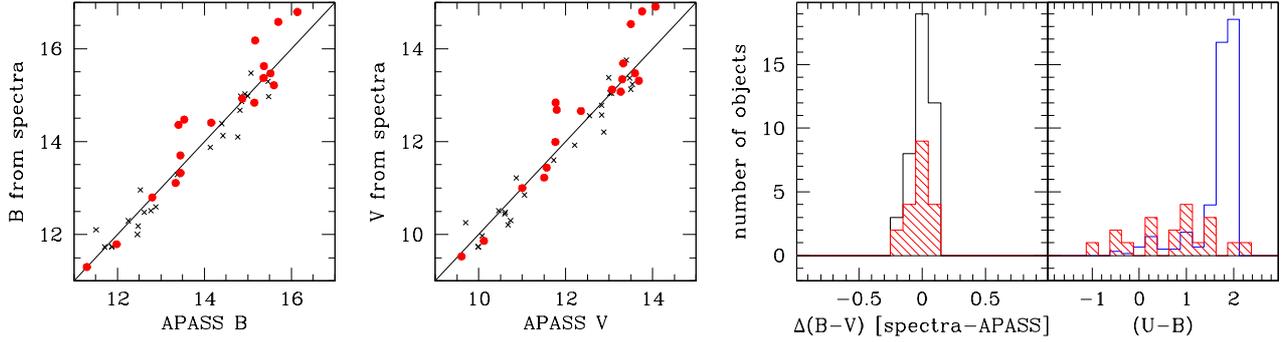}
	\caption{{\it Left-most panels}: comparison between the average $B$
         and $V$ magnitudes from APASS all-sky survey, and the corresponding
         values derived from Asiago 1.22m flux-calibrated spectra for a
         subset of the 53-sample.  The solid lines mark the 1:1 relation. 
         {\it Right-most panels}: histogram distribution of differences
         between Asiago and APASS derived $B$$-$$V$ colours, and distribution
         of $U$$-$$B$ colours derived from Asiago spectra.  The line in blue
         shows the distribution of a hundred M giants from the Solar
         neighborhood as measured by \citet{1994A&AS..105..311F}.  Red colour refers
         to symbiotic stars discovered in this paper, black to other GALAH M
         giants observed along with them (including also some radial
         pulsators).}
    \label{fig:BV_apass}
    \end{figure*}

First, M supergiants of VVCep stars are intrinsically much brighter than the
M giants characterizing our new SySt.  The two are compared in
Fig.~\ref{fig:syst_vs_vvcep_mk}, with values taken from
Table~\ref{tab:tabA2} for the new SySt, and Table~\ref{tab:vvcep} for the
VVCep stars.  The median absolute M(K$_S$) magnitude for VVCep stars is
$-$10.0, that of the new SySt only $-$5.9.  Secondly, their distance over
the Galactic plane is radically different.  Being very massive and therefore
very young, VVCep are examples of extreme Pop~I stars, laying close to the
Galactic plane, while our new SySt belong to a much older population as
discussed in Section~\ref{sec:evolutionary_stat} below.  Their distributions
in terms of height above the Galactic plane are compared in
Fig.~\ref{fig:syst_vs_vvcep_z}, with input data taken from
Table~\ref{tab:tabA2} and Table~\ref{tab:vvcep}: the median value is just
50pc for the VVCep stars and a much larger 1.8kpc for the new SySt.

\section{The finally selected 33 new symbiotic stars and their properties} \label{sec:selected33}

We have finally subjected the 53-sample to a series of follow-up
observations from the ground and the space to pick-up the most promising
objects.  Based on the ensuing results, our final list of 33 proposed {\it
acc}-SySt candidates is presented in Table~\ref{tab:main_table} (where GaSS
in their names stands for Galah Symbiotic Star); they are divided in two
groups, a primary and a supplementary sample.  The division is based
primarily on available follow-up observations (less for the supplementary
sample) and amplitude of photometric variability (larger for the
supplementary sample).  A plot similar to Fig.~\ref{fig:example_HaHb},
presenting the H$\alpha$, H$\beta$ profiles and the lightcurve, is provided
in the appendix for all 33 objects
(Fig.~\ref{fig:figA2}~to~\ref{fig:figA8}).

Gaia $e$DR3 lists parallaxes ($\pi$) for all the 33 new SySt, and they
are reported with their uncertainties ($\sigma(\pi)$) in
Table~\ref{tab:tabA2}.  In Table~\ref{tab:main_table} we provide distances,
by direct inversion of parallax, for only 20 of them, namely those satifying
the condition $\sigma(\pi)$/$\pi$$<$0.2.  This is the limit reccommended by
the Gaia team \citep{2015PASP..127..994B,2018A&A...616A...9L}.  The
parallaxes of the remaining program stars are too small in comparison to
their formal uncertainties and even generally smaller than the {\it mean}
0.019mas bias affecting Gaia $e$DR3 parallaxes as a whole
\citep{2020arXiv201201742L}.  Such bias depends in a non-trivial way on (at
least) the magnitude, colour, and ecliptic latitude of the sources, with no
firm correcting recipe available yet.  Improved parallaxes in future Gaia
data releases will allow to derive valuable distances to (most of) the
remaining program stars.

We now describe the follow-up observations that have led us to the
compilation of Table~\ref{tab:main_table}.  A complete description of such
extensive follow-up observations is far beyond the scope and breadth of the
present paper. Full details, augmented by the results of further
follow-up observations currently in progress, will be discussed elsewhere.

\subsection{Near UV up-turn} \label{sec:uv_upturn}

The presence of a bright accretion disk is betrayed by an emission excess at
wavelengths shorter than $\sim$4000~\AA, as illustrated for SU~Lyn in
Fig.~\ref{fig:sulyn_spec}.  With the long-slit B\&C spectrograph attached to
the Asiago 1.22m telescope, highly efficient down to the atmospheric cut-off
($\sim$3200 \AA\ at the 1000m above sea level location) we have observed
most of the objects in the 53-sample located north of $-$29$^\circ$ in
declination (Table~\ref{tab:lowres_log} in the Appendix provides a log-book
of these observations).  At the Asiago $+$46$^\circ$ latitude, this
corresponds to a horizon distance of 15 degrees.  The near-UV faintness of
the program stars and the high atmospheric extinction at such large
air-masses conspired to make such observations quite demanding to execute. 
A consistent number of spectro-photometric standards were observed in
parallel each night at similarly large air-masses.  Suitably high S/N at
3400-3600~\AA\ was obtained for only a fraction of the attempted objects,
and sometimes only after adding spectra collected at different revisits. 
The objects marked "spc" in the $U_{\rm exc}$ column of
Table~\ref{tab:main_table} are those showing excess emission on spectra over
the range 3400$\leq$$\lambda$$\leq$4000~\AA.  There are clear hints for some
of them that the amount of near-UV excess changes when spectra taken at
different epochs are compared.

\begin{table*}
        \centering
        \caption{The candidate accreting-only symbiotic stars ({\it
        acc}-SySt) discovered in this paper based on spectra from the GALAH
        survey.  E(B-V) is from \citet{2011ApJ...737..103S}.  The spectral
        type is on the MKK scale and $f$ is the template label/bin (see
        Table~\ref{tab:templates} and Fig.~\ref{fig:figA1}).  $V$ is from
        the APASS catalog.  $\Delta V$ is our estimate of the amplitude of
        variability on ASAS-SN sky patrol data
        (Section~\ref{sec:filtering_radialp}).  'P orb' lists possible
        orbital periods (for 2$\times$P alternative see
        Section~\ref{sec:orbital_periodicites}).  'dist' is the distance (in
        kpc) derived from Gaia~$e$DR3 parallax for objects satisfying the
        conditions $\sigma(\pi)$/$\pi$$<$0.2 (cf. 
        Section~\ref{sec:selected33}).  RV$_\odot$ is from GALAH DR3
        (rv\_guess).  Objects marked 'var' in the 'Asiago RV' column show a
        variability in radial velocity (cf.  Table~\ref{tab:rvs} and
        Section~\ref{sec:rv_var}).  The column 'Swift' marks with 'X' the
        objects counterpart to X-ray sources in the 2SXPS catalog
        \citep{2020yCat.9058....0E}, and with 'UV' those found emitting in
        the ultraviolet UVM2 band in our {\it Swift} follow-up observations
        (cf.  Table~\ref{tab:acc_lum} and Section~\ref{sec:xray_uv_obs}). 
        $U_{\rm exc}$ lists the presence of emission excess at $U$-band
        wavelengths: 'pht' from photometric UBV observations
        \citep{2002ApJS..141...81M}, 'spc' from our spectra
        (Section~\ref{sec:uv_upturn}).  An 'F' in the 'Flk' column marks the
        presence of flickering (Section~\ref{sec:flickering}).  The final
        two columns provide the velocity (w.r.t.  M giant) and the FWHM of
        the narrow absorption superimposed to the broader H$\alpha$ emission
        (Section~\ref{sec:filtering_radialp}).}.
        \includegraphics[angle=90,width=17.6cm]{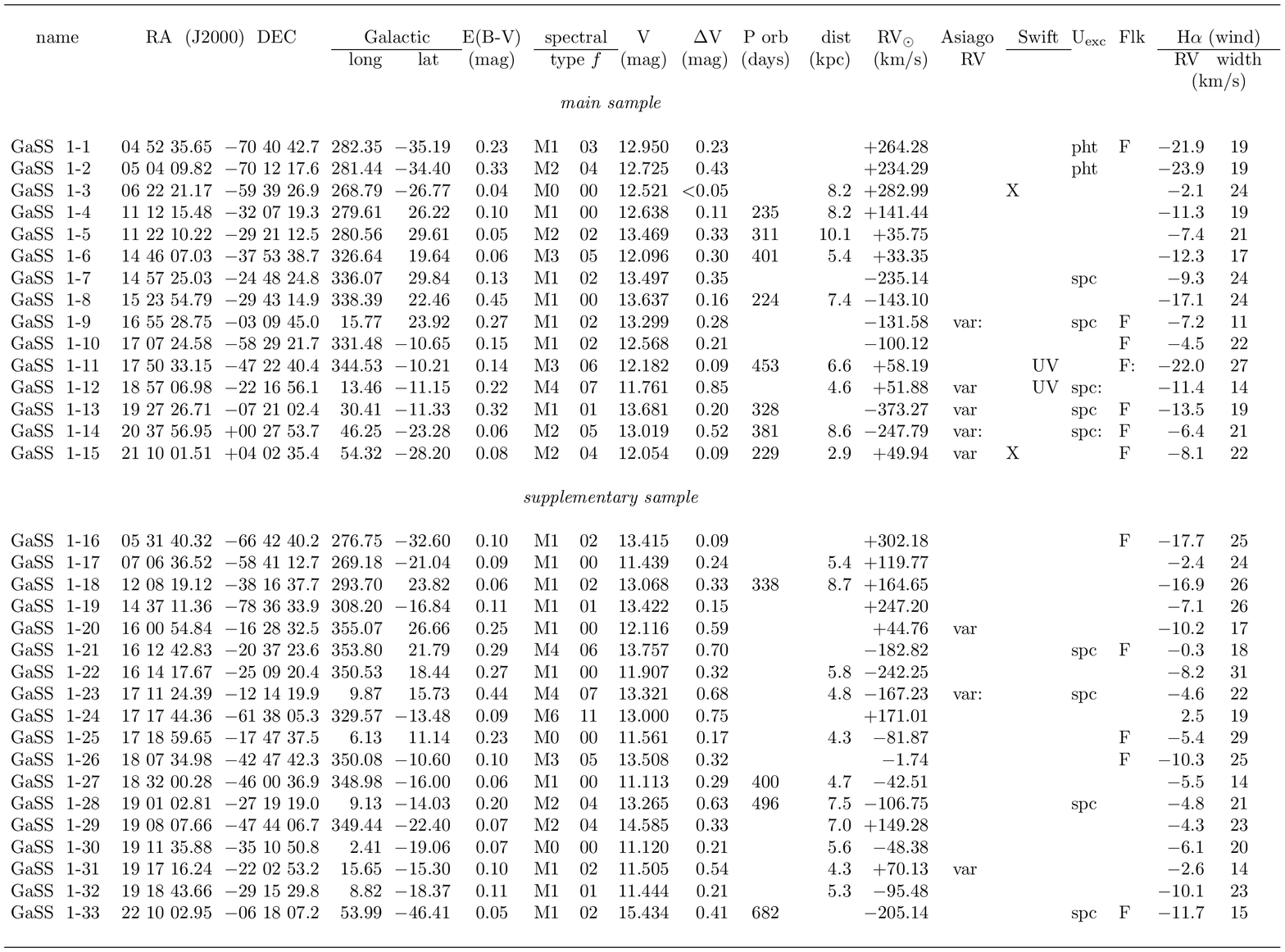}
\label{tab:main_table}
\end{table*}

\begin{table*}
\centering
\caption{Heliocentric radial velocity of candidate GALAH symbiotic stars
measured with the Asiago 1.82m telescope + Echelle spectrograph compared
with the corresponding values of rv\_guess listed in GALAH DR3.  The mean
radial velocity listed in Gaia DR2 is also given, with its formal error and
the number of epoch transits over which it has been computed.}
\includegraphics[angle=90,width=15.6cm]{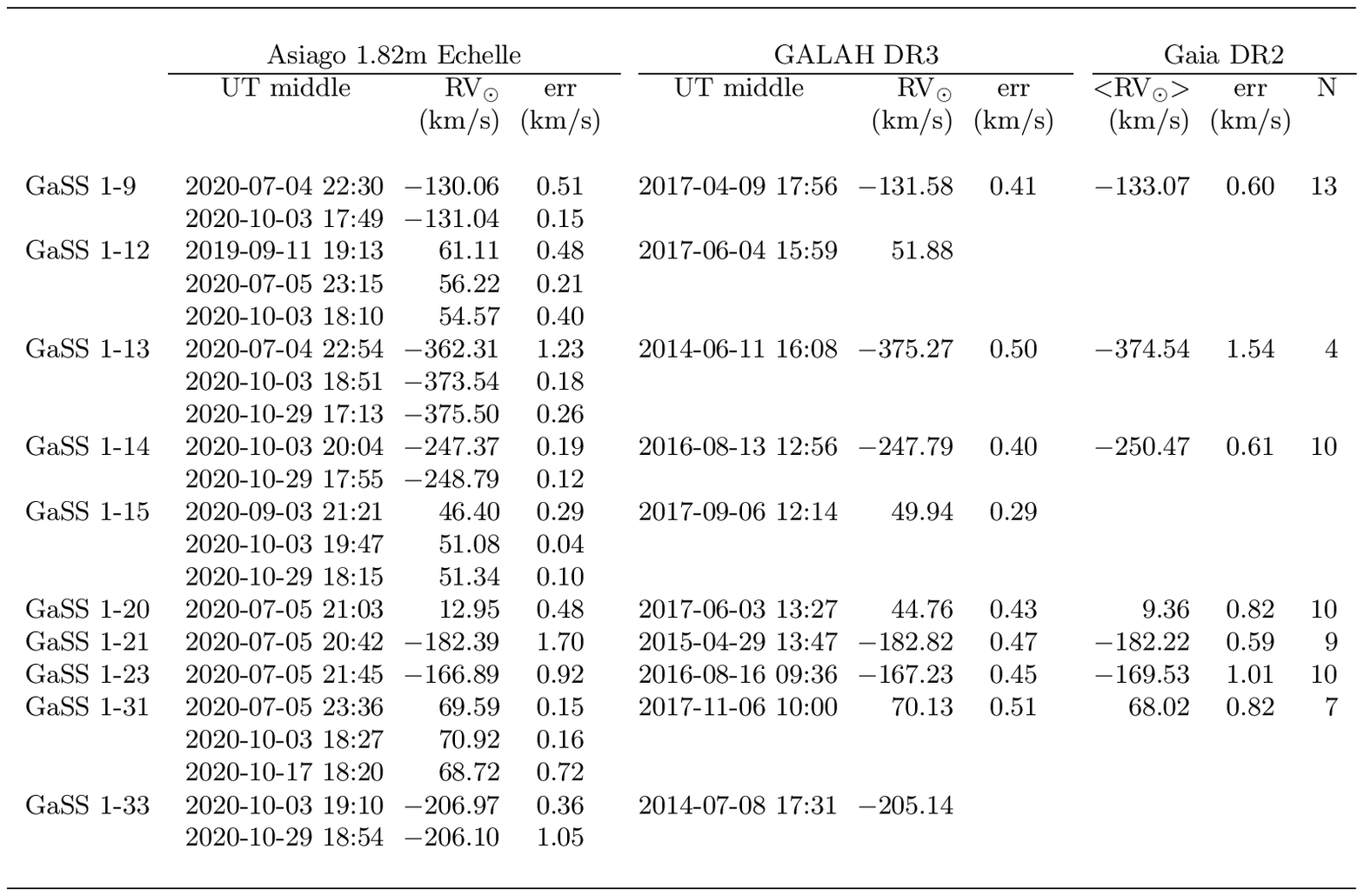}
\label{tab:rvs}
\end{table*}

    \begin{figure}
	\includegraphics[width=\columnwidth]{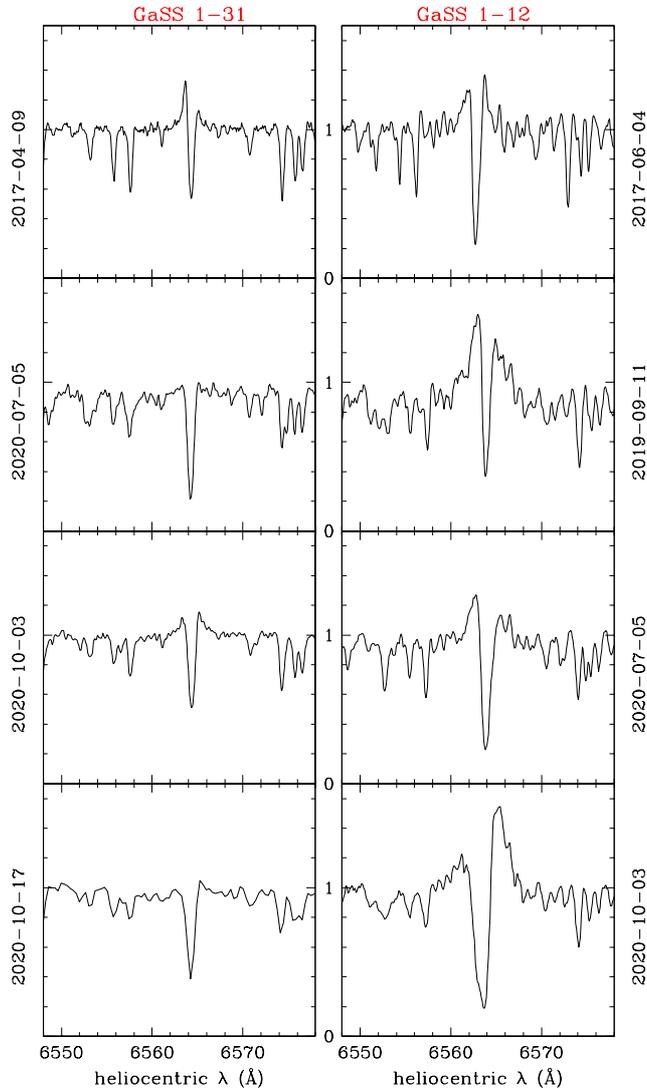}
        \caption{Example of variability observed on H$\alpha$  comparing
        the original GALAH spectra (top row) with Asiago 1.82m Echelle spectra
        obtained at later epochs (the left-bottom spectrum is from Varese
        84cm telescope). Only GALAH spectra are corrected for 
        telluric absorption lines.}
    \label{fig:Ha_var}
    \end{figure}

A quick summary of the photometric data extracted from spectra is presented
in Fig.~\ref{fig:BV_apass}.  The $B$ and $V$ data lie well on a 1:1 relation
with APASS mean data, the scatter being at least in part related to the
variability all objects display (as per column $\Delta V$ of
Table~\ref{tab:main_table}).  The difference in the $B$$-$$V$ colour
(indicative of the accuracy of the {\it slope} of fluxed spectra) between
APASS and spectra appears well distributed around 0.0 with a FWHM of about
0.1 mag.  The most interesting panel of Fig.~\ref{fig:BV_apass} is the last
one, showing the distribution of $U$$-$$B$ measured on spectra compared with
the distribution of $U$$-$$B$ for a hundred M giants of the Solar
neighborhood (suffering from negligible reddening) as measured by
\citet{1994A&AS..105..311F}.  It is clear how, on average, the $U$$-$$B$
colour of candidate {\it acc}-SySt is bluer that those of (supposedly)
single, normal M giants, supporting the notion that for a sizeable fraction
of them we have actually observed an excess in $U$ as coming from the
companion and the accretion disk.  The {\it acc}-SySt scoring an "spc" in
Table~\ref{tab:main_table} are those with an $U$$-$$B$ bluer than +0.4 in
Fig.~\ref{fig:BV_apass}, and "spc:" for those between +0.4 and +0.8.

\subsection{Radial velocity variability} \label{sec:rv_var}

Armed with evidence from Fig.~\ref{fig:sulyn_temp} that the low amplitude,
irregular variability affecting the M giant in SU~Lyn does not reverberate
into changes in radial velocity, we observed some of the 53-sample north of
$-$25$^\circ$ in declination with the Echelle spectrograph mounted on the
Asiago 1.82m telescope.  The results are summarized in Table~\ref{tab:rvs}. 
As a check, some normal stars from the initial 15k-sample and a few
radial pulsators from the 223-sample were also observed, with results
presented in Table~\ref{tab:helio_rv}. In both tables, the quoted error for
Asiago RVs is the {\it internal} error, i.e.  that derived by comparing the
RV obtained from different Echelle orders by cross-correlation with spectra
of IAU radial velocity standards of M spectral type.  We selected to limit
the measurement to 5 of the 32 orders covered by the Asiago Echelle
spectrograph, favoring those at redder wavelengths (where S/N is higher),
and avoiding the orders with TiO band-heads or telluric atmospheric
absorptions that would affect the cross-correlation results.  A comparison
of Asiago RVs with the GALAH rv\_guess and Gaia DR2 values in
Table~\ref{tab:helio_rv} for the GALAH M giants sub-sample returns a
negligible offset and an rms $\sim$1.5~km~s$^{-1}$.  The latter could in
part be accounted by variability of epoch radial velocities intrinsic to the
sources.

From the data in Table~\ref{tab:rvs}, a clear RV variability has been
observed for five objects, and borderline for an additional three.  They are
marked with "var" in column "Asiago RV" of Table~\ref{tab:main_table}.  We
favor an interpretation in terms of orbital motion for this RV variability.

\subsection{Emission line variability}

The {\it acc}-SySt of this paper show a clear variability of the intensity
and profile of H$\alpha$ emission line on spectra collected with the Asiago
1.22m and 1.82m telescopes.  An example is presented in
Fig.~\ref{fig:Ha_var}.  GaSS~1-31 showed a clear emission in H$\alpha$ on
the GALAH spectrum recorded on Apr 9, 2017, a weaker one on the Asiago
spectrum for Oct 2, 2020, and no emission on Jul 5, 2020 (see also
Section~\ref{sec:xray_uv_obs} below) and on Oct 17, 2020 (this last spectrum
has been obtained at a lower resolving power - 12,000 because of 2$\times$2
CCD binning - with an Echelle spectrograph mounted on the Varese 84cm
telescope).  The H$\alpha$ of GaSS 1-12 has been always observed in emission
on all three visits with the Asiago 1.82m, with a varying degree of
intensity and velocity of the emission relative to the M giant.  Most
interestingly, the last Asiago spectrum for Oct 3, 2020 shows the appearance
of a second narrow absorption, blue shifted with respect to the primary one,
an event reminiscent of what happened to SU~Lyn in 2017 (cf. 
Fig.~\ref{fig:sulyn_temp}).

\subsection{Optical flickering} \label{sec:flickering}

The accretion in binaries is characterized by chaotic processes (e.g. 
density fluctuations in the accretion flow from the donor to the accreting
star) that develop on time scales (seconds, minutes or a few hours) much
shorter than those of other sources of variability (orbital motion,
pulsation, rotation, outbursts, etc.).  The possibility to detect flickering
on optical photometric data depends primarily on its contrast with the
steady sources in the system: the fainter the two components of the binary,
the easier for flickering to become observable (e.g. 
\citealp{2001cvs..book.....H}).  Therefore, if flickering is easy to observe
in cataclysmic variable stars that radiate about 1~L$_\odot$ at optical
wavelengths, it is an entirely different matter for symbiotic binaries in
which the optical luminosity of the cool giant alone amounts to
10$^2$~L$_\odot$.

The regions in the disk responsible for flickering are hot
\citep{1995CAS....28.....W,2020AN....341..430Z}, and their emission is
consequently rising toward the blue; that of the giant rises instead toward
the red.  Therefore, to reduce the background glare of the giant it is
convenient to go as blue as possible \citep{2005A&A...440..995S}.  Detection
of flickering at satellite ultraviolet wavelengths is straightforward as
demonstrated by the {\it Swift} observations of SU~Lyn by Mk16.  On
ground-based observations, the bluest photometric band is $U$ (in its many
variants: Landolt's $U$, $u$ from Stromgren, or SLOAN $u'$).  $U$ band was
relatively easy to observe in the old times of photo-electric photometers
mounted on all-reflecting telescopes with bare-aluminum surfaces, thanks to
the high instrumental throughput (no coatings or refractive components) and
the excellent sensitivity of photo-multipliers like the standard RCA 31034A
and Hamamatsu R934-02.  Many searches for flickering in symbiotic stars were
performed at that time (e.g.  \citealp{1978MNRAS.185..591S,
1990AcA....40..129M, 1996A&AS..116....1T}.  The advent of CCDs, with their
much lower sensitivity at $U$-band wavelengths, badly affected the study of
flickering, with rarer attempts carried out (e.g. 
\citealp{1996AJ....111..414D, 2001MNRAS.326..553S, 2006AcA....56...97G,
2011IBVS.5995....1Z}).  To further complicate the matter, the sensitivity of
CCDs, that extends up and beyond 1~$\mu$m, allows signal to be recorded from
the red-leak affecting most $U$ filters, especially those of the modern
multi-layer dielectric type.  Such a red-leak may be so severe in the case
of cool giants, that the fraction of recorded photons coming through the
red-leak easily outnumbers those going through the proper transmission
profile of the $U$-band \citep{2012BaltA..21...22M}.

To search for flickering among our candidate {\it acc}-SySt we used a 50cm
telescope, with a 40arcmin well corrected field of view and quality
photometric filters, operated robotically for ANS Collaboration in Chile at
San Pedro de Atacama.  For the above described limitations about $U$
filters, we selected to search for flickering in $B$-band, with interspersed
observations in $V$-band serving to construct the $B$$-$$V$ colour base for
the transformation from the instantaneous local photometric system to the
Landolt's standard one.  Local photometric sequences were extracted from the
APASS multi-epoch, all-sky $B$$V$$g$$r$$i$ photometric survey
\citep{2016yCat.2336....0H} in its DR8 version.

Given the much brighter emission of the M giant in $B$ compared to $U$, the
expected amplitude of observable flickering decreases to (at most) a few
hundredths of a magnitude.  To be detected, such a tiny amount of variation
requires highly accurate observations and careful data reduction, under
clear and stable skies as usually enjoyable in Atacama.  We adopted
integration times in $B$ of 30sec for the brightest targets and 5min for the
faintest, with 60sec for the bulk of objects, thus ensuing that all
objects have been observed at high S/N ($>$100 on each single point) while
remaining short of the saturation thresholds and non-linear regime. Every
10 exposures in $B$, one was obtained in $V$ to compute the coefficients of
the transformation colour equations.  The procedure was repeated identically
for 70min for each program star.  The photometry was performed for most
objects in aperture mode, their high galactic latitudes implying sparsely
populated fields and therefore no need to revert to PSF-fitting, which was
instead necessary for three targets.  On average, the 50 field stars closest
on the image to the symbiotic star, of a similar magnitude and well isolated
from neighboring stars were also measured on all recorded images in exactly
the same way as the symbiotic star.  The photometry of these 50 field stars
was then inspected looking for those with a $B$$-$$V$ colour as close as
possible to the symbiotic star in the center.  Typically 5 such stars were
found, with a few cases scoring just 2 and others up to 8.  These field
stars with magnitude and colour closely matching those of the symbiotic star,
serve as samplers of the observational noise above which the flickering has
to be detected: the closely similar colour nulls the effects of
differential atmospheric transmission, and the closely similar brightness
nulls the non-linear effects in the differential statistical noise.

The program {\it acc}-SySt subjected to the search for flickering are listed
in Table~\ref{tab:hjd_b_v}, together with the HJD of the central image for
the 70min time series, and the average of $B$ and $V$ measured magnitudes
during such time series.  The error of such $B$ and $V$ data is totally
negligible in its Poissonian component, while the amount due to
transformation from the local system to the APASS one (closely adherent to
Landolt's) is systematic and constant for a given object and amounts on
average to 0.008 mag (which obviously cancels out when comparing flickering
data within a given time-series).
 
\begin{figure}
	\includegraphics[width=\columnwidth]{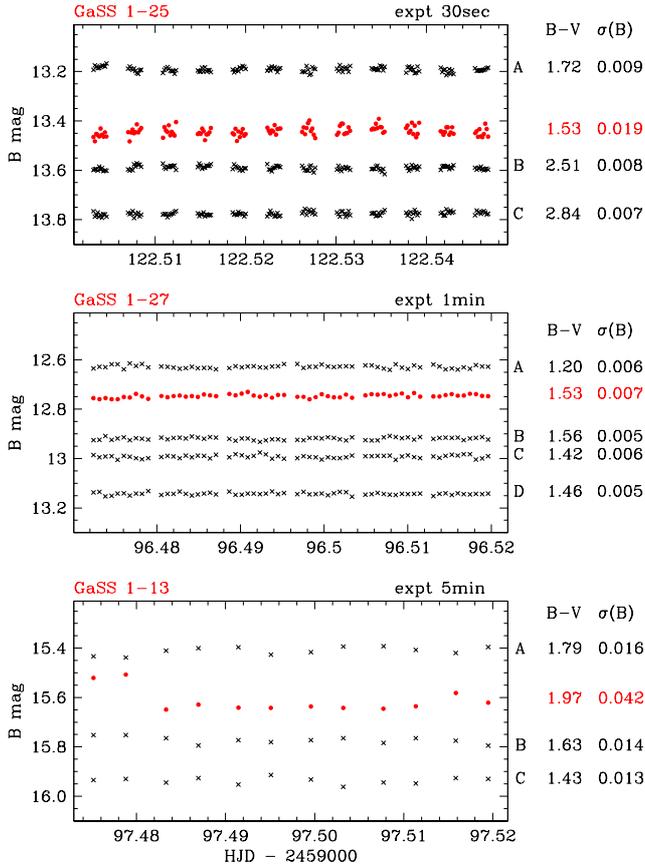}
        \caption{Examples of the $B$-band time series collected on the
        program stars, with 30sec, 1min and 5min sampling time (from top to
        bottom), in search for flickering from the accretion disk.  The
        regular gaps correspond to the acquisition of $V$-band frames for
        the calibration of colour equations to transform data to the standard
        Landolt system.  The measured $B$$-$$V$ colour of the symbiotic star
        (in red and filled circles) and of nearby field stars of similar
        brightness and colour (in black and crosses), is marked to the right,
        together with the dispersion around the median of the plotted
        $B$-band data.  Symbiotic stars on the top and bottom panels clearly
        show flickering well in excess of the noise affecting field stars,
        while that on the center panel is constant at the same noise level
        of the field stars.}
    \label{fig:flickering}
    \end{figure}

An example of the collected data is presented in Fig.~\ref{fig:flickering},
with one object each for the 30sec, 1min and 5min sample times.  At the top
and bottom panels we show objects with a clear detection of flickering well
in excess of the noise affecting the field stars as usually quantified
by their respective rms deviation from the mean
\citep{1996AJ....111..414D}, while in the center we present a
non-flickering symbiotic that remains stable at a few millimag-level through
the 70min monitoring. GaSS 1-25 at the top of
Fig.~\ref{fig:flickering} exemplifies the frequent case of incessant
flickering (e.g.  RS~Oph as observed by \citealp{1996AJ....111..414D}),
while the behavior displayed by GaSS 1-13 at the bottom of
Fig.~\ref{fig:flickering}, i.e.  isolated flarings over an otherwise
relative flat background, is a close match to that observed by
\citet{2017AN....338..680Z} in EF~Aql.

The objects for which flickering was detected well above observational noise
are marked with an 'F' in the 'Flk' column of Table~\ref{tab:main_table}. 
The detectability of flickering and its amplitude are usually dependent on
the actual epoch of observation \citep{2018BlgAJ..28...42S}.  Orbital phases
with the hot spot in plain view are favoured (WD at quadrature moving toward
superior conjunction), while others can be detrimental, e.g.  during
eclipses or passage at apoastron in a highly eccentric orbit (with
consequent reduction in mass transfer rate).  Therefore, stars in
Table~\ref{tab:main_table} that failed to show flickering at the single
epoch of our observation could do otherwise at a later revisit.  This
is also the case for SU~Lyn.  When it was observed in late 2015 by Mk16 its
was in a high accretion state and it flickered wildly, while it was in a low
accretion state and did not flicker when \citet{2018arXiv181103317D}
performed their observations in January 2018.  Similarly, while large
amplitude flickering has always been the rule for MWC~560 (eg. 
\citealp{2020MNRAS.492.3107L}), it has disappeared altogether during the
last couple of years \citep{2019ATel13236....1Z}.

We postpone a detailed analysis of the data collected during our search
for flickering for the program stars to a follow-up paper.  Derivation of
temperature and radius of the region(s) responsible for the observed
flickering (e.g.  \citealp{2020AN....341..430Z}) is straightforward but
largely out of the goals of this paper, as it is the search about possible
quasi-periodicities or signatures of the WD rotation (e.g. 
\citealp{1999ApJ...517..919S, 2019MNRAS.489.2930Z}).  Such analysis will
strongly benefit from further observations we have planned for the SySt
reported in this paper and are carrying out also for other SySt assigned to
later papers in this series.

\subsection{Satellite X--ray/UV observations} \label{sec:xray_uv_obs}

The presence and nature of the accreting source in {\it acc}-SySt systems,
either a WD or a NS, is best defined by satellite observations performed in
the ultraviolet and X--rays, where the peak of the emissivity for
accretion-induced processes is located (e.g.  \citealp{2007A&A...464..277M}
and references therein, and also \citealp{kuranov2015} for a recent review). 
The X--ray/UV properties of the {\it acc}-SySt prototype SU~Lyn have been
investigated by Mk16, \citet{lopes2018}, and \citet{2021MNRAS.500L..12K}
based on observations with {\it Swift}, {\it NuSTAR} and ASTROSAT
satellites.  They found SU~Lyn to show (i) a large (by a factor of a few
hundred in flux) UV excess with respect to a `normal' red giant; (ii)
long-term (months to years) hard X--ray variations by an order of magnitude
between low- and high-level states in the 15--35 keV hard X--ray emission;
and (iii) fast UV variability during the high X--ray states (flickering). 
In the latter ones, the average X--ray flux was $\sim$10$^{-11}$ erg
cm$^{-2}$ s$^{-1}$ in the 0.3--50 keV band, with the spectrum modelled with
a thermal plasma of temperature $kT \sim 20$ keV, plus a fluorescent iron
emission at 6.4 keV, absorbed by a hydrogen column N$_{\rm H}$ as large as
$\sim$3$\times$10$^{22}$ erg cm$^{-2}$.  This absorption has however little
effect on the UV excess observed from this source.  At a distance of 650 pc,
this corresponds to unabsorbed X-ray luminosities of up to $\sim$10$^{33}$
erg s$^{-1}$ in the high state, and a factor of 10 larger in the UV.  The
passage from low- to high-level states in the UV and X-ray emission is
attributed to large excursions in the transfer and accretion rates.

\begin{table*}
\centering
\caption{Results of the {\it Swift} XRT and UVOT observations of some of the
{\it acc}-SySt presented in this paper.  The fluxes are in units of
10$^{-13}$ erg cm$^{-2}$ s$^{-1}$.  The bottom three objects are radial
pulsators observed for comparison.  The data for 20 35 04.52 $-$05 47 35.3
include also additional 485sec exposure obtained on 8 Sep.  2019.}
\label{tab:swift}
\begin{tabular}{lcccccccccrc}
\hline
&&&\\
name~~/~~RA,DEC & Obs. & Start     & XRT obs.     & 0.3--10 keV & 0.3--10 keV & UVOT obs.    & UVM2 \\
     & date & time (UT) & duration (s) & count rate  & flux        & duration (s) & mag. \\
&&&\\
GaSS 1-11                  &	13 Jun. 2020 &	12:05 &	1885 &	$<$3.0e-3 &	$<$2.6e-13 &	1864 &	17.33$\pm$0.06 \\
GaSS 1-12                  &	16 Sep. 2019 &	00:02 &	1516 &	$<$3.2e-3 &	$<$2.8e-13 &	1701 &	16.21$\pm$0.04 \\
GaSS 1-31                  &	14 Jun. 2020 &	01:01 &	839  &	$<$3.7e-3 &	$<$3.3e-13 &	830  &	$>$19.92       \\
&&&\\
05 11 30.88 $-$61 29 03.6  &	10 Jun. 2020 &	06:31 &	1973 &	$<$1.7e-3 &	$<$1.5e-13 &	1948 &	$>$20.65       \\
20 06 56.32 $-$28 35 32.3  &   11 Oct. 2019 &	00:49 &	1536 &	$<$2.5e-3 &	$<$2.2e-13 &	1527 &	$>$20.71 \\ 
20 35 04.52 $-$05 47 35.3  &   04 Sep. 2019 &   12:23 &	1504 &	$<$3.6e-3 &	$<$3.2e-13 &	1485 &	$>$20.82 \\ 
&&&\\
\hline
\end{tabular}
\end{table*}

In anticipation of a devoted observing campaign, we have performed some
quick exploratory observations of three of the {\it acc}-SySt discovered in
this paper with the {\it Swift} satellite \citep{2004ApJ...611.1005G}, and
for comparison of three of the identified radial pulsators considered as a
control sample.  The observations have been carried out in
Target-of-Opportunity mode.  This type of observations is generally limited
to roughly 2000 seconds per object, so this can be considered as a quick and
rather shallow survey of our target sample.  The pointings were performed in
two time slots: the first one in September-October 2019, and the second one
in June 2020.  The results are summarized in Table~\ref{tab:swift}.

Our {\it Swift} observations were acquired with the on-board instruments
X-Ray Telescope (hereafter XRT, \citealp{2005SSRv..120..165B}) and
UltraViolet-Optical Telescope (hereafter UVOT,
\citealp{2005SSRv..120...95R}).  The XRT allows covering the X--ray band
between 0.3 and 10 keV band, whereas UVOT data were collected using the UV
filter $UVM2$ (reference wavelength: 2246 \AA; see
\citealp{2008MNRAS.383..627P} and \citealp{2011AIPC.1358..373B} for
details).  On-source pointings were simultaneously performed with the two
instruments and lasted between $\sim$800 and $\sim$2000 s.

All observations were reduced within the {\sc ftools} environment
\citep{1995ASPC...77..367B}.  The XRT data analysis was performed using the
{\sc xrtdas} standard pipeline package ({\sc xrtpipeline} v.  0.13.4) in
order to produce screened event files.  All X--ray data were acquired in
photon counting (PC) mode \citep{2004SPIE.5165..217H} adopting the standard
grade filtering (0-12 for PC) according to the XRT nomenclature.  For each
source, scientific data were extracted from the images using an extraction
radius of 47$''$ (20 pixels) centered at the optical coordinates of the
source, while the corresponding background was evaluated in a source-free
region of radius 94$''$ (40 pixels) within the same XRT acquisition.  In all
cases, no emission above a signal-to-noise threshold $S/N=3$ was detected. 
The XRT count rate upper limits in the 0.3--10 keV range were then measured
within the {\sc xspec} package.

X--ray flux limits were determined using the {\sc webpimms} online
tool\footnote{{\tt
https://heasarc.gsfc.nasa.gov/cgi-bin/Tools/w3pimms/w3pimms.pl}} by assuming
a spectral model similar to that of Mk16 for SU~Lyn, i.e.  a thermal plasma
emission with temperature $kT$ = 17 keV absorbed by a column density N$_{\rm
H}$ = 2.9$\times$10$^{22}$ cm$^{-2}$, which implies a count rate-to-flux
conversion factor of $\sim$8.8$\times$10$^{-11}$ erg cm$^{-2}$ s$^{-1}$
cts$^{-1}$.  We note that, with this model, the unabsorbed fluxes (in
Table~\ref{tab:acc_lum}) in this same band are about 50\% larger than the
absorbed ones.

Count rates on Level 2 (i.e.  calibrated and containing astrometric
information) UVOT images at the position of the objects of our sample were
measured through aperture photometry using 5$''$ apertures, whereas the
corresponding background was evaluated for each image using a combination of
several circular regions in source-free nearby areas.  Magnitudes and upper
limits were measured with the {\sc uvotsource} task.  The data were then
calibrated using the UVOT photometric system described by
\citet{2008MNRAS.383..627P}, and we included the recent (November 2020)
fixings recommended by the UVOT team.

Of the three {\it acc}-SySt observed with {\it Swift}, two were found to be
strong UV emitters, confirming their symbiotic nature.  The third, GaSS 1-31
was not detected.  There may be a clear reason for that, though.  The {\it
Swift} pointing was carried out on 14 June 2020.  Three weeks later, on 5
July 2020, we observed the same star with the Asiago 1.82m and the Echelle
spectrograph.  The H$\alpha$ profile for that date is presented in
Fig.~\ref{fig:Ha_var}, and doesn't show the faintest trace of an emission. 
In addition, the equivalent width of the H$\alpha$ absorption is same as for
field stars of the same spectral type (and greater than for the other dates
in Fig.~\ref{fig:Ha_var}), indicating no partial filling from even very
faint emission.  In short, it looks like the emission from the accretion
disk was 'switched-off' at the time of the {\it Swift} pointing, with
consequent non-detection.  An eclipse behind the M-giant seems improbable in
view of the limited variability of the epoch GALAH and Asiago radial
velocities listed in Table~\ref{tab:rvs} (and their similarity with the
average velocity provided by Gaia DR2), precluding a high orbital
inclination for GaSS 1-31.  It rather seems a drastic reduction in the mass
accretion rate was taking place at the time of {\it Swift} observations. 
This could have been caused by a reduction in the wind blown-off by the M
giant or passage at apoastron in a highly eccentric orbit.  The spectrum
taken a hundred days later (that for 3 Oct 2020 in Fig.~\ref{fig:Ha_var})
shows that accretion has weakly resumed, with both a reduction in the
equivalent width of H$\alpha$ absorption (partially filled-in) as well as
emission wings extending above the local continuum.

None of the three radial pulsators we tried with {\it Swift} were detected,
in spite of having been selected among those showing the strongest H$\alpha$
and H$\beta$ emission on GALAH spectra (far stronger than typically observed
in {\it acc}-SySt).  This supports the clear-cut role of satellite UV
observations in segregating {\it acc}-SySt \citep{2015ApJ...810...77S}, in
particoular from radial pulsators of similar spectral appearance on optical
spectra.

A few final words are in order to explain the non-detection of GaSS 1-11 and
1-12 in X-rays in our shallow observations.  The upper limit to their flux
is about 0.01$\times$ the flux Mk16 recorded from SU~Lyn during their {\it
Swift} observations.  Those observations of SU~Lyn were obtained during an
exceptionally bright state of the accretion disk (cf. 
Fig.~\ref{fig:sulyn_temp} and the discussion at the end of
Section~\ref{sec:sulyn}), while the flux at other times (when the H$\alpha$
shown by SU~Lyn is much weaker and more similar in appearance to that
revealed by GALAH spectra for our sample of 33 {\it acc}-SySt) was just 1/10
of that (cf.  \citealp{lopes2018}).  There are two other factors playing
against GaSS 1-11 and 1-12.  The most important is the distance.  SU~Lyn is
just 0.65 kpc away, while GaSS~1-11 and 1-12 are respectively at 6.6
and 4.6 kpc distance according to Gaia $e$DR3 parallaxes, corresponding to
dilution factors of 100 and 50. In addition, SU Lyn is seen face-on, while
the unknown orbital inclination of GaSS 1-11 and 1-12 introduces a reduction
in flux proportional to $\cos i$ and a further reduction for limb-darkening
reasons.  Considering all factors together, it is no wonder that GaSS 1-11
and 1-12 fell below the X--ray detection threshold of snapshot ToO with {\it
Swift}.

\subsection{Accretion luminosities} \label{sec:acc_lum}

\begin{table}
\centering
\caption{Unabsorbed accretion luminosity from the {\it Swift} data in
Table~\ref{tab:swift} (see Section~\ref{sec:acc_lum} for details).}
\label{tab:acc_lum}
\begin{tabular}{cccc}
\hline
          &                &                    &                        \\
symbiotic & L$_{\rm UV}$   & L$_{\rm UV}$       &  M$_{\rm acc}$         \\
star      & (erg s$^{-1}$) &  (L$_\odot$)       & (M$_\odot$ yr$^{-1}$)  \\
          &                &                    &                        \\
GaSS 1-11 &    1.43e34     &       3.7          &          1.7e-9        \\ 
GaSS 1-12 &    3.30e34     &       8.5          &          4.0e-9        \\ 
GaSS 1-31 & $<$4.29e32     &    $<$0.11         &      $<$5.1e-11        \\ 
          &                &                    &                        \\
\hline
\end{tabular}
\end{table}

\begin{table}
\centering
\caption{Flux recorded in H$\alpha$ and corrected for reddening
(F(H$\alpha$)), isotropic luminosity radiated in H$\alpha$ at Gaia $e$DR3
parallax (L(H$\alpha$)), and accretion luminosity (L$_{\rm acc}$) derived as
described in Section~\ref{sec:acc_lum} for the program star with available
Asiago spectra to flux-calibrate GALAH spectra.}
\label{tab:Ha_flux}
\begin{tabular}{cccc}
\hline
           &           &              &         \\
symbiotic  & F(H$\alpha$)             & L(H$\alpha$)   &  L$_{\rm acc}$ \\
star       & (erg cm$^{-2}$ s$^{-1}$) & (erg s$^{-1}$) &  (L$_\odot$) \\
           &           &              &         \\
 GaSS 1-12 & 1.391E-13 & 3.52e+32     &   14    \\
 GaSS 1-14 & 3.307E-14 & 2.93e+32     &   12    \\
 GaSS 1-15 & 3.205E-14 & 3.23e+31     &    1    \\
 GaSS 1-23 & 7.257E-14 & 2.00e+32     &    8    \\
 GaSS 1-25 & 1.179E-13 & 2.61e+32     &   10    \\
 GaSS 1-28 & 2.785E-14 & 1.88e+32     &    7    \\
 GaSS 1-31 & 2.367E-13 & 5.24e+32     &   22    \\
           &           &              &         \\
\hline
\end{tabular}
\end{table}

We have previously stated that the aim of the present paper is to
search for new SySt accreting at low rates similar to SU~Lyn.  Mass
accretion rate ($M_{\rm acc}$) and accretion luminosity ($L_{\rm acc}$) are
related by the usual
\begin{equation} L_{\rm acc} = G \frac{M_\ast
M_{\rm acc}}{R_\ast} 
\end{equation} 
where $M_\ast$ and $R_\ast$ are the mass and radius of the accreting object,
respectively.  The accretion on a WD, as it is the case for SU~Lyn, radiates
most of its output in the UV and much less into X-rays, while the reverse
applies to neutron stars and their much deeper potential well.

The UVM2 magnitudes of the new SySt in Table~\ref{tab:swift} correspond to
flux densities (in units of 10$^{-15}$ erg cm$^{-2}$ s$^{-1}$ \AA$^{-1}$) of
0.54, 1.52 and $<$0.05, for GaSS 1-11, 1-12 and 1-31, respectively (from the
countrate-to-flux calibration given in \citealp{2008MNRAS.383..627P} and
\citealp{2011AIPC.1358..373B}).  Then, considering the distances and
$E(B-V)$ colour excesses reported in Table~\ref{tab:main_table}, correcting
for the Galactic reddening according to the prescription in section 3.2 of
\citet{2008ApJ...672..787K} and the UV correction coefficients in their
table 5, and assuming for the three objects a flat spectral distribution in
the UV as Mk16 did for SU~Lyn, we obtain the unabsorbed UV luminosities
listed in Table~\ref{tab:acc_lum}.  In the hypothesis of accretion onto a WD
of mass 1 M$_\odot$, as Mk16 has inferred for SU~Lyn, in
Table~\ref{tab:acc_lum} we report also the corresponding accretion rates.

For the other new SySt that were not observed by {\it Swift}, an estimate of
$L_{\rm acc}$ can be derived from the flux recorded in the H$\alpha$
emission.  If an emission line forms in an accretion disk, the flux radiated
by the line and the flux emitted by the disk as a whole are related.  The
correlation can be tight, with temporal variations in $L_{\rm acc}$ that may
be accurately tracked and replicated by $L$(H$\alpha$) (cf. 
\citealp{2019MNRAS.488.5536M}, their fig.  14).

GALAH spectra are unfortunately not flux-calibrated.  To circumvent the
problem, we look for assistance from the fluxed Asiago spectra.  After
de-reddening the Asiago spectra for the $E_{B-V}$ values listed in
Table~\ref{tab:main_table}, we used them to flux-calibrate the GALAH spectra
of the corresponding objects.  Being primarily interested in deriving just
an order-of-magnitude, and noting the new SySt exhibit only a limited
photometric variability (cf.  Section~\ref{sec:filtering_radialp} above), we
ignore the fact that Asiago and Galah spectra are not simultaneous.  On the
template-subtracted version of the flux-calibrated GALAH spectra, we
measured the integrated flux of H$\alpha$ emission line $F$(H$\alpha$)
(therefore already corrected for reddening), and list it in
Table~\ref{tab:Ha_flux}, where we consider only the new SySt with
Gaia~$e$DR3 distances in Table~\ref{tab:main_table}.  If $d$ is the
distance, the luminosity $L$(H$\alpha$) radiated isotropically by the disk
in H$\alpha$ is obviously
\begin{equation} 
L{\rm (H\alpha)} = 4 \pi d^2 F{\rm (H\alpha)} 
\end{equation} 
As customary practice, we may write a power-law proportionality between 
the luminosity radiated in the line and by the disk as a whole:
\begin{equation}
\log (L_{\rm acc}) = \alpha_{\rm H\alpha} + \beta_{\rm H\alpha}\times \log(L{\rm (H\alpha)})
\end{equation}
where $\alpha_{\rm H\alpha}$ and $\beta_{\rm H\alpha}$ are coefficients
specific to H$\alpha$.  Adopting their value from
\citet{2011A&A...535A..99M}, we have derived the $L_{\rm acc}$ listed in the
last column of Table~\ref{tab:Ha_flux}.  The accretion luminosities derived
directly from {\it Swift} ultraviolet data or estimated from the flux
radiated in H$\alpha$ are similar, and confined to the 1$-$20~L$_\odot$
range, the same roughly spanned by SU~Lyn while varying between high- and
low-states.  Interestingly, there is an estimate of $L_{\rm acc}$ for
GaSS~1-12 both from the {\it Swift} UV observations and also from H$\alpha$
emission line flux, and the two differ by less than a factor of 2.  While
this can be read as an indirect confirmation of the reliability of $L_{\rm
acc}$ estimated from H$\alpha$, it also points to the constant variability
of $L_{\rm acc}$ observed in {\it acc}-SySt.  In fact, the $F$(H$\alpha$) of
GaSS~1-12, measured on the fluxed version of the Echelle spectra presented
in Fig.~\ref{fig:Ha_var}, varies by more than the above 2$\times$ factor.

\subsection{Orbital-like periodicities} \label{sec:orbital_periodicites}

    \begin{figure}
	\includegraphics[width=\columnwidth]{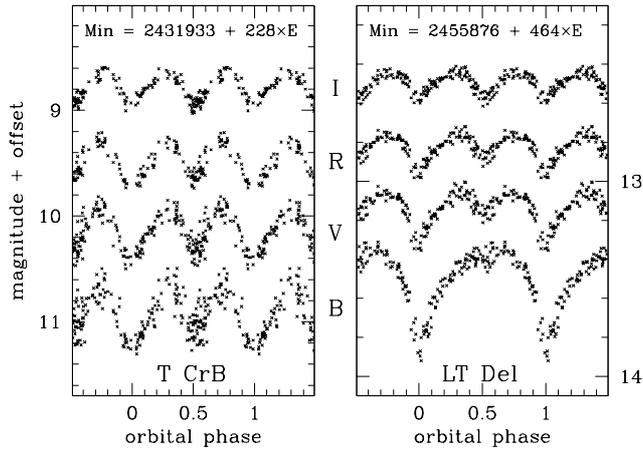}
        \caption{Examples of the modulation of the $B$$V$$R$$I$ lightcurves
        induced by orbital motion in non-eclipsing symbiotic stars (see
        Section~\ref{sec:orbital_periodicites}).  Note the increasing impact
        of flickering going toward bluer bands for the accreting-only T CrB,
        and its absence for the burning-type LT Del.  The lightcurves of
        both objects are obtained from data extending over at least 10
        consecutive orbital cycles.  Adapted from
        \citet{2019MNRAS.488.5536M}.}
    \label{fig:orb_modul}
    \end{figure}

The long-term lightcurves of SySt are usually reconstructed from
multi-band photometric campaigns (e.g.  \citealp{2007AN....328..909S,
2012AN....333..242S, 2019CoSka..49...19S}), from digging historical plate
archives (e.g.  \citealp{2002A&A...386..237M}), or searching the public
databases of all-sky patrol surveys (e.g.  \citealp{2009AcA....59..169G,
2013AcA....63..405G, 2021CoSka..51..103M}).  Aside from the obvious effect
of eclipses for systems seen at high inclination, the orbital motion may
affect also the lightcurve of SySt seen at intermediate inclinations.  The
basic type of orbital-induced modulation of the lightcurve of non-eclipsing
SySt is shown in Fig.~\ref{fig:orb_modul} (adapted from
\citealp{2019arXiv190901389M}), aiming to highlight the interplay between
the irradiation effect (better visible in the bluest photometric bands and
for {\it burn}-SySt) and the ellipsoidal distortion (which is more prominent
in the reddest bands and in {\it acc}-SySt).

We have already mentioned that the M3III giant of T CrB fills its Roche
lobe, causing an ellipsoidal deformation of the surface, that along an
orbital period presents to observers twice the larger projected area (the
two maxima in the lightcurve) and twice the smaller one (the two minima). 
The reason for the change in brightness is primarily geometric, so its shape
is similar at all wavelengths as the left panel of Fig.~\ref{fig:orb_modul}
shows.  In that figure, the presence of flickering superimposed to the
regular orbital motion is quite obvious, in the form of a scatter of points
of an increasing amplitude moving from redder to bluer bands (cf. 
Section~\ref{sec:flickering} above).  A search for periodicity on such a
lightcurve would return {\it half} the value of the true period: 114 days
instead of 228.  It is the spectroscopic orbit that would solve the
uncertainty in favor of 228 days.

    \begin{figure*}
	\includegraphics[width=15.8cm]{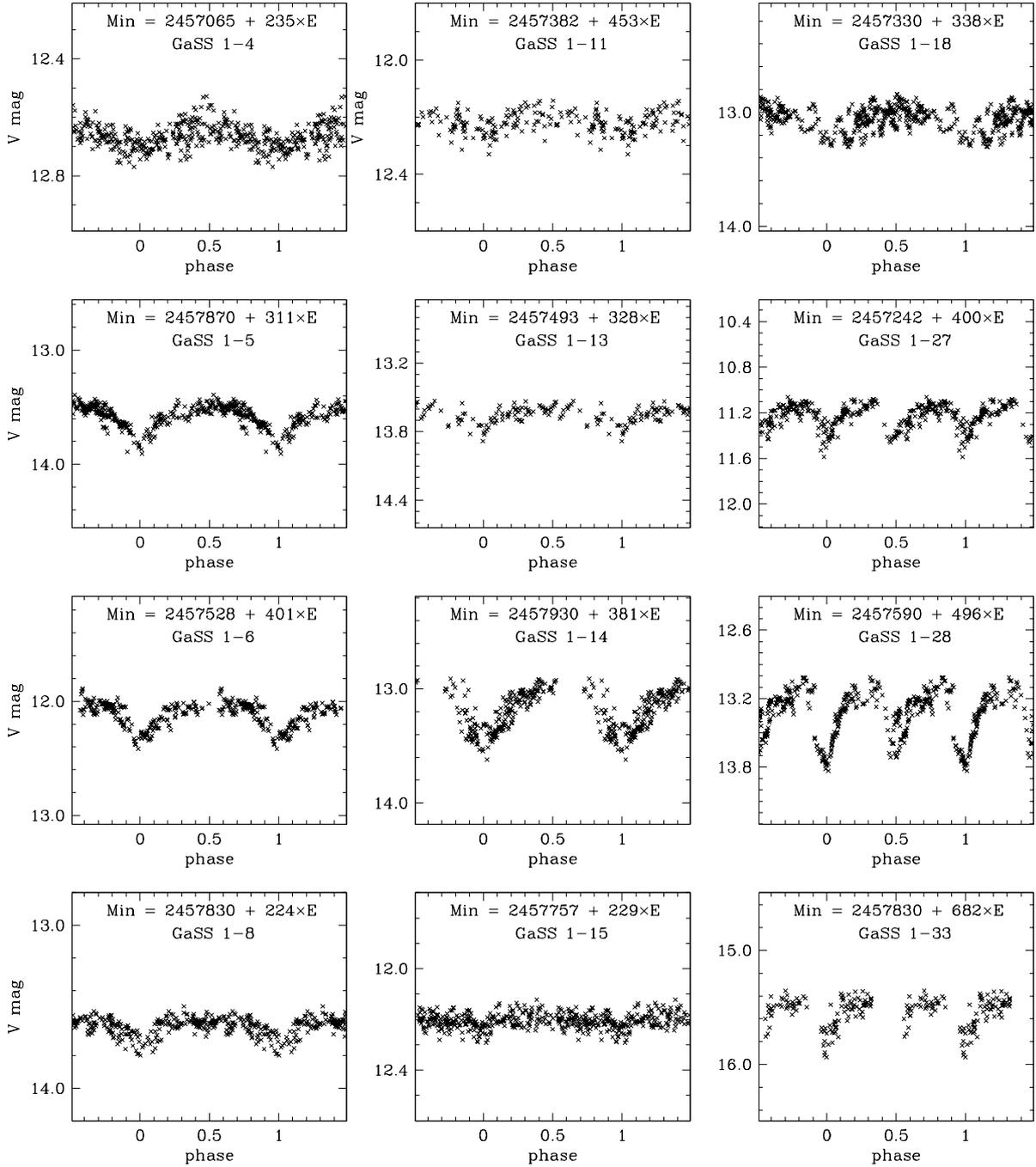}
        \caption{Phased light-curves (see ephemeris at the top of each
         panel) for the twelve {\it acc}-SySt marked in
         Table~\ref{tab:main_table} as showing a modulation reminiscent of
         orbital motion.}
    \label{fig:phased_lcs}
    \end{figure*}

The cool giant in LT Del, the other symbiotic star in
Fig.~\ref{fig:orb_modul}, equally fills its Roche lobe, but contrary to T
CrB, its WD is burning on its surface.  The hot and bright companion
irradiates the facing side of the cool giant, thus heating its photosphere
and partially ionizing the atmosphere above it, with the consequence to make
the system brighter when the irradiated side is in full view
\citep{1986syst.book.....K}.  The effect is larger going to bluer
wavelengths.  Thus, while the behavior of T CrB and LT Del is the same in
$I$-band where the ellipsoidal distortion of the cool giant dominates,
differences begin to emerge moving toward shorter wavelengths: the
irradiation effect takes the lead in shaping the $B$-band lightcurve of LT
Del.  If the support of radial velocities is required for T CrB, the orbital
period of LT Del can be fixed without uncertainty on the base of multi-band
photometry alone.  Unfortunately, most all-sky patrol surveys observe in
just one band, and a disentangling like that for LT~Del is not possible with
their data alone, leaving on the desk the ambiguity between P and 2$\times$P
for the true orbital period.

We have searched the ASAS-SN photometric data for the 33 candidate {\it
acc}-SySt looking for a type of lightcurve modulation that could be
orbitally induced along the lines just outlined for T CrB and LT Del in
Fig.~\ref{fig:orb_modul}.  We found a suitable shape for 12 of them, and
they are presented in Fig.~\ref{fig:phased_lcs}, with the corresponding
periods listed in Table~\ref{tab:main_table}.  Their length is rather
typical for symbiotic stars, the majority of known cases going from 7 months
to 2.5 years \citep{2003ASPC..303....9M}.

In two cases, GaSS 1-27 and GaSS 1-28, a slight difference in the depth of
minima seems enough to resolve the ambiguity between P and 2$\times$P. 
Given the limited time span of ASAS-SN coverage, such a difference could
also result from other causes for photometric variability, and needs
confirmation over more orbital cycles.  Even more so for a possible third
case for unequal minima, that of GaSS 1-33, and its sparsely covered
lightcurve (cf Fig.~\ref{fig:phased_lcs}).  A borderline case is that of
GaSS 1-8, with just a hint of a secondary minimum around phase 0.5 which
could be spurious. The modulation displayed by GaSS 1-28, while
somewhat larger than usual, is a fine match in both shape and amplitude to
that of Hen~3-1591, whose modulation \citet{2007BaltA..16...37G} attribute
to orbital modulation and report that it is confirmed by variations in
radial velocity.

The remaining objects in Fig.~\ref{fig:phased_lcs} show equal-depth minima,
so the actual orbital period could be twice the value listed in
Table~\ref{tab:main_table} and on the ephemerides in their respective panels
of Fig.~\ref{fig:phased_lcs}.  To help resolve the controversy, we have
started a $B$$V$$R$$I$ monitoring of them all, but the fruits of this effort
will be collected only in a few years time, when at least a few orbital
cycles will be covered.

\subsection{Lithium}

\citet{2008PASP..120..492W}, in their chemical analysis of high resolution
spectra of M giants in symbiotic stars, found that the recurrent novae and
accreting-only systems T CrB and RS Oph show an over-abundance of $^7$Li in
their spectra.  None of the other symbiotic stars in their large sample
showed $^7$Li enhancement and none was known to have undergone nova
outbursts similar to T CrB and RS Oph.  The over-abundance of $^7$Li in T
CrB has been recently revised upward to A(Li) = 2.4$\pm$0.1 by
\citet{2020AJ....159..231W}.  The apparent link between $^7$Li and nova
outbursts has been further reinforced when the symbiotic star V407 Cyg
erupted in 2010.  $^7$Li over-abundance was discovered earlier in V407 Cyg
by \citet{2003MNRAS.344.1233T} at a time when the object was simply known as
a Mira variable with a possible companion, and therefore attributed to
indigenous production via the \citet{1971ApJ...164..111C} mechanism
activated in the Mira variable. We know now that also V407 Cyg
undergoes nova outbursts, and this sheds an alternative explanation for its
$^7$Li overabundance.

The production of $^7$Li during nova eruptions has been predicted
theoretically (e.g.  \citealp{1998ApJ...494..680J, 2017PASP..129g4201R,
2020ApJ...895...70S}) and confirmed observationally (e.g. 
\citealp{2015Natur.518..381T, 2015ApJ...808L..14I, 2016MNRAS.463L.117M}). 
Up to $\sim$1$\times$10$^{-5}$ of the whole mass of the ejecta can be in the
form of $^7$Li, actually $^7$Be that will decay into $^7$Li.  The large
amounts produced during outbursts (especially those occurring on white
dwarfs of the Carbon-Oxygen type), suggests that novae could be the main
producer of $^7$Li in the Milky Way \citep{2020MNRAS.492.4975M}.

    \begin{figure}
	\includegraphics[width=\columnwidth]{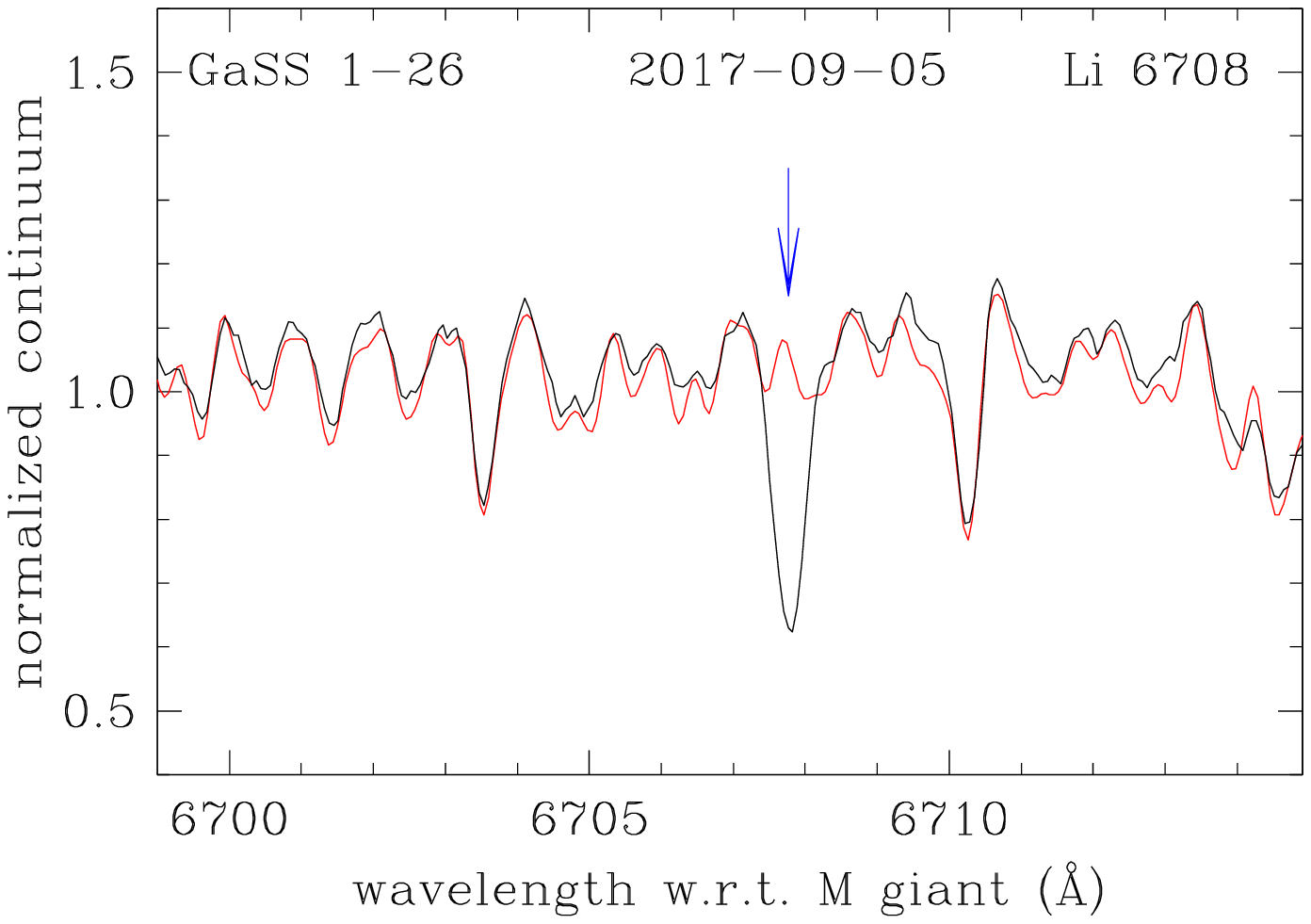}
        \caption{The GALAH spectrum of GaSS 1-26 (black) around the position of LiI
        6707.8 \AA\ compared with that (red) of the template for the same
        $f$-ratio bin. The overabundance of LiI is quite obvious.}
    \label{fig:Gass1-26_li}
    \end{figure}

It is tempting to attribute the over-abundance of $^7$Li observed in T CrB,
RS Oph and V407 Cyg to pollution of the cool giant by the ejecta of the nova
outbursts experienced by the WD companion.  Their eruptions occur at a high
frequency (eight have been observed in historic times for RS Oph) such that
they could replenish the surface of the cool giant with $^7$Li faster than
convection can dilute $^7$Li into the interior of the star.  Alternatively,
$^7$Li could instead come from the interior of the cool giant, and
conditions leading to an efficient transport of $^7$Li to the surface could
also lead more probably to nova outbursts as a consequence of a higher mass
loss rate and/or an enrichment in CNO nuclei.

In analogy with T CrB, RS Oph and V407 Cyg, we have searched for $^7$Li
overabundance in our 33 candidate new SySt by comparing with the template
spectrum for their $f$-ratio, and found a clear case of enhancement in GaSS
1-26, as shown in Fig.~\ref{fig:Gass1-26_li}.  The object is completely
anonymous (no matching entry in SIMBAD, VSX and similar catalogs) and its
optical variability is inconspicuous (cf Fig.~\ref{fig:figA7} in the
Appendix).

We have subjected GaSS 1-26 to a search for flickering and got a robust
detection (cf Table~\ref{tab:main_table}), which supports its binary nature
and the presence of an accreting companion in the form of a WD.

Armed with that, we turned to the Harvard plate stack in search for an
unnoticed nova outburst in the distant past.  DASCH project provides access
to calibrated scans of the huge collection of sky patrol photographic plates
obtained over a century with Harvard astrographs located in both hemispheres
\citep{2009ASPC..410..101G}.  The area of the sky where GaSS 1-26 is located
has not yet been included in the available DASCH data releases (1 to 6).  We
have been however granted access to the data for the area of sky around GaSS
1-26 prior to their publication in DASCH DR7, currently planned for July
2021.  This led to some initial excitement when two plates for June 6, 1908
and September 17, 1916 were logged as reporting GaSS 1-26 much brighter than
at other epochs and at a blue magnitude compatible with a nova outburst. 
This pair of observations belongs to a time interval with very few Harvard
plates covering the GaSS 1-26 position and being deep enough to have
recorded it in quiescence (we measured it at $B$=15.4 during our search for
flickering, cf.  Table~\ref{tab:hjd_b_v}, thus quite faint for the
photographic means of the time).  To assess if any nova outburst had really
affected GaSS 1-26 in 1908 and/or 1916, we asked for and were provided with
a scan of both plates (Edward Los, private communication).  Inspection of
these scans reveals the presence of artefacts at the astrometric position
for GaSS 1-26, in the form of ink from a pen marking for the 1908 plate and
a scratch on the emulsion for the 1916 one.  Therefore, no nova outburst
affected GaSS 1-26 in either 1908 and 1916.  Even if the search on Harvard
plates for unnoticed past nova outbursts has not been fruitful, it should be
extended to other photographic plate archives, although not many cover the
southern hemisphere at faint magnitudes and over protracted periods of
times, and even less have preserved the plates in good conditions and
provide access to them in person or via digitized scans.

    \begin{figure}
	\includegraphics[width=\columnwidth]{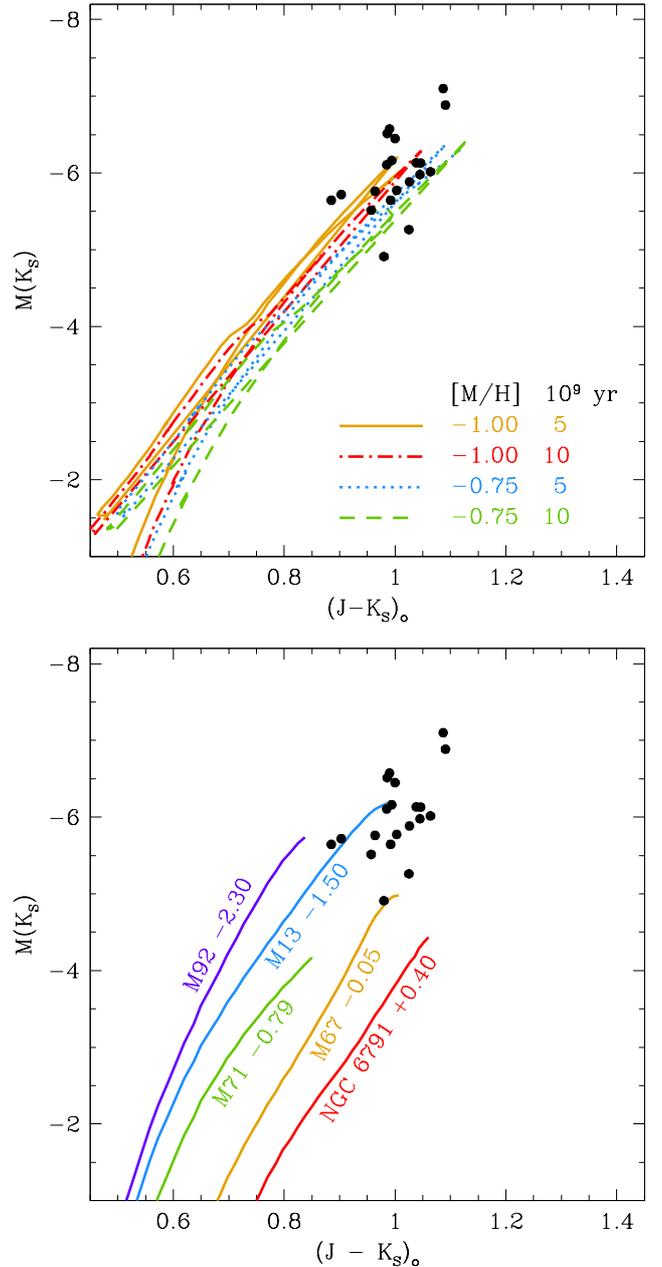}
        \caption{Location of the new SySt on the near-IR isochrone
        plane, compared with Padova PARSEC isochrones (top) and observed
        ones (bottom) for clusters spanning a range in metallicity (see
        Section~\ref{sec:evolutionary_stat} for details).}
    \label{fig:syst_parsec}
    \end{figure}

    \begin{figure}
	\includegraphics[width=\columnwidth]{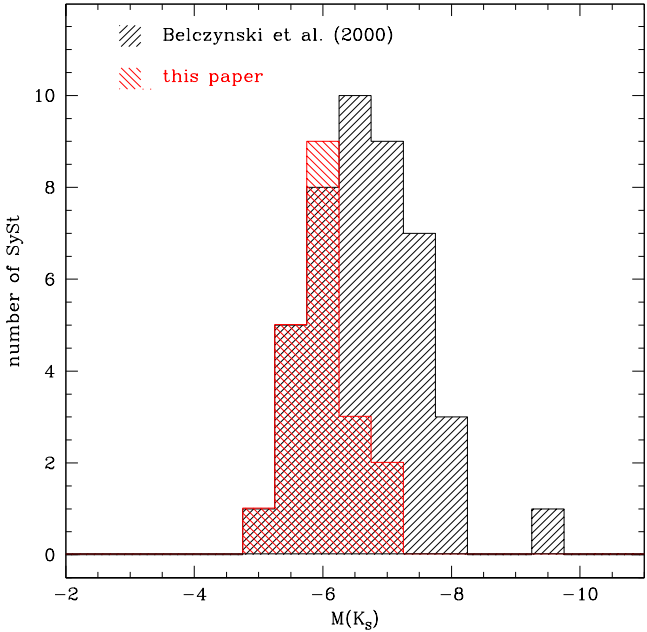}
        \caption{Distribution in absolute 2MASS $K_S$ magnitude of
        \citet{2000A&AS..146..407B} symbiotic stars with an M spectral type,
        and those from the present paper.  Only objects with a Gaia $e$DR3
        parallax satisfying the accuracy condition $\sigma(\pi)$/$\pi$$<$0.2
        are plotted.}
    \label{fig:belczynski}
    \end{figure}

In favor of a possible pollution scenario for the $^7$Li in GaSS 1-26 is the
fact that indigenous $^7$Li production via the \citet{1971ApJ...164..111C}
mechanism is expected to occur during hot-bottom burning conditions (HBB),
when the outer part of the burning H-shell is included in the envelope
convection.  A key aspect of HBB is the existence of a lower mass limit for
it to occur, which depends on metallicity \citep{2005ARA&A..43..435H}.  The
minimum initial mass for HBB is M$_{\rm ini}$$\geq$5~M$_\odot$ at solar
metallicity \citep{1997A&AS..123..241F}, which can decrease down to
3~M$_\odot$ for very metal-poor stars \citep{2002ApJ...570..329S}.  The
latter is obviously not the case given the otherwise normal spectrum
displayed by GaSS 1-26.  The faintness and negligible reddening place GaSS
1-26 at a great distance from us; its Gaia $e$DR3 parallax is affected by a
large error, but it seems safe to argue that a lower limit to the distance
is 10 kpc.  At a galactic latitude $-$10$^{\circ}_{.}$6, it means a height
over the galactic plane of $z$$\geq$1.9 kpc.  It would be rather atypical to
find a young and massive M$_{\rm ini}$$\geq$5~M$_\odot$ star at such a large
height above the Galactic plane.

\subsection{Evolutionary status} \label{sec:evolutionary_stat}

The nature of the new SySt will be investigated in detail elsewhere,
including their chemical partition and Galactic orbits.  However, some
preliminary comments seem in order here.

To asses the evolutionary status of the new SySt, we have taken their
$J$$H$$K_S$ near-IR photometry from 2MASS survey \citep{2003yCat.2246....0C}
and after correcting for reddening and distance given in
Table~\ref{tab:main_table}, we have plotted their positions on the absolute
$M(K_S)$/de-reddened $(J-K_S)_{\rm o}$ plane in Fig.~\ref{fig:syst_parsec}.

On the top panel of Fig.~\ref{fig:syst_parsec}, the comparison is carried
out against the Padova PARSEC isochrones \citep{2012MNRAS.427..127B,
2017ApJ...835...77M, 2020MNRAS.498.3283P}.  It provides evidence that the
new SySt are all located close to the tip of the Giant Branch, and that they
are relatively old and metal poor objects.  This conforms with their
significant height $z$ above the Galactic plane as given in
Table~\ref{tab:tabA2}.  For that age and metallicity, their mass estimated
from the isochrones ranges from 1.03 to 1.22~M$_\odot$.  Similar conclusions
are drawn by comparison with the observed isochrones compiled by
\citet{2013PASA...30...11K} for a sample of clusters spanning a large range
of observed metallicities, as illustrated in the bottom panel of
Fig.~\ref{fig:syst_parsec}.

How do the red giants of the new SySt compare to well established SySt ? 
This is addressed in Fig.~\ref{fig:belczynski}, where their distributions in
$M(K_S)$ are compared.  For the well established SySt we adopted the catalog
by \citet{2000A&AS..146..407B}, and retained only the 44 objects with a
giant of M spectral type and with a Gaia~$e$DR3 parallax accurate enough to
satisfy the inversion condition $\sigma(\pi)$/$\pi$$<$0.2.  To compute their
$M(K_S)$ we adopted for them the interstellar extinction estimated from the
Bayestar~2019 3D model of the Galaxy \citep{2019ApJ...887...93G}.

The distribution in Fig.~\ref{fig:belczynski} of the new and the well
established SySt is the same.  The fact that GALAH is a magnitude-limited
survey ($12 < V_{JK} < 14$) means that intrinsically brighter objects need
to be at larger distances to fall between the limits in magnitude.  The lack
of GALAH new SySt at the brighter $M(K_S)$ values in
Fig.~\ref{fig:belczynski} simply reflects the fact that current Gaia~$e$DR3
parallaxes are not accurate enough to satisfy the condition
$\sigma(\pi)$/$\pi$$<$0.2 for the most distant, and therefore intrinsically
brightest objects.  It may be guessed that future Gaia data releases will
push forward the horizon of parallax accuracy and allow to include the
remaining new SySt and let them populate the histogram of
Fig.~\ref{fig:belczynski}.

\section{Conclusions and future prospects} \label{sec:conclusion}

We present the discovery and characterization of 33 bona-fide candidates of
the accreting-only type of symbiotic stars.  So far, this class of binary
systems has evaded a thorough scientific scrutiny, otherwise received by its
counterpart, the nuclear-burning class of symbiotic stars.  Both describe
the same underlying configuration, a binary system composed of a red giant
and a degenerate companion, however, they are distinguished by the physical
processes under way inside the system and the consequent observational
implications, favoring the study of {\it burn}-SySt.  Because {\it
acc}-SySt are believed to be much more abundant in the Galaxy than {\it
burn}-SySt, while the ratio of known (investigated) objects is just the
reverse, it is important to uncover this hidden population of interacting
binary stars in order to assess the viability of symbiotic stars as
progenitors of type Ia supernovae, their impact on the pollution of the ISM,
and to better understand the cycle of exchange between the accreting-only
and the nuclear-burning phases.

To this end, we have investigated a large sample of 600\,255 stars from the
GALAH spectroscopic all-sky survey, which, given its unbiased observational
strategy, promises to enable the envisaged statistical analysis, while
supported by a suite of follow-up observations, provides a rich source of
information on individual symbiotic stars.  The first 33 candidates
presented in this paper were selected using relatively conservative
criteria: parallax and colour cut, strength of molecular (TiO) absorption
bands, prominence of H$\alpha$ emission line, rejection based on the
presence of various observational and reduction artefacts.  Furthermore, the
photometric light curves were checked to filter out radial pulsators, and
great care was taken to reinforce our diagnostics for the symbiotic nature
of program stars, building on the knowledge provided by the prototype
{\it acc}-SySt SU Lyn and by conducting follow-up ground and space based
observations examining the near-UV excess, RV and emission line variability,
optical flickering, and X--ray/UV luminosities.

One of the 33 objects, Gass 1-26, shows a clear enhancement in $^7$Li, which
makes it an excellent candidate for follow-up observations aiming to
illuminate the possible connection between pollution by novae outbursts and
the galactic lithium abundance.  Future studies will extend the analysis
presented in this work in diverse directions, including:
\begin{itemize}

\item weaker emission features in spectra will be considered, as we
currently only accept H$\alpha$ profiles which, after subtraction of the
template spectrum, reach higher than 0.5 above the adjacent continuum

\item other spectral diagnostic features will be established to identify
{\it acc}-SySt despite a possible absence of emission lines, for example a
peculiar chemical signature in the spectrum of a giant star that was
polluted by eruptions from the compact companion and thus by nuclearly
processed material

\item earlier spectral types of the primary stars will be investigated, also
to account for the more metal-poor systems which do not achieve the same
large radius and low surface temperatures as M giants

\item automatic and manual inspection of legacy observations is being carried
out, such as digging through diverse photometric archives and historical
plates stacks in order to find e.g.  past active phases or nova outbursts

\item existing and future high-energy instruments/surveys (e.g.  SWIFT,
$e$-ROSITA) can be exploited for follow-up observations or cross-matching of
catalogued detections

\item combining the chemical abundances and galactic orbits with the aid of
the next Gaia releases will provide the means to confirm/reject the belief
of symbiotic stars being an old population, primarily belonging to the
bulge, thick disk, and halo.  This view is built from the infrared
characteristics of their cool giants, but it is heavily biased by how (and
where in the Galaxy) the classical symbiotic stars were discovered by
objective prism surveys of half a century ago.  Determining the parent
stellar population of symbiotic stars has a profound implication considering
that Ia are the only type of supernovae known to erupt in elliptical
galaxies, whose stellar populations resemble the bulge of our Galaxy.

\end{itemize}

The proposed future work will greatly expand the number of characterized
{\it acc}-SySt and thus provide the community with large enough samples for a
statistical study of the population of symbiotic stars as a whole.  In the
meantime, we encourage the readers to further investigate the 33 candidates 
presented here and help confirm their symbiotic nature with further follow-up
observations.

\section{Data availability} \label{sec:data}

The data underlying this article will be shared on reasonable request to
the corresponding author.

\section*{Acknowledgements}

This research has made use of the ASI Science Data Center Multimission
Archive; it also used the NASA Astrophysics Data System Abstract Service,
which is operated by the Jet Propulsion Laboratory, California Institute of
Technology, under contract with the National Aeronautics and Space
Administration.  This research has also made extensive use of the SIMBAD and
VIZIER databases operated at CDS, Strasbourg, France.  We thank Dr.  Jamie
Kennea and the {\it Swift} team for the quick approval and the rapid
acquisition of the observations we requested.  We thank R.  Jurdana for the
initial contact, and J.  Grindlay and E.  Los for grating access to DASCH
data and scans prior to publication.  NM acknowledges financial support
through ASI-INAF agreement 2017-14-H.0, and UM through MAINSTREAM - PRIN
INAF 2017 "The origin of lithium: a key element in astronomy" (P.I.  Paolo
Molaro).  We would also like to thank S.  Dallaporta, G.  Cherini and F. 
Castellani (ANS Collaboration) for contributing to the photometric
observations of SU Lyn.  This work was supported by the Swedish strategic
research programme eSSENCE.  G.T.  was supported by the project grant ''The
New Milky Way'' from the Knut and Alice Wallenberg foundation and by the
grant 2016-03412 from the Swedish Research Council.  G.T.  acknowledges
financial support of the Slovenian Research Agency (research core funding
No.  P1-0188 and project N1-0040).  This research made use of Astropy
(http://www.astropy.org) a community-developed core Python package for
Astronomy \citep{2013A&A...558A..33A,2018AJ....156..123A}.

\clearpage

\appendix
\section{Additional tables and Figures} \label{sec:additional}

\begin{figure*}
	\includegraphics[width=16.8cm]{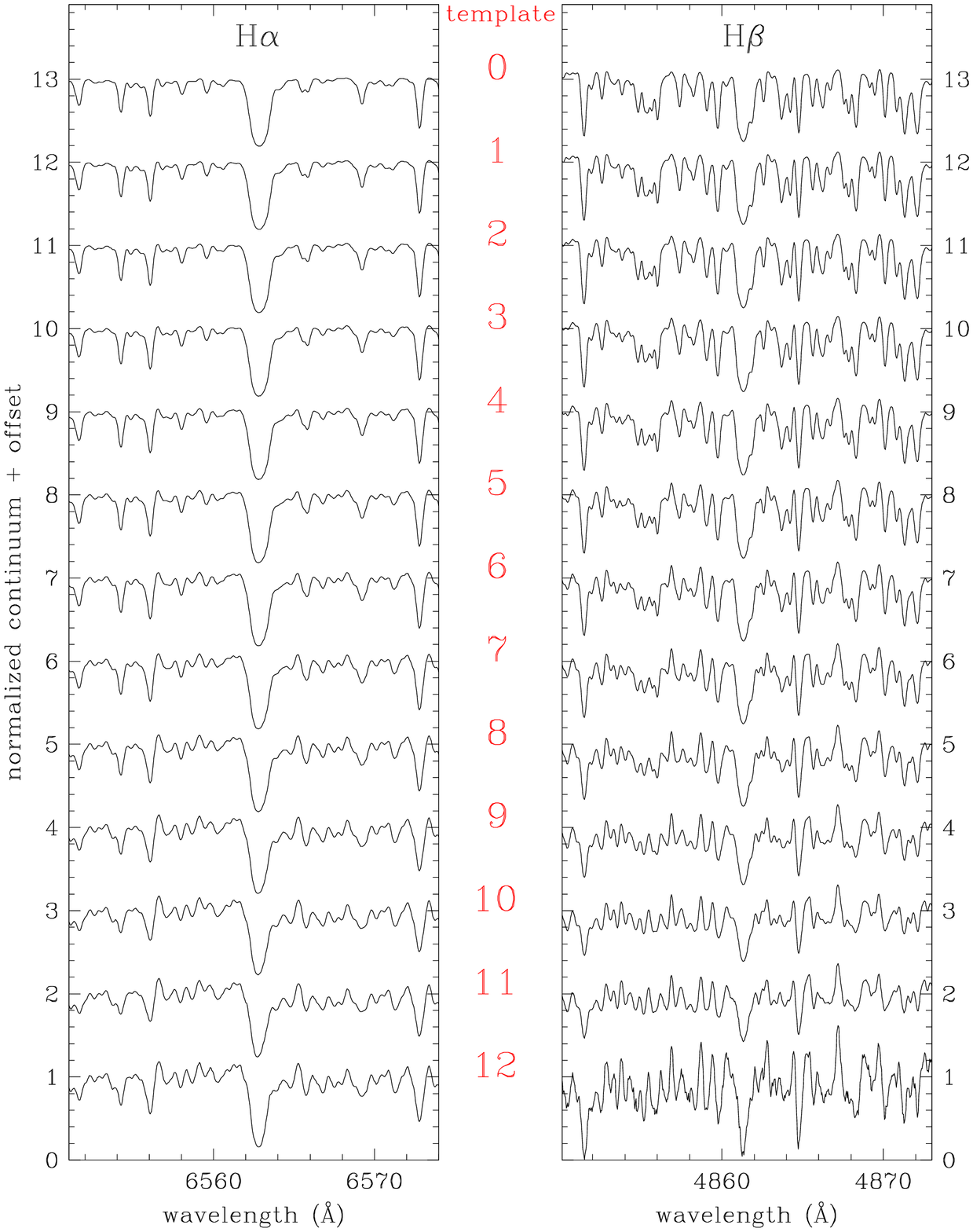}
	\caption{Sequence of templates as listed in
	Table~\ref{tab:templates}.  The template label/bin is indicated in red, and
	the template spectra are ordered from earlier to later M-type going from top
	to bottom.  The bottom-most spectrum is of relatively lower S/N due to the
	small number of spectra that were used in constructing this template (see
	Table~\ref{tab:templates}).}
    \label{fig:figA1}
    \end{figure*}

\begin{figure*}
	\includegraphics[width=16.8cm]{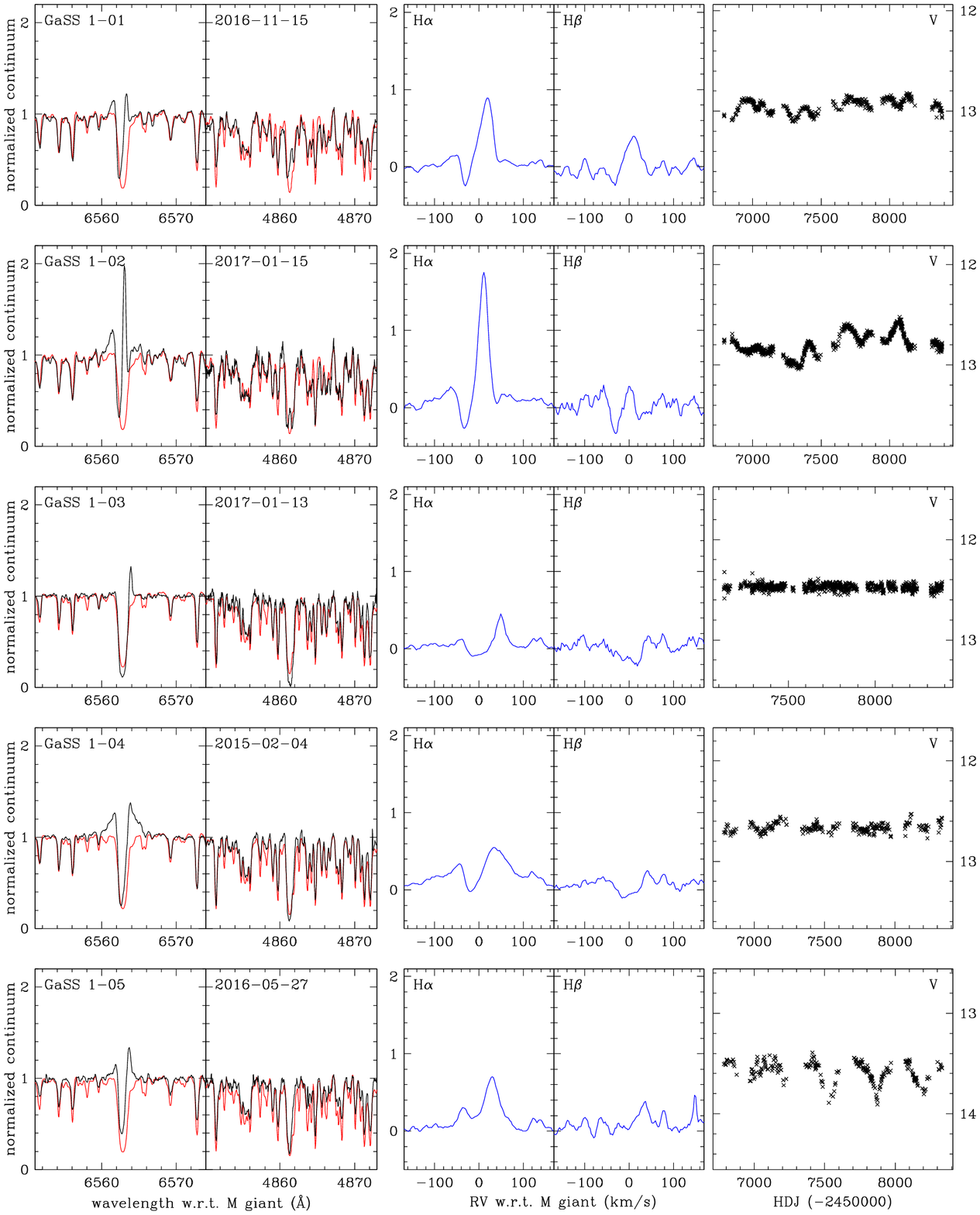}
	\caption{{\it Left-most panels}: H$\alpha$ and H$\beta$ profiles (in
        black) compared to respective templates (in red) for program stars
        GaSS 1-1 to GaSS 1-5.  {\it Center panels}: the result of
        subtracting the template from the object spectrum.  {\it Right-most
        panel}: $V$-filter lightcurve from ASAS-SN sky patrol data.}
    \label{fig:figA2}
    \end{figure*}

\begin{figure*}
	\includegraphics[width=16.8cm]{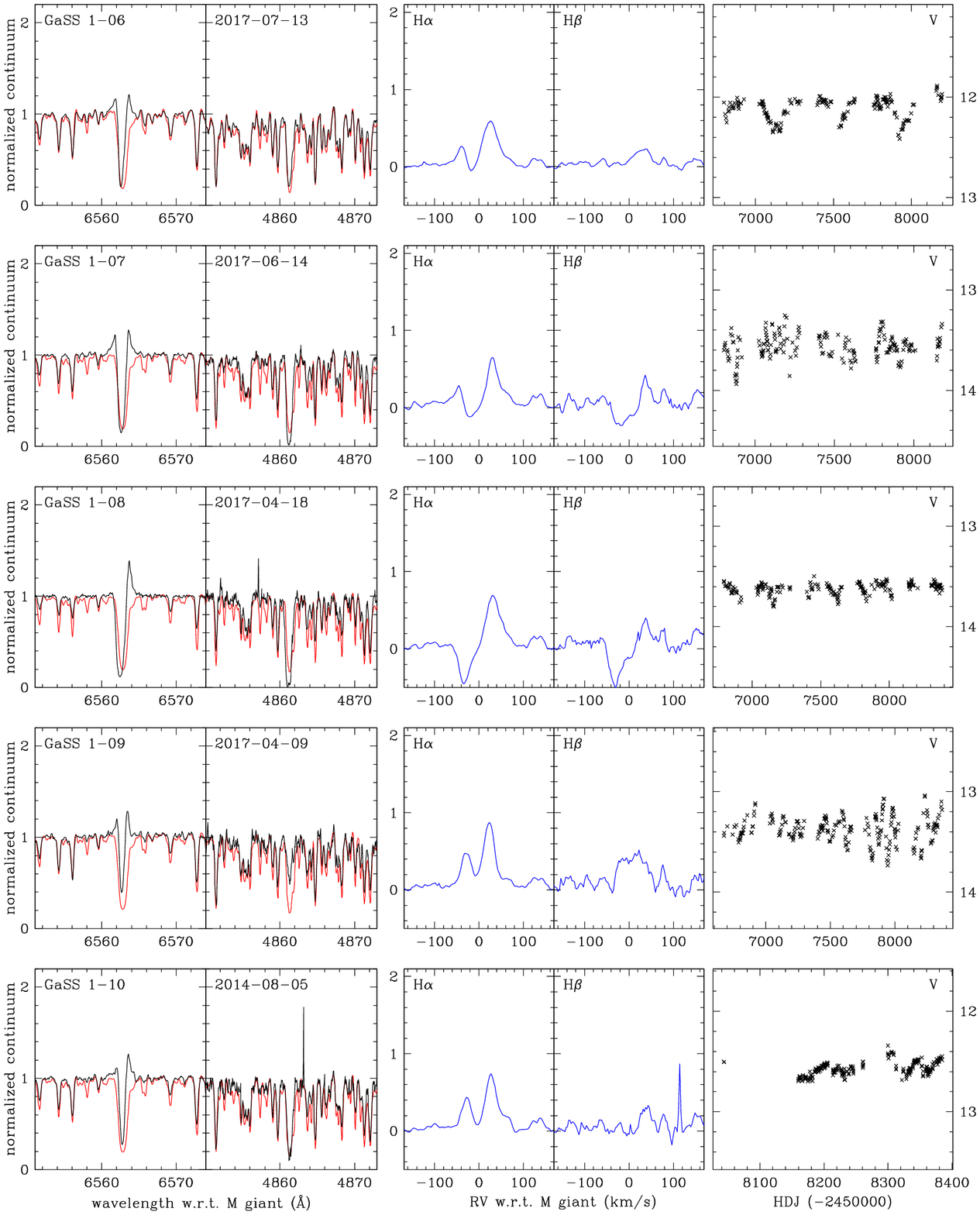}
	\caption{Similar to Fig.~\ref{fig:figA2} for program stars GaSS 1-6 to GaSS 1-10.}
    \label{fig:figA3}
    \end{figure*}

\begin{figure*}
	\includegraphics[width=16.8cm]{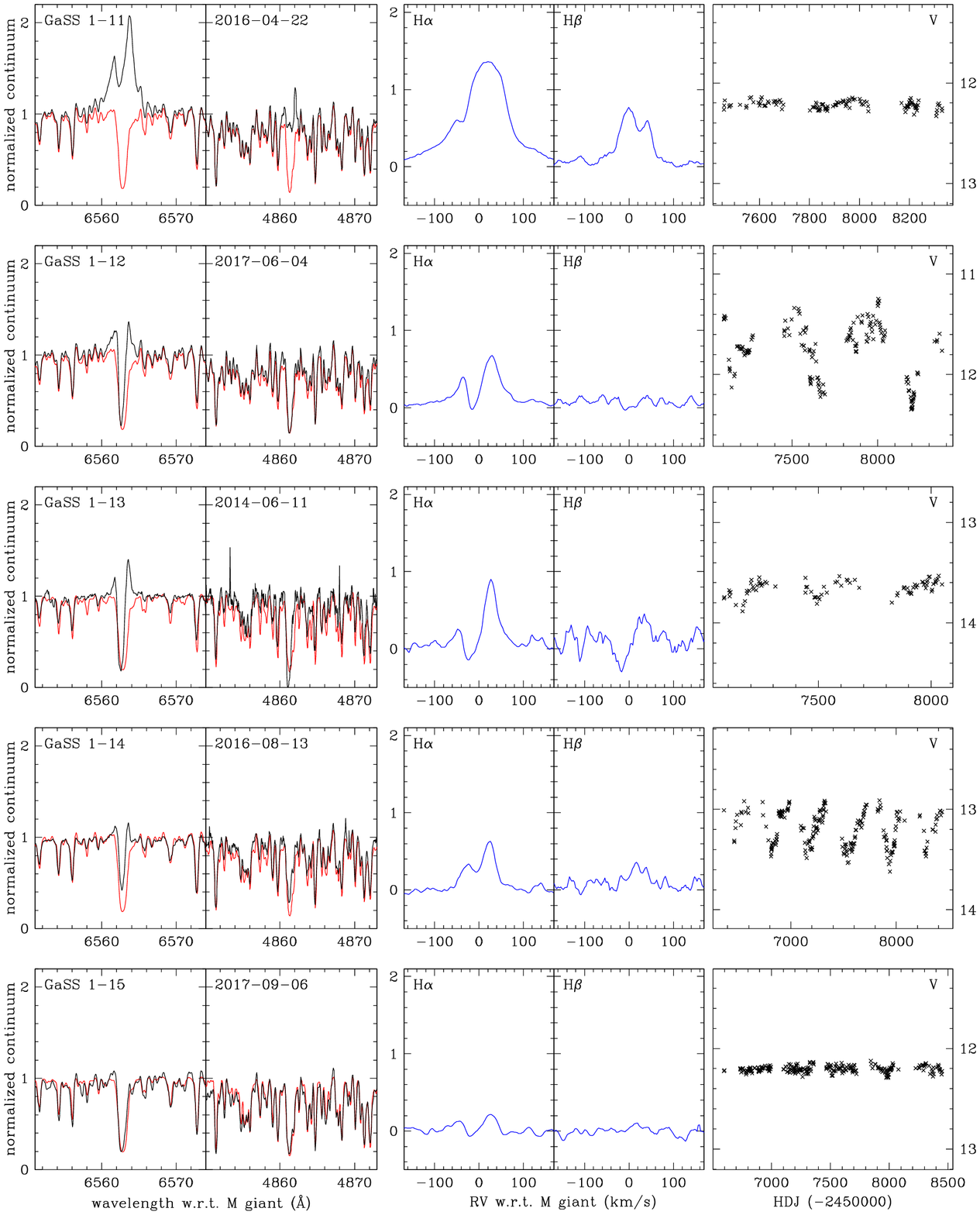}
	\caption{Similar to Fig.~\ref{fig:figA2} for program stars GaSS 1-11 to GaSS 1-16.}
    \label{fig:figA4}
    \end{figure*}

\begin{figure*}
	\includegraphics[width=16.8cm]{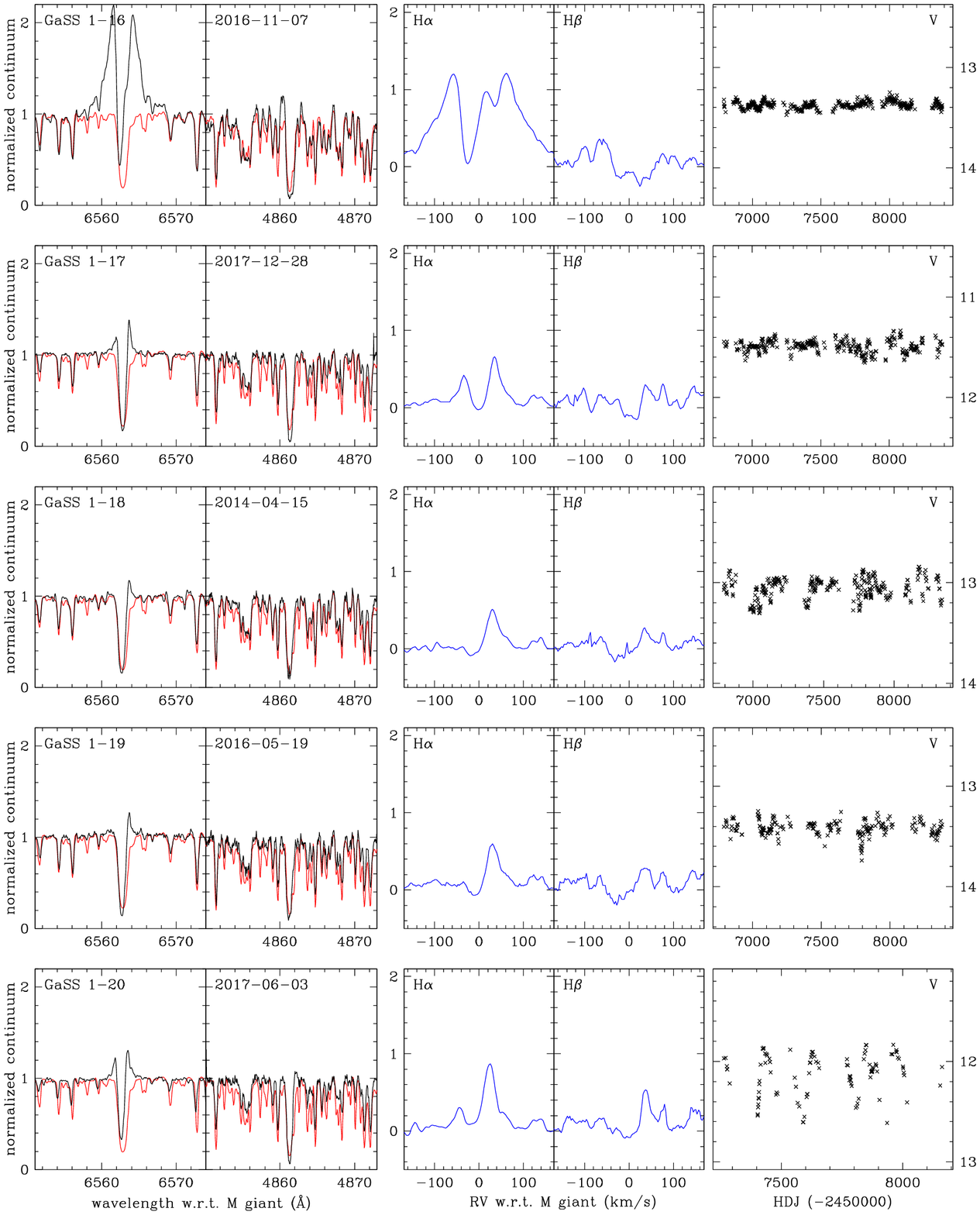}
	\caption{Similar to Fig.~\ref{fig:figA2} for program stars GaSS 1-16 to GaSS 1-20.}
    \label{fig:figA5}
    \end{figure*}

\begin{figure*}
	\includegraphics[width=16.8cm]{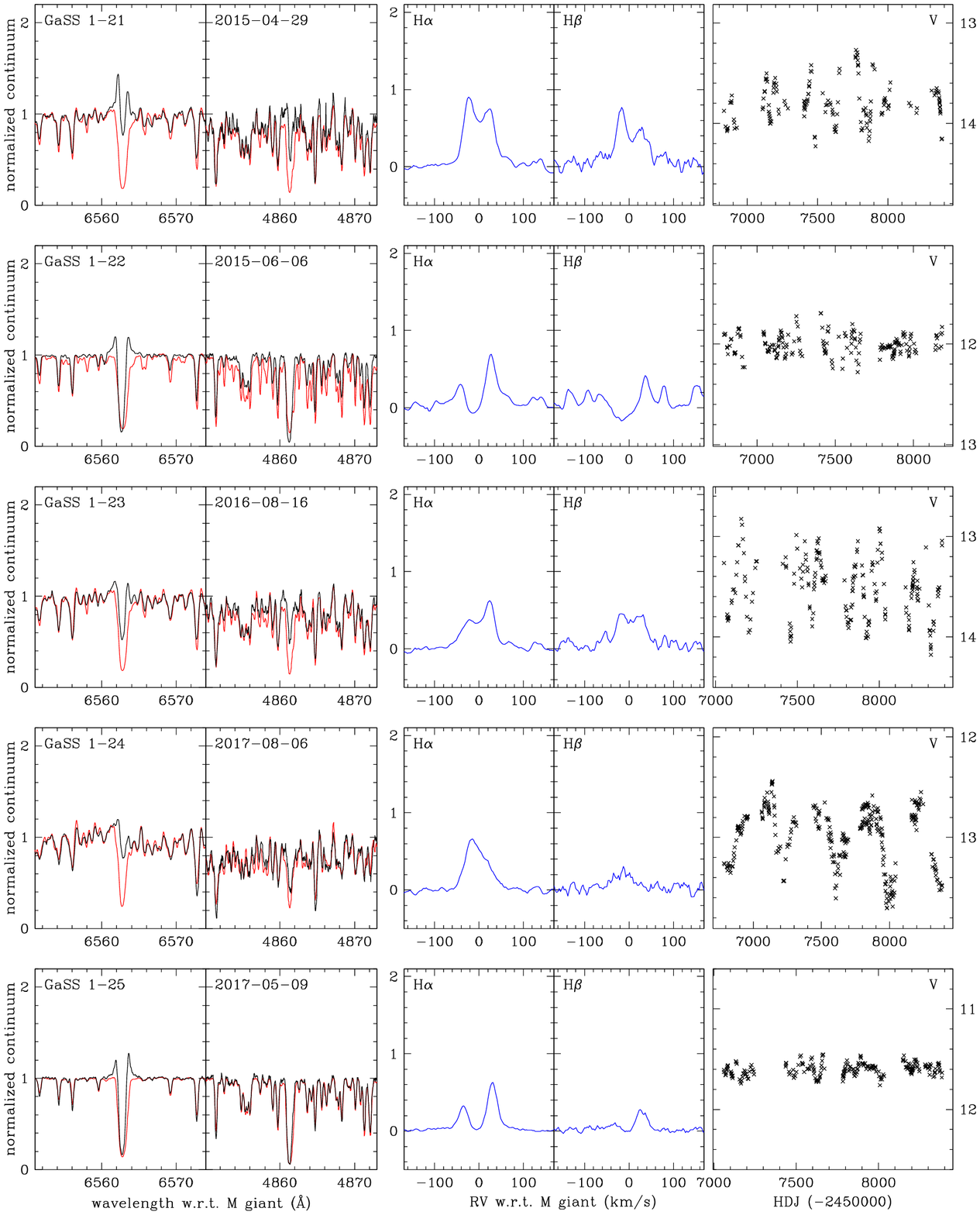}
	\caption{Similar to Fig.~\ref{fig:figA2} for program stars GaSS 1-21 to GaSS 1-25.}
    \label{fig:figA6}
    \end{figure*}

\begin{figure*}
	\includegraphics[width=16.8cm]{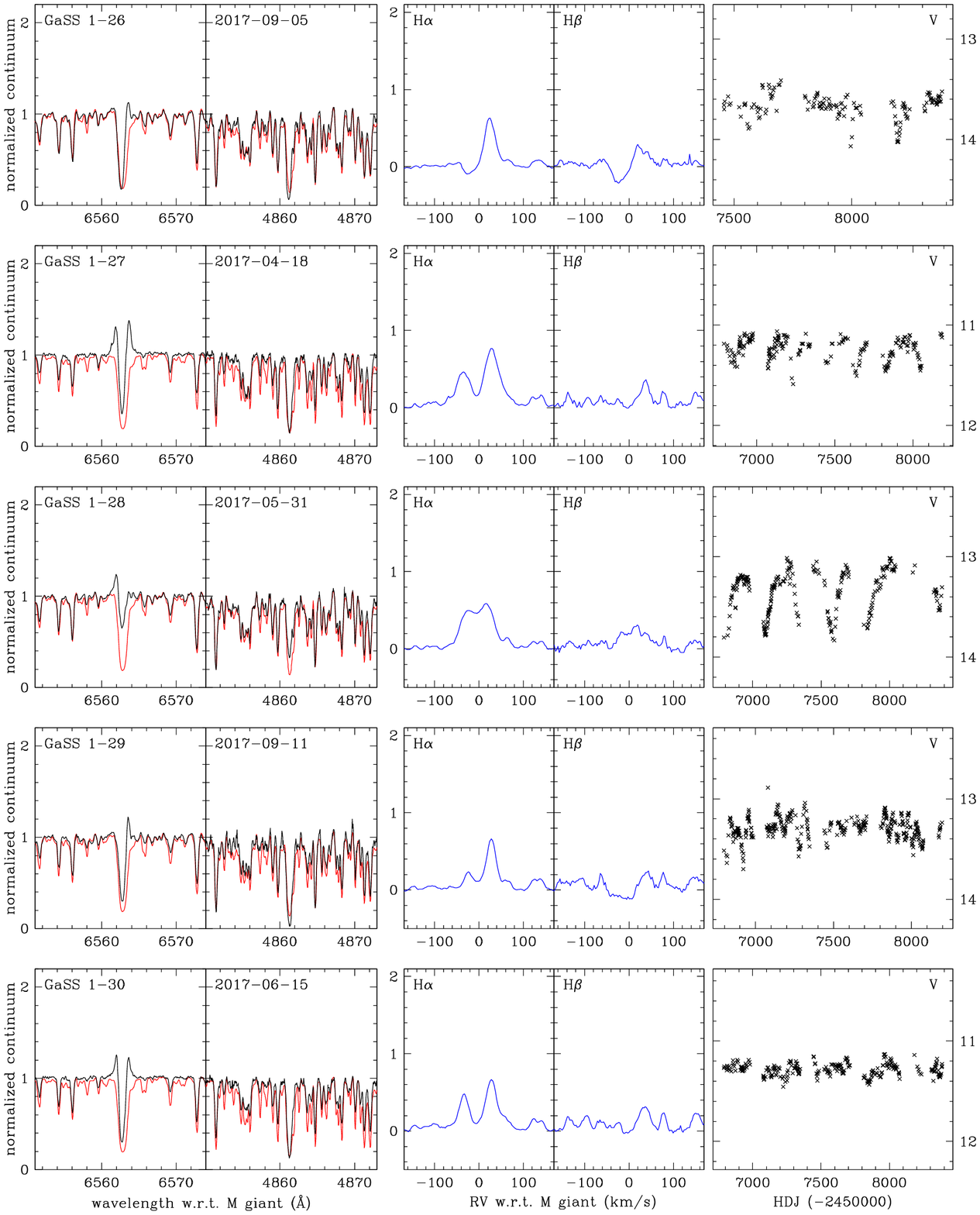}
	\caption{Similar to Fig.~\ref{fig:figA2} for program stars GaSS 1-26 to GaSS 1-30.}
    \label{fig:figA7}
    \end{figure*}

\begin{figure*}
	\includegraphics[width=16.8cm]{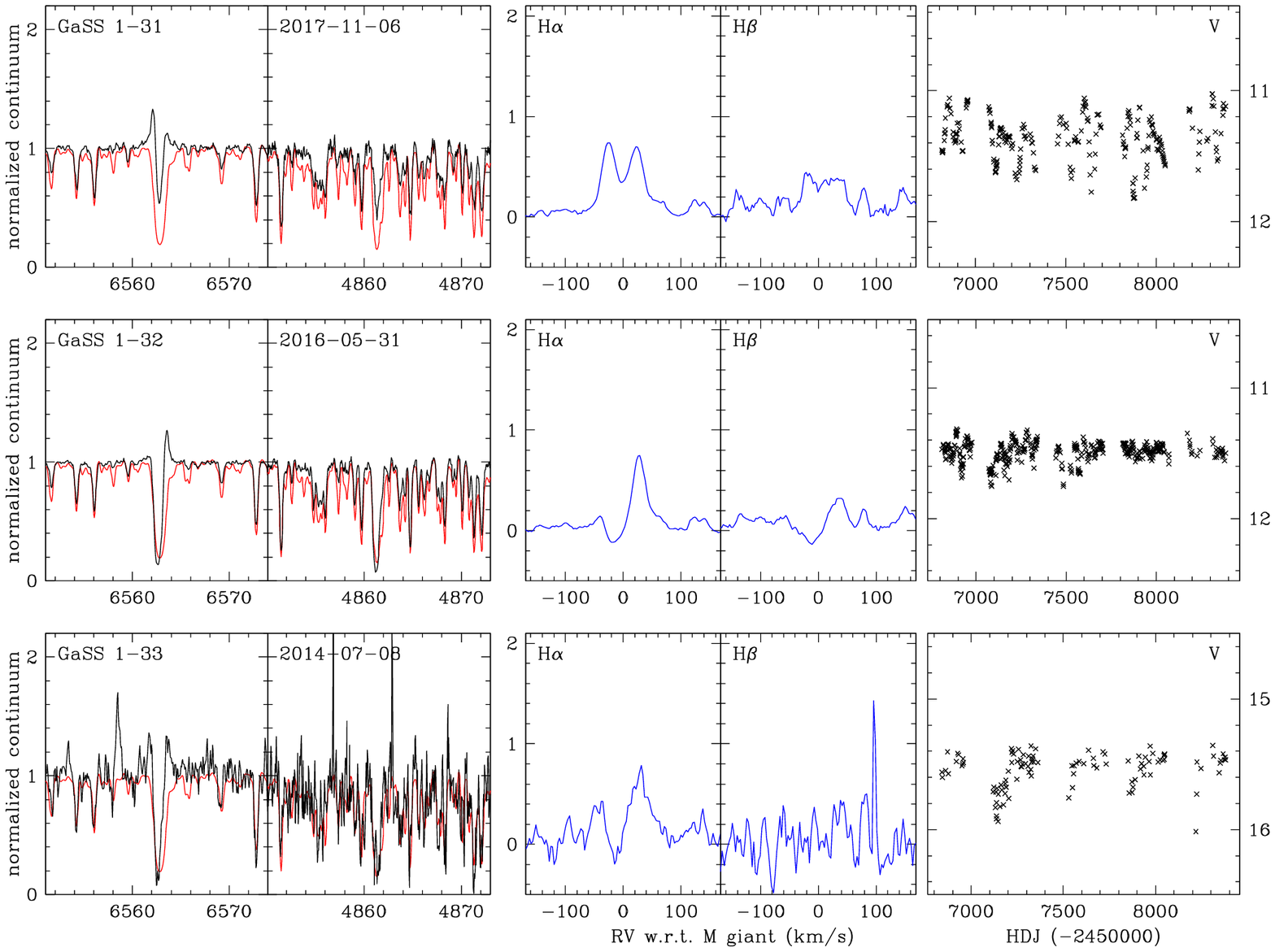}
	\caption{Similar to Fig.~\ref{fig:figA2} for program stars GaSS 1-31 to GaSS 1-33.}
    \label{fig:figA8}
    \end{figure*}

In this section, we present ancillary information that might be useful to
the interested reader.  First, we provide some details that concern
follow-up observations of 33 candidate {\it acc}-SySt when searching for
signs of the near-UV upturn (Table~\ref{tab:lowres_log}) and flickering
(Table~\ref{tab:hjd_b_v}).  Furthermore, we compute values for the position
and luminosity for the 33 systems based on information given by the
cross-match with external catalogues in Table~\ref{tab:tabA2}.  The radial
velocity information measured or collected for some of the 33 program stars
in Table~\ref{tab:rvs} is expanded for comparison reasons to include some
GALAH radially pulsating and field M giants in Table~\ref{tab:helio_rv}. 
The list of VVCep stars discussed in sect.  4.8 is provided in
Table~\ref{tab:vvcep}.  Finally, Fig.~\ref{fig:figA1} displays the region of
high S/N normalised template spectra around H$\alpha$ and H$\beta$ lines. 
These templates were used for subtraction from program star spectra in order
to detect prominent H$\alpha$ emission features, with details of the
subtraction results complemented by the ASAS-SN $V$-filter lightcurves
plotted in Fig.~\ref{fig:figA2}~to~\ref{fig:figA8}.

\newpage
\begin{table}
        \centering
        \caption{Log of low-resolution spectroscopic observations obtained
        with the Asiago 1.22m + B\&C + 300 ln/mm (3300-8000 \AA, 2.31 \AA/pix)
        in search for the near-UV upturn. The exposure time is in seconds.}
        \label{tab:lowres_log}
\begin{tabular}{lccr}
\hline
&&\\
SySt & \multicolumn{2}{c}{UT middle} & expt \\
&&\\   & \\ 
 GaSS 1-7   & 2020-06-18 & 20:40 &  120     \\
 GaSS 1-7   & 2020-06-27 & 20:32 &  960     \\
 GaSS 1-9   & 2019-08-09 & 19:59 &  480     \\
 GaSS 1-9   & 2019-08-16 & 19:43 &  480     \\
 GaSS 1-9   & 2019-09-01 & 19:48 & 1200     \\
 GaSS 1-9   & 2020-06-27 & 22:31 &  960     \\
 GaSS 1-9   & 2020-08-25 & 19:18 &  480     \\
 GaSS 1-12  & 2019-08-16 & 20:27 &  720     \\
 GaSS 1-12  & 2019-08-19 & 20:54 &  960     \\
 GaSS 1-12  & 2019-09-01 & 19:27 & 1200     \\
 GaSS 1-12  & 2020-08-25 & 20:11 &  480     \\
 GaSS 1-13  & 2020-08-25 & 21:37 &  480     \\
 GaSS 1-13  & 2020-10-13 & 18:05 & 1800     \\
 GaSS 1-14  & 2019-08-30 & 18:59 &  480     \\
 GaSS 1-14  & 2019-09-01 & 20:18 &  960     \\
 GaSS 1-14  & 2019-12-06 & 17:24 & 1200     \\
 GaSS 1-15  & 2020-09-02 & 21:58 &  960     \\
 GaSS 1-15  & 2020-10-13 & 19:49 & 1800     \\
 GaSS 1-21  & 2020-06-19 & 21:24 &  720     \\
 GaSS 1-21  & 2020-06-27 & 21:54 & 1200     \\
 GaSS 1-23  & 2019-08-16 & 19:57 & 1200     \\
 GaSS 1-23  & 2019-08-19 & 19:41 &  960     \\
 GaSS 1-23  & 2019-09-01 & 19:58 & 1200     \\
 GaSS 1-23  & 2020-08-25 & 19:37 &  480     \\
 GaSS 1-25  & 2020-08-25 & 19:27 &  480     \\
 GaSS 1-28  & 2019-09-03 & 19:44 & 1200     \\
 GaSS 1-31  & 2020-08-25 & 20:20 &  480     \\
 GaSS 1-31  & 2020-10-13 & 17:30 & 1800     \\
 GaSS 1-33  & 2020-08-27 & 22:18 & 1440     \\
 GaSS 1-33  & 2020-10-13 & 20:28 & 3600     \\
&&\\
\hline
\end{tabular}
\end{table}

\begin{table}
\centering
\caption{Gaia $e$DR3 parallax and its uncertainty for the program stars,
distance and height above the Galactic plane for stars with
$\sigma$($\pi$)/$\pi$$<$0.2, and corresponding absolute $V$ and $K_s$
magnitudes from APASS and 2MASS values.}
\label{tab:tabA2}
\begin{tabular}{lccc@{~~}c@{~~}c@{~~}c}
\hline
&&\\
&\multicolumn{2}{c}{$e$DR3}&d&z&M(V)&M(Ks)\\ \cline{2-3}
&$\pi$& err($\pi$)& (kpc) & (kpc) & (mag)&(mag)\\
&&\\                                                                                 
  GaSS~ 1-1  &  0.0149 & 0.0125  &       &      &		&	   	  \\ 
  GaSS~ 1-2  &  0.0249 & 0.0114  &       &      &		&	   	  \\ 
  GaSS~ 1-3  &  0.1216 & 0.0100  &  8.2  &  3.7 &       $-$2.2	&       $-$5.8    \\ 
  GaSS~ 1-4  &  0.1226 & 0.0169  &  8.2  &  3.6 &	$-$2.2	&	$-$6.1    \\ 
  GaSS~ 1-5  &  0.0988 & 0.0157  & 10.1  &  5.0 &	$-$1.7	&	$-$5.5    \\ 
  GaSS~ 1-6  &  0.1843 & 0.0218  &  5.4  &  1.8 &	$-$1.7	&	$-$5.9    \\ 
  GaSS~ 1-7  &  0.0476 & 0.0216  &       &      &	   	&	          \\ 
  GaSS~ 1-8  &  0.1358 & 0.0153  &  7.4  &  2.8 &	$-$2.1	&	$-$5.6    \\ 
  GaSS~ 1-9  &  0.0974 & 0.0311  &       &      &	   	&	          \\ 
  GaSS~ 1-10 &  0.0715 & 0.0159  &       &      &	       	&	          \\ 
  GaSS~ 1-11 &  0.1523 & 0.0263  &  6.6  &  1.2 &	$-$2.3	&	$-$6.9    \\ 
  GaSS~ 1-12 &  0.2183 & 0.0321  &  4.6  &  0.9 &	$-$1.9	&	$-$7.1    \\ 
  GaSS~ 1-13 &  0.0423 & 0.0223  &       &      &	   	&	          \\ 
  GaSS~ 1-14 &  0.1163 & 0.0170  &  8.6  &  3.4 &	$-$1.8	&	$-$6.0    \\ 
  GaSS~ 1-15 &  0.3503 & 0.0257  &  2.9  &  1.4 &	$-$0.5	&	$-$4.5    \\ 
  GaSS~ 1-16 &  0.0219 & 0.0137  &       &      &	   	&	          \\ 
  GaSS~ 1-17 &  0.1861 & 0.0111  &  5.4  &  1.9 &	$-$2.5	&	$-$6.2    \\ 
  GaSS~ 1-18 &  0.1154 & 0.0157  &  8.7  &  3.5 &	$-$1.8	&	$-$5.6    \\ 
  GaSS~ 1-19 &  0.0589 & 0.0125  &       &      &	   	&	          \\ 
  GaSS~ 1-20 &  0.1327 & 0.0343  &       &      &	       	&	          \\ 
  GaSS~ 1-21 &  0.0391 & 0.0247  &       &      &	   	&	          \\ 
  GaSS~ 1-22 &  0.1711 & 0.0218  &  5.8  &  1.8 &	$-$2.7	&	$-$6.6    \\ 
  GaSS~ 1-23 &  0.2094 & 0.0245  &  4.8  &  1.3 &	$-$1.5	&	$-$6.0    \\ 
  GaSS~ 1-24 &  0.0902 & 0.0233  &       &      &	   	&	          \\ 
  GaSS~ 1-25 &  0.2306 & 0.0183  &  4.3  &  0.8 &	$-$2.3	&	$-$6.1    \\ 
  GaSS~ 1-26 &  0.0633 & 0.0171  &       &      &	   	&	          \\ 
  GaSS~ 1-27 &  0.2143 & 0.0177  &  4.7  &  1.3 &	$-$2.3	&	$-$5.7    \\ 
  GaSS~ 1-28 &  0.1326 & 0.0187  &  7.5  &  1.8 &	$-$1.4	&	$-$5.8    \\ 
  GaSS~ 1-29 &  0.1422 & 0.0157  &  7.0  &  2.7 &	$+$0.1	&	$-$5.3    \\ 
  GaSS~ 1-30 &  0.1799 & 0.0194  &  5.6  &  1.8 &	$-$2.7	&	$-$6.5    \\ 
  GaSS~ 1-31 &  0.2344 & 0.0170  &  4.3  &  1.1 &	$-$2.3	&	$-$6.1    \\ 
  GaSS~ 1-32 &  0.1877 & 0.0228  &  5.3  &  1.7 &	$-$2.5	&	$-$6.4    \\ 
  GaSS~ 1-33 &  0.0665 & 0.0327  &       &      &		&	   	  \\ 
&&\\
\hline
\end{tabular}
\end{table}

\begin{table*}
\centering
\caption{Heliocentric radial velocity of some GALAH radially pulsating and
field M giants observed along with the candidate symbiotic stars of
Table~\ref{tab:rvs}.  The values measured with the Asiago 1.82m telescope +
Echelle spectrograph are compared with the corresponding values of rv\_guess
listed in GALAH DR3.  The mean radial velocity listed in Gaia DR2 is also
given, with its formal uncertainty and the number of epoch transits over
which it has been computed.}
\label{tab:helio_rv}
\begin{tabular}{@{}c@{~~}rc@{~~}r@{~~}c@{~~}c@{~~}c@{~~}r@{~~}c@{~~}c@{~~}r@{~~}c@{~~}r@{}}
\hline
              &                 &                       &            &        &&                  &            &         &&            &        &      \\              
              &                 &\multicolumn{3}{c}{Asiago 1.82m Echelle}     &&    \multicolumn{3}{c}{GALAH DR3}        &&\multicolumn{3}{c}{Gaia DR2} \\    \cline{3-5} \cline{7-9} \cline{11-13}
\multicolumn{2}{c}{}            &    UT middle          & RV$_\odot$ &  err   &&  UT middle          & RV$_\odot$ & err  && $<$RV$_\odot$$>$& err & N\\
              &                 &                       & (km/s)     &(km/s)  &&                     & (km/s)      &(km/s)&&  (km/s)    &(km/s)  &      \\
              &                 &                       &            &        &&                     &            &      &&            &        &      \\
     RA       &       DEC~~~~   &\multicolumn{8}{c}{\it radial pulsators}\\              
              &                 &                       &            &        &&                     &            &      &&            &        &      \\
 19 17 30.06  & $-$18 47 02.1   & 2020-10-30   17:16    & $-$28.14   &  0.51  && 2017-05-07 18:28    &  $-$34.63  & 0.03 &&   $-$32.82 &   2.87 &  2   \\
 20 35 04.52  & $-$05 47 35.3   & 2020-10-29   17:32    &$-$120.82   &  1.13  && 2014-07-09 15:39    & $-$121.61  & 0.21 &&            &        &      \\
 23 31 17.83  & $-$00 30 39.4   & 2019-12-09   18:30    & $-$34.67   &  0.49  && 2017-07-24 17:27    &  $-$37.11  & 0.37 &&   $-$35.81 &   0.93 &  6   \\                                      
              &                 & 2020-01-15   17:06    & $-$36.17   &  0.33  &&                     &            &      &&            &        &      \\
              &                 & 2020-10-29   19:59    & $-$39.47   &  0.31  &&                     &            &      &&            &        &      \\
              &                 &                       &            &        &&                     &            &      &&            &        &      \\
&&\multicolumn{8}{c}{\it other GALAH M giants}\\
              &                 &                       &            &        &&                     &            &      &&            &        &      \\
 03 40 25.30  &    14 30 00.6   & 2020-01-13   20:42    &    11.46   &  0.21  && 2016-01-10 10:47    &      9.94  & 0.48 &&       9.45 &   0.70 &  6   \\                                      
 05 21 50.50  &    01 51 26.8   & 2019-12-07   01:50    & $-$16.54   &  0.29  && 2015-08-30~~~~~~~~  &  $-$14.65  &      &&   $-$16.44 &   1.84 &  6   \\                                      
              &                 & 2020-01-13   20:56    & $-$17.47   &  0.10  && 2015-12-27 13:09    &  $-$12.42  & 0.31 &&            &        &      \\                                      
 06 19 12.30  & $-$17 35 20.8   & 2020-01-13   22:17    &   121.14   &  0.43  && 2017-01-06 13:35    &    120.96  & 0.40 &&     121.47 &   0.80 &  7   \\                                      
 10 00 41.60  & $-$21 35 54.7   & 2020-04-10   20:13    &   188.45   &  0.58  && 2016-04-24 09:55    &    189.81  & 0.61 &&     191.32 &   1.02 &  6   \\                                      
 10 32 55.50  &    12 51 30.4   & 2020-04-10   20:34    &   132.32   &  0.12  && 2017-02-05 15:42    &    131.48  & 0.39 &&     135.27 &   1.79 &  3   \\                                      
 10 51 30.19  &    00 43 59.5   & 2020-04-10   21:04    &   165.91   &  0.36  && 2017-01-27 14:10    &    163.82  & 0.33 &&     163.32 &   0.82 &  2   \\                                      
 12 08 10.13  & $-$09 07 52.8   & 2020-04-10   21:50    &    45.34   &  0.22  && 2016-04-03 12:48    &     47.24  & 0.33 &&      45.68 &   0.43 &  20  \\                                      
 12 16 50.58  & $-$06 31 19.3   & 2020-04-10   22:04    &    83.14   &  0.19  && 2017-01-31 15:22    &     81.32  & 0.63 &&      83.23 &   0.32 &  28  \\                                      
 17 19 00.39  & $-$17 45 42.9   & 2020-07-05   21:21    & $-$23.56   &  0.91  && 2017-05-09 15:29    &  $-$23.62  & 0.61 &&            &        &      \\                                      
 19 59 58.75  &    08 04 07.3   & 2020-10-29   16:57    & $-$57.45   &  0.05  && 2016-10-08 09:34    &  $-$57.27  & 0.02 &&   $-$57.85 &   0.15 &   3  \\
 20 00 03.82  &    07 22 39.7   & 2020-10-31   16:52    &     3.28   &  0.02  && 2016-10-08 09:34    &      2.65  & 0.15 &&       2.14 &   0.14 &  14  \\
 20 05 15.00  &    08 54 02.5   & 2020-10-31   16:59    &  $-$0.49   &  0.05  && 2016-05-31 19:34    &   $-$0.07  & 0.24 &&    $-$1.14 &   0.42 &   6  \\
 22 04 33.89  & $-$08 14 49.9   & 2020-10-29   20:43    &    20.63   &  0.18  && 2014-07-11 17:59    &     19.40  & 0.01 &&            &        &      \\
 22 34 17.16  & $-$05 04 57.7   & 2020-10-29   20:55    & $-$20.77   &  0.52  && 2015-11-11 10:06    &  $-$19.89  & 0.09 &&            &        &      \\
              &                 &                       &            &        &&                     &            &      &&            &        &      \\
\hline
\end{tabular}
\end{table*}

\begin{table}
        \centering
        \caption{$B$ and $V$ photometry of the candidate {\it acc}-SySt as
        measured during the search for flickering.  The given HJD is the
        middle UT of the 70min duration of the time-series photometry.}
        \label{tab:hjd_b_v}
	\begin{tabular}{lrcc}
		\hline
&&\\
             &     HJD    &     $B$    &   $V$      \\    
             & (-2459000) &     (mag)  &  (mag)     \\
&&\\
GaSS  1-1    &   117.834  &    14.248  &   13.078   \\
GaSS  1-2    &   117.878  &    14.374  &   12.899   \\
GaSS  1-3    &   102.882  &    14.067  &   12.497   \\
GaSS  1-9    &    94.496  &    15.429  &   13.321   \\
GaSS  1-11   &    95.543  &    13.825  &   12.083   \\
GaSS  1-12   &    96.546  &    13.874  &   12.122   \\
GaSS  1-13   &    97.497  &    15.605  &   13.641   \\
GaSS  1-14   &   129.530  &    14.792  &   12.976   \\
GaSS  1-15   &    97.549  &    13.838  &   12.092   \\
GaSS  1-16   &   124.865  &    15.125  &   13.385   \\
GaSS  1-17   &   120.852  &    13.137  &   11.419   \\
GaSS  1-21   &   130.986  &    15.921  &   14.010   \\
GaSS  1-23   &    94.546  &    15.419  &   13.301   \\
GaSS  1-25   &   122.525  &    13.441  &   11.541   \\
GaSS  1-26   &   146.505  &    15.427  &   13.723   \\
GaSS  1-27   &    96.496  &    12.746  &   11.213   \\
GaSS  1-28   &   128.526  &    15.415  &   13.566   \\
GaSS  1-30   &   124.526  &    12.964  &   11.213   \\
GaSS  1-31   &   125.504  &    13.007  &   11.193   \\
GaSS  1-32   &   126.502  &    13.336  &   11.473   \\
GaSS  1-33   &   130.531  &    17.076  &   15.403   \\
&&\\
\hline
\end{tabular}
\end{table}

\begin{table*}
        \centering
        \caption{The VVCep stars selected from the catalog of
	\citet{2020RNAAS...4...12P}.  The distance, height over the galactic
	plane and absolute infrared magnitude M(K$_S$) are computed from
	Gaia~$e$DR3 and 2MASS values.}
        \label{tab:vvcep}
\begin{tabular}{lcccrrrll}
\hline
&&\\
\multicolumn{1}{c}{name}&RA&DEC&\multicolumn{1}{c}{dist}&\multicolumn{1}{c}{z}&\multicolumn{1}{c}{K$_S$} 
&\multicolumn{1}{c}{M(K$_S$)} & \multicolumn{2}{l}{spectrum}\\
&&&(kpc)&(pc)&(mag)&(mag)&&\\
&&\\
     V641 Cas &  00 09 26.3 &    +63 57 14  &    2.9 &      75 &     1.7 &  $-$10.8 &    M3 & Iab      \\
       KN Cas &  00 09 36.4 &    +62 40 04  &    4.6 &      15 &     4.3 &   $-$9.3 &    M1 & Iab      \\
     V554 Cas &  01 10 20.1 &    +62 30 40  &    2.6 &   $-$13 &     2.7 &   $-$9.8 &    M2 & I        \\
       AZ Cas &  01 42 16.5 &    +61 25 16  &    3.3 &   $-$49 &     4.1 &   $-$8.9 &    M0 & Ib       \\
       XX Per &  02 03 09.4 &    +55 13 57  &    2.5 &  $-$274 &     2.0 &  $-$10.1 &    M4 & Ib       \\
   HDE 237006 &  02 49 08.8 &    +58 00 48  &    2.5 &   $-$59 &     3.1 &   $-$9.2 &    M1 & Ib       \\
       WY Gem &  06 11 56.2 &    +23 12 25  &    1.9 &      76 &     1.9 &   $-$9.7 &    M2 & Iab      \\
     V926 Mon &  07 02 06.7 &  $-$03 45 17  &    1.1 &      12 &     2.0 &   $-$8.2 &    M2 & Ib       \\
       KQ Pup &  07 33 48.0 &  $-$14 31 26  &    0.7 &      33 &     0.1 &   $-$9.3 &    M2 & Iab      \\
     V624 Pup &  08 00 41.4 &  $-$32 50 25  &    5.0 &  $-$125 &     4.2 &   $-$9.5 &    M2 & Iab      \\
       WY Vel &  09 21 59.1 &  $-$52 33 52  &    1.9 &   $-$61 &     0.4 &  $-$11.3 &    M3 & Ib:      \\
   HDE 300933 &  10 38 03.0 &  $-$56 49 02  &    3.1 &      79 &     1.8 &  $-$10.8 &    M2 & Iab/Ib   \\
     V730 Car &  10 44 57.3 &  $-$59 56 06  &    2.4 &   $-$36 &     2.5 &   $-$9.6 &    M1 & Iab      \\
    HD 101007 &  11 36 56.9 &  $-$61 10 58  &    2.1 &      15 &     2.0 &   $-$9.8 &    M3 & Ib       \\
     V772 Cen &  11 41 49.4 &  $-$63 24 52  &    2.3 &   $-$65 &     2.1 &   $-$9.9 &    M2 & Ib       \\
   CD-61.3575 &  12 44 16.1 &  $-$61 56 21  &    2.2 &      35 &     1.6 &  $-$10.3 &    M2 & Ia       \\
   CD-58.6089 &  15 34 41.4 &  $-$58 42 40  &    4.1 &  $-$163 &     3.4 &   $-$9.8 &    M2 & Ib       \\
 $\alpha$ Sco &  16 29 24.5 &  $-$26 25 55  &    0.2 &      46 &  $-$4.1 &  $-$10.3 &  M0.5 & Iab      \\
       FR Sct &  18 23 22.8 &  $-$12 40 52  &    2.4 &      14 &     2.1 &  $-$10.4 &  M2.5 & Iab      \\
     V381 Cep &  21 19 15.7 &    +58 37 25  &    1.8 &     201 &     0.8 &  $-$10.8 &    M1 & Ib       \\
       VV Cep &  21 56 39.1 &    +63 37 32  &    1.0 &     123 &  $-$0.0 &  $-$10.2 &    M2 & Ia-Iab   \\
   HDE 235749 &  22 11 35.7 &    +55 16 04  &    4.0 &   $-$55 &     3.3 &   $-$9.9 &    M2 & Ib       \\
        U Lac &  22 47 43.4 &    +55 09 30  &    4.2 &  $-$263 &     1.9 &  $-$11.4 &    M4 & Iab      \\
&&\\
\hline
\end{tabular}
\end{table*}

\clearpage


\clearpage
\bsp	
\label{lastpage}
\end{document}